
\magnification=\magstep1
\hoffset=-.5cm
\voffset=.7cm
\overfullrule=0pt
\def\no{{\noindent}}
\def\half{{1\over 2}}
\def\cn{{\{n_i\nu_i\}}}
\def\Re{{\rm Re}~}
\def\Im{{\rm Im}~}
\def\q{\quad}
\def\qq{\qquad}
\def\qqq{\qquad}
\def\vp{\varphi}

\rightline{Columbia/UCLA/94/TEP/39}

\bigskip

\centerline{{\bf  THE BOX GRAPH IN SUPERSTRING THEORY}
\footnote{$\dagger$}{ Research Supported in part by NSF grants PHY-92-18990
and DMS-92-04196.}}
\bigskip

\centerline{{\bf Eric D'Hoker}
\footnote{*} {E-Mail Address: DHOKER@UCLAHEP.PHYSICS.UCLA.EDU}}

\centerline{{\it Physics Department}}
\centerline{{\it University of California, Los Angeles}}
\centerline{{\it Los Angeles, California 90024-1547}}

\bigskip

\centerline{{\bf D. H. Phong}
\footnote{\#} {E-Mail Address: PHONG@MATH.COLUMBIA.EDU}}
\centerline{{\it Mathematics Department}}
\centerline{{\it Columbia University}}
\centerline{{\it New York, N.Y. 10027, USA}}

\bigskip

\centerline{\bf ABSTRACT}

In theories of closed oriented superstrings, the one loop amplitude is
given by a single diagram, with the topology of a torus. Its interpretation
had remained obscure, because it was formally real, converged only for purely
imaginary values of the Mandelstam variables, and had to account for the
singularities of both the box graph and the one particle reducible graphs
in field theories. We present in detail an analytic continuation method
which resolves all these difficulties. It is based on a reduction to
certain minimal amplitudes which can themselves be expressed in terms of
double and single dispersion relations, with explicit spectral densities.
The minimal amplitudes correspond formally to an infinite superposition
of box graphs on $\phi ^3$ like field theories, whose divergence is
responsible for the poles in the string amplitudes.
This paper is a considerable simplification and generalization of our
earlier proposal published in Phys. Rev. Lett. 70 (1993) p 3692.

\vfill\eject

\centerline{\bf I.  INTRODUCTION}

\bigskip

The quantization of the force of gravity consistent with causality,
Lorentz invariance, unitarity and renormalizability or finiteness is one
of the outstanding problems of contemporary theoretical physics.  The string
picture of elementary particles in the critical dimension has provided exciting
developments towards such a goal.  Critical superstring theory
automatically contains a massless graviton, as well as other particles,
which interact consistently with general coordinate invariance.
\footnote{$\dagger$} {For general reviews see [1,2].}
The rules of superstring perturbation theory ensure causality, unitarity and
Lorentz covariance [3,4], and there are strong indications that the scattering
amplitudes are finite to all orders in perturbation theory [1,2,3,4,5].

Unlike in quantum field theory though, the rules of perturbation theory do not
``manifestly'' exhibit finiteness or renormalizability.  In
quantum field theory, simple scaling arguments and recursive
combinatorics guarantee a simple physical picture of the property of
renormalizability in perturbation theory.  Feynman's $i\epsilon$
prescription on the particle propagators, together with Cutkowsky and
Landau cutting rules guarantee a simple picture of the property of
unitarity [6].  In string theory, unitarity of the amplitudes could be
established only indirectly, by showing equivalence of the Lorentz covariant
formulation with a manifestly unitary, but not manifestly Lorentz invariant
formulation in the lightcone gauge [3].  For the convergence issue, no such
indirect understanding is available at present.

In fact, even the nature of the most basic one-loop amplitude, say in
the Type II theory, for the scattering of four massless bosons, such as
the graviton, antisymmetric tensor and dilaton had not, until recently [7],
been completely elucidated.  Following the rules of Lorentz covariant
superstring perturbation theory, one finds a one-loop amplitude
represented by an integral over the positions of vertex operators of
incoming and outgoing states and moduli on the torus [1,2,8,9].  This
representation is modular invariant, but the integral appears to be
absolutely convergent only when the Mandelstam variables $s$, $t$ and
$u$ are all purely imaginary [7,10,11].
For real external momenta and in fact for any values of $s,~t$ and $u$
that are not purely imaginary, it is easy to see that the integral
is real and infinite
[7,12].

Both the reality and the divergence of the one-loop amplitude are
physically unacceptable.  The imaginary part of the (forward) one-loop
amplitude is related -- by the optical theorem -- to the absolute value
square of the tree level four point function.  Reality of the one-loop
amplitude for real momenta would imply the vanishing of the tree level
four point amplitude, which is in contradiction with its known
non-trivial expression.  In fact, reality and divergence of the integral
representation are manifestations of the same illness.  The integral
representation derived from the rules of string perturbation theory has
not been properly analytically continued in its dependence on external
momenta.  One would arrive at the same circumstance in quantum field
theory if one were to omit the $i\epsilon$ prescription in the Feynman
propagators.  Amplitudes constructed in standard perturbation theory
rules would now be real, but cease to be convergent as every frequency
integration would encounter a pole exactly on the real axis.  In effect,
the $i\epsilon$ prescription in quantum field theory instructs one on how
to perform the analytic continuation.

In string theory, the equivalent of the $i\epsilon$ prescription is
understood only at a formal level through the connection with the
light-cone gauge formulation [3,4].  Basically, the difficulty resides in the
fact that it is not so natural to exhibit all intermediate string
propagators in a dual diagram, though string field theory may
well be able to produce such a representation [13].

Another peculiar feature of string theory is that at each loop level,
there is a single diagram, which must then account for the singularities
of both the box and the one particle reducible diagrams in field theory.
This feature is a manifestation of duality and points to the subtleties
inherent to an $i\epsilon$ prescription in string theory.

In two recent letters [7], we have presented the crucial steps and basic
arguments for the construction of the analytic continuation of these
integral representations.  Specifically, we considered the amplitude for
the scattering of four external massless bosons, including the graviton,
dilaton and anti-symmetric tensor to 1-loop in the Type II superstring.
This is the simplest non-vanishing on-shell loop amplitude, and the
first non-trivial quantum loop amplitude which is both finite and
unitary.

In the present paper, we shall generalize the discussions and proof of
[7], derive the double spectral density for all branch cuts, and
present some of the proofs of the results stated in the second reference of
[7].   Arguments are presented in a simplified manner, which should also
allow for a reasonably straightforward generalization to the
case of higher point functions to one loop and to the case
of higher loops, though the latter is further complicated by
the presence of intricate spin structure issues.
We show that an analytic continuation exists, and we
presented explicit formulas for the singularities in the complex
momentum plane, in the form of poles in $s,~t$ and $u$ and branch cuts
along the positive real axis.  Specifically, these topics are organized as
follows.

\bigskip

In \S II, we recall standard results about tree-level and one-loop
Type II superstring amplitudes. We discuss the convergence
properties of the integral representation and stress that
absolute convergence is achieved only for purely imaginary
values of the Mandelstam parameters. We argue that assuming
Lorentz invariance and causality of the amplitudes,
S-matrix elements should be analytic functions of the
external momenta.
\footnote{*} {This hypothesis was elevated to an axiom
in analytic S-matrix theory [6], but as we shall establish here,
follows from string theory.}
The problem of properly extending the definition of the
amplitudes to physical momenta -- for which the Mandelstam
variables are real -- thus reduces to constructing an
analytic continuation in external momenta of the integral
representations.
The essence of this paper is in the proof that such a
continuation exists and in the explicit construction of
the analytically continued amplitudes.
By contrast, we point out that the integral representation
for the analogous amplitude in the bosonic  string is nowhere
convergent, and thus cannot allow for a unique analytic extension.
{}From this point of view, the bosonic string amplitudes remain ill-defined.

In \S III, we divide up the integration over moduli of the
torus -- including the positions of the vertex operators --
into 3 inequivalent regions which transform into one another
under duality [11,14]. We exhibit a formal equivalence of the
amplitudes with an infinite sum over $\phi ^3$ quantum field
theory box diagrams in which each propagator can have
a different mass. This formal equivalence is invalidated though by the
fact that the sum is not uniformly convergent,
\footnote{**} {A different proposal for constructing
the on-shell amplitudes is given in [15], but this difficulty is not addressed
there.}
which will be intimately related to the stringy nature of the
amplitudes and to duality. Next, we carefully isolate the
{\it minimal factors} that provide an obstruction to a
uniform expansion, so that the remaining factors admit
a uniform Taylor series expansion. This separation
is one of the crucial steps in our constructions.

In \S IV, the integral of each {\it minimal factor} -- as defined
above -- in turn is analytically continued. We do this by
re-expressing the integral in terms of a double dispersion
relation -- very analogous to the Mandelstam representation of [16] --
for which the spectral density and its support are
explicitly known. Once such a dispersion relation is
obtained and proper convergence has been established,
the analytic continuation problem becomes a straightforward
one, and the analytic properties of the {\it minimal amplitudes}
may be read off directly from the dispersion representation.
The analytic continuation of these {\it minimal factors} and
the construction of the double dispersion representations
and the explicit formulas for the spectral densities
provides the next crucial step in obtaining the
analytically continued amplitudes.

In \S V, we show that all one particle intermediate state
contributions -- which are well-known to arise in string
theory -- are contained in the double dispersion
representation of the minimal factors.
Their appearance is shown to be directly tied into
the lack of absolute convergence of the series in terms
of field theory box graphs, which is resolved by
retaining the {\it minimal factors}. Explicit formulas for
residues are obtained along the way.

In \S VI, we summarize the analogous analytic continuation results
for Heterotic superstrings [17].

In \S VII, we present some immediate applications of our
results which follow from direct use of the analytically continued
scattering amplitudes.
These include a precise and consistent $i\epsilon$ prescription
for the scattering amplitudes, a complete determination of the
decay rate of massive superstrings that couple in the four-point
function into 2 string states of lesser mass, and finally,
an explicit dicussion of the mass shift to one loop order
for the lowest mass state.
\footnote{$\dagger$} {Early dual model calculations of asymptotics
of decay rates can be found in [18], while more recent
results on decay widths and mass shifts are in [19].}

\medskip

There are six appendices, organized as follows.
In \S A, we present and prove a number of results on
analytic continuation of integrals on fixed tori, with
integrations over the positions of the vertex operators only.
The analytic behavior of this kind of contribution is
meromorphic in the Mandelstam variables.
In \S B, we present a compendium of formulas related
to hypergeometric functions and their integral transforms,
which are crucial to the analyticity properties of the
{\it minimal factors}. In \S C, we exhibit explicit formulas
for the expansion of the terms in a uniformly convergent
series.
In \S D, we review the analytic structure of the $\phi ^3$
field theory box graph with arbitrary masses, and relate
the support of the double spectral density to the ones
we find in string amplitudes.
In \S E, we present a more direct analytic continuation method
that is particularly convenient for the continuation
of the forward amplitudes with $t=0$.
Finally, in \S F, we give some useful analytic continuation
formulas, not yet covered by the preceding appendices.

The case of amplitudes with more than four external particles
is currently under investigation, and proceeds along completely
analogous lines [20].

\bigskip
\bigskip
\bigskip

\centerline
{\bf II.  STRING AMPLITUDES THROUGH ANALYTIC CONTINUATION}

\bigskip

In this section we present the integral representations of the superstring
amplitudes to one-loop provided by perturbation theory. For the 4-point
function, they are given by integrals over the moduli space of Riemann
surfaces with four punctures and the number of handles equal to the loop
order of perturbation theory. The integrand is built out of Green's functions.
We discuss the convergence of these representations,
and explain why analytic continuation is needed for physical values of the
external momenta. We also compare the issue of causality and analyticity
of the S-matrix in string theory with that of field theory.

\medskip
The amplitudes $A_\ell (k_i, \epsilon_ i)$ for the
scattering of four external massless bosons of momenta $k_i^\mu$ and
polarization tensor $\epsilon_i^{\mu\nu}(k_i)$ with $\ell = 0$ or $\ell = 1$
loops in the Type II superstring are of the form [8]
$$
A_\ell  (k_i, \epsilon_i ) =
(2\pi)^{10} \delta(k)g^4 A_\ell (s,t,u)
K_{\mu_1\mu_2 \mu_3 \mu_4} K_{\bar \mu_1 \bar \mu_2 \bar \mu_3 \bar \mu_4}
\prod _{i=1} ^4 \epsilon_i ^{\mu_i\bar\mu_i} (k_i)
\eqno(2.1)
$$
Here $g$ is the string coupling constant, $k=k_1+k_2+k_3+k_4$ and
the external states are characterized by the on-shell conditions
\footnote{*}{Repeated Lorentz indices are to be
summed over throughout, whereas repeated particle identification
indices should not be summed over.}
$$
k_i^\mu k_i^\mu = 0 \qq k_i^\mu \epsilon_i^{\mu\nu} (k_i) = k_i^\nu
\epsilon_i^{\mu\nu} (k_i) = 0
\eqno(2.2)
$$
The remaining kinematical invariants can be expressed in terms of
the Mandelstam variables (with $s+t+u = 0$)
$$
\eqalign{
& s = s_{12} = s_{34} = - (k_1 + k_2 )^2\cr
& t = s_{23} = s_{14} = - (k_2 + k_3 )^2\cr
& u = s_{13} = s_{24} = - (k_1 + k_3 )^2 \cr}
\eqno(2.3)
$$
The kinematical factor  $K$
is a polynomial in momenta and given by
$$
\eqalign{
K_{\mu_1 \mu_2 \mu_3 \mu_4} =
& -(st \eta _{13} \eta _{24} + su \eta _{14} \eta _{23}
   + tu \eta _{12}\eta _{34}) \cr
& +s(k_1^4 k_3^2 \eta _{24} + k_2^3 k_4^1 \eta _{13}
   + k_1^3 k_4^2 \eta _{23} + k_2^4 k_3^1 \eta _{14} ) \cr
& +t(k_2^1 k_4^3 \eta _{13} + k_3^4 k_1^2 \eta _{24}
   + k_2^4 k_1^3 \eta _{34} + k_3^1 k_4^2 \eta _{12} )\cr
& +u(k_1^2 k_4^3 \eta _{23} + k_3^4 k_2^1 \eta _{14}
   + k_1^4 k_2^3 \eta _{34} + k_3^2 k_4^1 \eta _{12} )\cr}
\eqno(2.4)
$$
Here superscripts on the momenta label the external string states,
and Lorentz indices $\mu_i$ have been abbreviated by subscripts $i$ only.
This amplitude is schematically shown in Fig. 1.
The crucial factors in (2.1) are thus $A_{\ell}(s,t,u)$, which we describe
next.

\medskip
\noindent
{\it Tree Level Amplitudes}

To tree-level, the amplitude is given by an integral
representation derived in [8] which can be expressed in terms
of $\Gamma$-functions :
$$
\eqalign{
A_0(s,t,u)  =& - {2\over u^2} \int\nolimits  d^2z |z|^{-s-2}
|1-z|^{-u}\cr
            =& \pi  {\Gamma(-s/2)\Gamma(-t/2)\Gamma(-u/2)\over
\Gamma(1+s/2) \Gamma(1+t/2)\Gamma(1+u/2)}
\cr }
\eqno(2.5)
$$
Meromorphicity of the $\Gamma$ function with simple poles at
negative or zero integers implies the appearance of simple poles
in $s$, $t$ or $u$ at positive even integers in $A_0$.
Physically, these poles correspond to intermediate physical
massive string states and residues are given by
$$
A_0(s,t,u) {{{~}\atop{\displaystyle \sim \atop\displaystyle
s\rightarrow 2n }}}
{1\over s-2n} {8\pi\over t^2} \bigl \{ C_n(t)\bigr \} ^2 \qquad\qquad
C_n(t)  \equiv  {\Gamma(t/2 + n)\over \Gamma(t/2) \Gamma(n+1)}
\eqno(2.6)
$$
The function ${2\over t}C_n(t)$ corresponds to the tree-level
three-point amplitude of an intermediate state of mass $2n$, and two
external massless states, and will play an important role also in
our study of the one-loop amplitudes.

It is important to realize though that even this simplest situation
required an analytic continuation in $s,t$ and $u$.
As it stands, the integral in (2.5) has singularities at
$z=0,1,\infty$, and is absolutely convergent only for
$$
\Re(s) <0, \q \Re(t) < 0\q \Re(s) + \Re(t) > -2
\eqno(2.7)
$$
In this region, $A_0(s,t,u)$ is holomorphic. The poles
and the definition of $A_0$ beyond this region must be obtained
through analytic continuation.
The expression (2.5)
in terms of $\Gamma$-functions can be viewed as a solution to
this analytic continuation problem.

Without the explicit $\Gamma$-function formulas, we can still construct an
analytic continuation for the integral (2.5) in the following way.
By setting $x=|z|$ in (2.5), we can reduce the problem
to the analytic continuation
of a Mellin transform, given by the integral
$$
\int_0^{\delta}dx \,x^{-1-s}f(x^2)
\eqno(2.8)
$$
where $f$ is a smooth function, and $\delta$ is a fixed positive cut-off.
The integral (2.8) is absolutely convergent for $\Re (s) <0$ and is holomorphic
in this region. If we expand $f(x^2)$ in a finite Taylor expansion
with remainder $x^{2N} R_N(x^2)$, the analytic
continuation of this integral to a larger region $\Re (s) <2N$ is given by
$$
\int_0^{\delta}dx\, x^{-1-s}~f(x^2)=-\sum_{n=0}^{N-1}{f^{(n)}(0)\over n!}
{\delta^{2n-s} \over
s-2n}+ \int_0^{\delta}dx~ x^{-1+2N-s}R_N(x^2)
\eqno(2.9)
$$
The remainder integral is absolutely convergent for  $\Re (s)<2N$ and thus
holomorphic in this region. The pole terms are the only
singularities of $A_0(s,t,u)$ in $s$ for $\Re (s) <2N$.
Their residues are $-f^{(n)}(0)/n!$.
The number $N$ is arbitrary in this argument and so
given a region, we can find all singularities in this region,
plus a holomorphic remainder term.
This simple example will also serve as a prototype for the more
involved analytic continuations to be carried out in this paper.

\medskip
\noindent
{\it One loop Type II Amplitudes}

To one-loop level, the amplitude was first derived by Green and Schwarz [8,9],
and admits the integral representation
$$
A_1(s,t,u) = \int\nolimits_F {d^2\tau\over \tau_2^2}
\int\nolimits_{M_{\tau}} \prod_{i=1}^4 {d^2z_i\over \tau_2} \prod_{i<j}
\exp \big ( \half s_{ij} G(z_i, z_j) \big )
\eqno(2.10)
$$
This integral arises from the insertion of four vertex operators at
points $z_i$ on the torus $M_\tau$, with corners at $0,1,\tau$ and
$1+\tau$.  Here $\tau = \tau_1 + i \tau_2$ is the complex modulus of the torus,
and runs through the fundamental domain
$$
F = \biggl \{ \tau \in {\bf C}, \q \tau_2 > 0, \q |\tau_1|\leq
\half,\q |\tau|\geq 1\biggr \}
\eqno(2.11)
$$
The Mandelstam variables $s_{ij}$ were defined in (2.3) and $G(z,w)$ is
the Green function for the Euclidean metric on the torus $M_\tau$,
defined by
$$
-2\partial_z \partial_{\bar z} G(z,w) = 4\pi \delta(z-w) - {4\pi\over
\tau_2}
$$
and expressed in terms of Jacobi $\vartheta$-functions
$$
G(z, w) = - {\rm ln}\,\biggl |{\vartheta_1(z-w,\tau)\over
\vartheta_1'(0,\tau)}\biggr |^2 +
{2\pi\over \tau_2} \bigl \{ \Im (z-w) \bigr \} ^2
\eqno(2.12)
$$
An explicit evaluation of the integral (2.10) in terms of standard
special functions is not known, and thus the integral representation
is our only way to define the one loop amplitude.

\medskip
\noindent
{\it Qualitative Analysis of Convergence}

There are two types of singularities in the integral (2.10), which are
schematically depicted in Fig. 2.

\item{(a)} When $z_i \sim z_j$, the Green function diverges :
$G(z_i,z_j) \sim -\ln |z_i -z_j|^2 \rightarrow +\infty$ and produces a
non-integrable singularity when $\Re (s_{ij}) \geq 2$. This type of
singularity occurs locally on the worldsheet when two vertex
operators come close together, and -- like at tree level --
corresponds to a massive intermediate string state.

\item{(b)} When $z_i$ and $z_j$ are well separated, but $\tau _2
\rightarrow \infty$ the Green function diverges : $G(z_i, z_j) \sim
-\tau _2 \rightarrow - \infty$ and produces a non-integrable
singularity when $\Re (s_{ij}) <0$. This type of singularity
occurs when the torus degenerates to a thin wire where
intermediate string states go on mass-shell. It is ultimately
responsible for branch cuts in the amplitude.
\medskip
More subtle singularities occur when simultaneously $z_i\sim z_j$
and $\tau_2\rightarrow\infty$, but we shall see shortly that the constraints
imposed by the basic types of singularities already determine completely
the range of values $s_{ij}$ for which we have convergence.
In fact convergence of the singularities (a) and (b) requires respectively that
$\Re (s_{ij}) <2$ and $\Re (s_{ij}) \geq 0$
for all $s_{ij}$. But $s+t+u=0$, and thus we must have
$$
\Re (s) = \Re (t) = \Re (u) =0
\eqno(2.13)
$$
Conversely, the integral converges for purely imaginary
$s,t,u$. Indeed the exponentials become pure phases since $G(z,w)$
is real, so that the full amplitude $A_1$ is bounded by the
volume of the moduli space for the torus, which is finite.
Thus the condition (2.13) completely specifies the values of $s_{ij}$ for
which $A_1(s,t,u)$ is well-defined through the integral representation of
(2.10).

\medskip

\noindent
{\it Comparison with the Bosonic String}

It is instructive to compare the above convergence analysis
with that for the bosonic string.
Actually, expression (2.10) closely resembles the 1-loop
four tachyon amplitude in the bosonic string
$$
A_1^{{\rm bosonic}}(s,t,u) = \int\nolimits_F {d^2\tau\over \tau_2^2} ~Z_1~
\int\nolimits_{M_{\tau}} \prod_{i=1}^4 {d^2z_i\over \tau_2}
\exp \big ( \half s_{ij} G(z_i, z_j) \big )
\eqno(2.14)
$$
where $Z_1$ is the modular invariant combination of scalar
and ghost determinants with their zero mode contributions included :
$$
\eqalign{
Z_1 (\tau ) = & \tau _2 ^{-12} |\eta (\tau )|^{-48} \cr
\eta (\tau ) = & e^{{i\pi \tau \over 12}} \prod _{n=1} ^\infty
(1 - e^{2\pi i \tau} ) \cr }
\eqno (2.15)
$$
The tachyon on-shell conditions imply $s+t+u=-8$ in the bosonic case.
The analysis for singularities of type (a) above is unchanged.
For singularities of type (b), we have two additional contributions.
First, as $\tau _2 \rightarrow \infty$, $Z_1$ diverges exponentially :
$$
Z_1 \sim \tau _2 ^{-12} e^{4 \pi \tau _2} ( 1 + {\cal O} (e^{-2\pi \tau _2}))
\eqno (2.16)
$$
and second, since now $s+t+u=-8$, at least some $s_{ij}$ has to
be negative, contributing an extra exponential divergence in $\tau _2$.
We know of course that these exponential divergences are the
result of the presence of the tachyon in the spectrum.
It follows that $A_1 ^{{\rm bosonic}} (s,t,u)$ is divergent
for {\it all} values of $s,t$ throughout the complex plane.
There is no starting point for an
analytic continuation in $s,t,u$, and no acceptable definition for the on-shell
scattering amplitude of lowest mass particles in the bosonic string.

\bigskip
\noindent
{\it Lack of Proper Definition of Amplitudes near Physical Momenta}

The main problem of superstring loop amplitudes addressed in this
paper is now clear.
The integral representation (2.10) converges and thus properly
defines the loop amplitude $A_1$ only for unphysical values
of the external momenta : when $s,t,u$ are purely imaginary.
Physical momenta are when $s,t,u$ are real, negative for
the deep inelastic or Euclidean regime and positive for the
physical state pair creation regime.
But around real $s,t,u$ the integral representation is divergent
and does not define the amplitude $A_1$.

A second related symptom of the same problem is that for real
$s,t,u$ the entire integral representation is real, and
$A_1$ would be real.
But assuming that the optical theorem holds, the imaginary part
of the amplitude equals the sum over states of the absolute value
squared of the tree level amplitude, and the latter is non-zero.
In general, in an interacting theory, the four point function
must have an imaginary part. For this reason as well, the
integral representation cannot properly define the one-loop
amplitude near physical momenta.

\bigskip
\noindent
{\it Causality and Analytic Structure of Amplitudes}

In view of the shortcomings of the integral representation,
pointed out above, our task is to properly define and construct
a one-loop amplitude throughout the complex $s,t,u$ planes
($u=-s-t$), including around physical values of $s,t,u$.

The only possibility is to analytically continue the integral
representation throughout the complex planes, starting from
purely imaginary $s,t,u$. This is what was done at tree-level.
To appreciate the physical meaning of such analytic continuations,
we compare the situation with that of local quantum field theory.
Here causality is equivalent to locality of observable
fields and this property combined with Lorentz invariance
implies analytic dependence of the Green functions of the
local fields on the external momentum variables.
The standard way to establish analyticity is to use
Lorentz invariance and locality to derive a spectral
representation for the amplitudes. The simplest case is the
K\"allen-Lehmann representation for the off-shell
two point function (repesented in Fig.3), say in a scalar field theory [21] :
$$
G(s) = \int _0 ^\infty d M^2 { \rho (M^2) \over s-M^2 + i \epsilon}
$$
The integration is only over $M^2\geq 0$, as the existence
of tachyonic intermediate states with $M^2 <0$ would automatically violate
causality. Provided this spectral representation is convergent,
it immediately follows that the two point function is analytic
in $s$, with singularities only on the positive real axis.
These singularities may be poles resulting from one-particle
intermediate states or branch cuts arising from multi-particle
states. In field theory, analyticity may be shown for other
amplitudes as well, and in analytic S-matrix theory, analyticity
has been elevated to an axiom [6].

In string theory, a formulation of the amplitudes in terms of
local field observables, commuting (or anti-commuting)
at space-like separations
is not at present available. It is not known whether analyticity
of the amplitudes can be derived solely from the requirements
of Lorentz invariance and causality (or locality of the
string interactions). Thus, analyticity of the string amplitudes
must be established, and our goal in this paper is to do so,
starting from the integral representation.

More specifically, we must address the following important questions :

\item{(a)} Does an analytic continuation exist ?
\item{(b)} Is it unique ?
\item{(c)} Is the analytic continuation physically acceptable ?

We have outlined in [7] the basic methods for the construction of
an analytic continuation in the cut plane
$s,t,u \in {\bf C}\setminus {\bf R}_+$
with $s+t+u=0$ and shown that there are no gaps in the domain of
holomorphicity.
In this paper we present all the details of the construction as well.
Uniqueness is guaranteed by the fact that two analytic
continuations have to agree on purely imaginary $s,t$ and thus
must be the same.

Even if the unique analytic continuation has been constructed, there
still remains the question as to whether it is consistent
with all physical principles. On physical grounds, we expect
the singularities of the 4-point function in the Type II
superstring to consist of branch cuts along the positive
real axis starting at positive even integers.
In addition, there may be simple and double poles in $s,t,u$ at
positive even integers, possibly on top of branch cuts.
These singularities  and their field theory analogues are
presented in Fig. 4.
The appearence of any other type of singularity or gap
would violate unitarity, causality or Lorentz inariance.

\bigskip
\bigskip
\bigskip

\centerline{\bf III. CONTRIBUTIONS TO THE BRANCH CUTS}

\bigskip

We describe now the method of analytic continuation for the 4-point
graviton amplitude $A_1(s,t,u)$, which we reproduce here for convenience
$$
A_1(s,t,u) = \int\nolimits_F {d^2\tau\over \tau_2^2}
\int\nolimits_{M_{\tau}} \prod_{i=1}^4 {d^2z_i\over \tau_2} \prod_{i<j}
exp \big ( \half s_{ij} G(z_i, z_j) \big )
\eqno(3.1)
$$

\bigskip

\noindent
{\it Real Time Evolution Coordinates for the 4-Point Function}

By translation invariance, we set $z_4 = 0$.  The integration region in
the $z_i$'s naturally decomposes into 6 regions according to the various
orderings of $0 \leq  \Im (z_1), ~\Im (z_2)$, $\Im (z_3) \leq \tau _2 $. The
contributions to the amplitude are the same for pairs of regions, and we
obtain [7,11,14]
$$
A_1(s,t,u) = 2 A_1(s,t) + 2A_1(t,u) + 2A_1 (u,s)
\eqno (3.2)
$$
with $A_1(s,t)$ given by (3.1), but with the restriction
$$
0 \leq \Im (z_1) \leq  \Im (z_2) \leq \Im (z_3) \leq \tau_2
\eqno(3.3)
$$
and $u = - s-t$. Thus it suffices to analytically continue $A_1(s,t)$, and
this is the problem to which we can now restrict ourselves
{}.
\medskip

We need to express the integrand of
$A_1(s,t)$ in a form where its singularities as $z_i-z_j\rightarrow 0$
and as $\tau_2\rightarrow\infty$ can be readily extracted. Now $G(z,w)$ is
given by (2.12)
and $\vartheta_1(z,\tau)$ admits the following product expansion (here we
set as usual $q = exp \{ 2\pi i \tau \}$)
$$
\vartheta_1(z,\tau)
=-iq^{1/8}e^{-i\pi z} \prod _{n=0}^{\infty}(1-q^ne^{2\pi iz})
(1-q^{n+1}e^{-2\pi iz}) (1-q^n)
\eqno(3.4)
$$
In view of overall momentum conservation, we may ultimately ignore all $z_i$
independent factors arising from (3.4) in the product
$\prod_{i<j}exp(-\half s_{ij}G(z_i,z_j))$. Thus the Green's function $G(z,0)$
can effectively be rewritten as
$$
exp(-G(z,0))
=exp\bigg(-{2\pi\over\tau_2}(\Im (z))^2- 2 \pi \Im (z) \bigg )
\biggl |\prod_{n=0}^{\infty}(1-q^ne^{2\pi iz})(1-q^{n+1}e^{-2\pi iz})
\biggr |^2
\eqno(3.5)
$$
It is convenient to introduce the following variables $w_{ij}$
$$
w_{ij}=\cases{e^{2\pi i(z_i-z_j)}, &Im$(z_i-z_j)>0$\cr
qe^{2\pi i(z_i-z_j)}, &Im$(z_i-z_j)<0$\cr}\eqno(3.6)
$$
for $i\not=j$. This definition has been arranged so that $|w_{ij}|\leq1$
in the region (3.3) and $w_{ij}w_{ji}=q$ for any $i\not=j$.
The infinite products in $\prod_{i<j}exp(-\half s_{ij}G(z_i,z_j))$
resulting from the infinite product in (3.5) can now be gathered in a single
expression ${\cal R}(w_{ij})$
$$
{\cal R}(w_{ij})
=\prod_{i\not=j}^4\prod_{n=0}^{\infty}|1-w_{ij}q^n|^{-s_{ij}}\eqno(3.7)
$$
To proceed further we need the real vertex coordinates $(x,y)$ defined by
$$
z=x+\tau y\eqno(3.8)
$$
Thus $0\leq x,y\leq 1$, and we parametrize all tori $M_{\tau}$ by
the same square of unit size. Set now
$$
\eqalign{
&u_1 = y_1\cr
&u_2  = y_2-y_1\cr
& u_3 = y_3 - y_2\cr
&u_4 = 1 - y_3 \cr}\qq
\eqalign{
&\alpha _1 = 2 \pi (x_1       + u_1 \tau_1) \cr
&\alpha _2 = 2 \pi (x_2 - x_1 + u_2 \tau_1) \cr
&\alpha _3 = 2 \pi (x_3 - x_2 + u_3 \tau_1) \cr
&\alpha _4 = 2 \pi \tau_1 - \alpha_1 - \alpha_2 - \alpha_3\cr}\qq
\eqno(3.9)
$$
Evidently $0\leq u_i\leq 1$, $i=1,\cdots,4$ in the region of integration
(3.3) for $A_1(s,t)$. The advantage of the variables $u_i$'s is that
the exponentials on the right hand side of (3.5)
combine into a remarkably simple expression when we form
$\prod_{i<j}exp(-\half s_{ij}G(z_i,z_j))$
$$
\prod_{i<j}exp(-\half s_{ij}G(z_i,z_j))=|q|^{-(su_1u_3+tu_2u_4)}{\cal
R}(w_{ij})
\eqno(3.10)
$$
Since the original complex variables $z_i$'s, $i=1,2,3$, can be replaced
by any three of the $u_i$'s and any three of the $\alpha_i$'s, we arrive
at the following expression for the amplitude $A_1(s,t)$ which we
take as starting point for our analytic continuation method:
$$
\eqalign{
A_1(s,t) = \int\nolimits_F {d^2\tau\over \tau_2^2} & \int\nolimits_0^{2\pi}
\prod_{i=1}^4 {d\alpha_i\over 2\pi} \cr
&\delta(2\pi\tau_1-\sum_{i=1}^4\alpha_i) \int\nolimits_0^1 \prod_{i=1}^4 du_i
  |q|^{-(su_1u_3 + tu_2u_4)} {\cal R} (w_{ij}) \cr }
\eqno (3.11)
$$
\bigskip
\noindent{\it Isolating Branch Cuts}

We have seen in the discussion of the tree-level string amplitudes \S II
that it is not difficult to construct analytic continuations when they are just
meromorphic in $s$ and $t$.
Thus our method for constructing an analytic continuation for $A_1(s,t)$
consists of two steps. In the first, we strip away all meromorphic terms,
reducing the integrand of $A_1(s,t)$ to a much simpler expression; in the
second, an explicit analytic continuation for the remaining integral is
provided
under the form of a double dispersion relation. This dispersion relation is
the only expression exhibiting branch cuts. The stripping process has to be
done with some care, since there are subtle convergence problems which
reflect precisely the stringy aspects of the scattering process.

\medskip

The precise statement that we need is the following. First we note that
$$
q=\prod_{i=1}^4|q|^{u_i}e^{i\alpha_i}
\eqno(3.12)
$$
Similarly all the $w_{ij}$'s can be expressed in terms of products of
$|q|^{u_i}e^{i\alpha_i}$
$$
\eqalignno{
w_{21}&=|q|^{u_2}e^{i\alpha_2},\quad
w_{14}=|q|^{u_1}e^{i\alpha_1},\quad
w_{32}=|q|^{u_3}e^{i\alpha_3},\quad
w_{43}=|q|^{u_4}e^{i\alpha_4}\cr
w_{13}&=|q|^{u_1+u_4}e^{i\alpha_1+\alpha_4},\quad
w_{24}=|q|^{u_2+u_4}e^{i\alpha_2+\alpha_4}&(3.13)\cr}
$$
and we may view ${\cal R}(w_{ij})$ as a function on $|q|^{u_i}$,
 $e^{i\alpha_i}$, $i=1,\cdots,4$
$$
{\cal R}(w_{ij})={\cal R}(|q|^{u_i},\alpha_i;s,t)\eqno(3.14)
$$
Of particular importance are the following four factors occurring in the
infinite product expansion for ${\cal R}(|q|^{u_i},\alpha_i;s,t)$
$$
\prod_{i=1}^4|1-e^{i\alpha_i}|q|^{u_i}|^{-s_i}
\eqno(3.15)
$$
where we have introduced the notation $s_i=s$ for $i$ even and $s_i=t$ for
$i$ odd. All other factors besides (3.15) contain $w_{ij}q^n$ which
contain at least two factors $|q|^{u_i}e^{i\alpha_i}$ when reexpressed
in terms of these variables. We now define the coefficients
$P_{\{n_i\nu_i\}}^{(4)}(s,t)$ as the ones occurring
in the series expansion for all these remaining factors
\footnote*{The upper index (4) is a reminder of how many factors
in ${\cal R}$ are not expanded in the Taylor series.
Later on we shall also use coefficients
with a different number of factors retained.}
$$
{\cal R}(|q|^{u_i},\alpha_i;s,t)
=\prod_{i=1}^4 \bigl |1-e^{i\alpha_i}|q|^{u_i} \bigr
 |^{-s_i}\sum_{n_i=0}^{\infty}\sum_{|\nu_i|\leq
n_i}P_{\{n_i\nu_i\}}^{(4)}(s,t)\prod_{i=1}^4|q|^{n_iu_i}e^{i\nu_i\alpha_i}
\eqno(3.16)
$$
The coefficients
$P_{\{n_i\nu_i\}}^{(4)}(s,t)$ are polynomials in $s$ and $t$, and
can be generated recursively by the relations given in Appendix C.

\bigskip

\noindent{\bf Theorem 1}. {\it For any positive integer} $N$, {\it we can
write the partial amplitude as}
$$
A_1(s,t)
= M_N(s,t) +
\sum_{n_1+\cdots+n_4\leq 4N} ~ \sum_{|\nu_i|\leq
 n_i}P_{\{n_i\nu_i\}}^{(4)}(s,t)A_{\{n_i\nu_i\}}(s,t)
\eqno(3.17)
$$
{\it where} $M_N(s,t)$ {\it is a meromorphic function in the region}
$Re(s),Re(t)<N$,
{\it and the
amplitudes} $A_{\{n_i\nu_i\}}(s,t)$ {\it are defined by}
$$
\eqalignno{A_{\{n_i\nu_i\}}(s,t)
=\int_1^{\infty}{d\tau_2\over\tau_2^2}&\int_0^{2\pi}\int_0^1
\prod_{i=1}^4{d\alpha_i
du_i\over 2\pi}
\delta(1-\sum_{i=1}^4u_i)|q|^{-su_1u_3-tu_2u_4}\cr
&\times\prod_{i=1}^4 \bigl |1-e^{i\alpha_i}|q|^{u_i}\bigr |^{-s_i}|q|^{n_iu_i}
e^{i\nu_i\alpha_i}&(3.18)\cr}
$$
\bigskip
Thus Theorem 1 reduces the analytic continuation of the amplitude $A_1(s,t)$
to the analytic continuation of the much simpler amplitudes
$A_{\{n_i\nu_i\}}(s,t)$, where there are no longer infinite products, and
where all angle $\alpha_i$-dependence is decoupled.

\bigskip

\noindent{\it Box Diagrams in Field Theory and String Theory}

Before giving
the proof of Theorem 1, we pause to discuss carefully the
convergence issues behind the $partial$ expansion (3.16), issues which
prevent us from using the $full$ expansion of all the factors in ${\cal
 R}(|q|^{u_i},\alpha_i;
s,t)$. The situation can be illustrated clearly by a comparison with
the box diagrams in a $\phi^3$-like field theory.
Suppose we had expanded all factors in ${\cal R}(|q|^{u_i},\alpha_i;s,t)$
into a series in $|q|^{u_i}e^{i\alpha_i}$
$$
{\cal R}(|q|^{u_i},\alpha_i;s,t)
=\sum_{n_i=0}^{\infty}\sum_{|\nu_i|\leq n_i}P^{(0)}_{\{n_i\nu_i\}}(s,t)
\prod_{i=1}^4|q|^{n_iu_i}e^{i\nu_i\alpha_i}
\eqno(3.19)
$$
The coefficients $P^{(0)}_{\{n_i\nu_i\}}$ are still polynomials in $s$ and $t$.
Formally, this expansion leads to an expansion for $A_1(s,t)$ into terms
of the form
$$
\eqalignno{A^{(0)}_{\{n_i\nu_i\}}(s,t)
=\int_F{d^2\tau\over\tau_2^2}\int_0^{2\pi}\prod_{i=1}^4
{d\alpha_i\over 2\pi}&\delta(2\pi\tau_1-\sum_{i=1}^4\alpha_i)
\int_0^{1}
\prod_{i=1}^4du_i\delta(1-\sum_{i=1}^4u_i)\cr
&\times|q|^{-su_1u_3-tu_2u_4}
\prod_{i=1}^4 |q|^{n_iu_i}
e^{i\nu_i\alpha_i}&(3.20)\cr}
$$
We shall see later that we may truncate the $\tau$ domain of
integration $F$ in this integral
to the simpler domain $\{\tau_2\geq 1, |\tau_1|\leq 1/2\}$ without affecting
the branch cuts nor the poles lying on top of them.
The variables $\tau_1$ and $\alpha_i$, $i=1,\cdots,4$ can then be all
integrated
out, leaving an integral of the form
$$
\int_1^{\infty}{d\tau_2\over\tau_2^2}
\int_0^1\prod_{i=1}^4du_i\delta(1-\sum_{i=1}^4u_i)
e^{2\pi\tau_2(su_1u_3+tu_2u_4-\sum_{i=1}^4n_iu_i)}
\eqno(3.21)
$$
where the integers $n_i$ are positive and even.

Consider now a $\phi^3$-like box diagram in $d$ dimensional space-time
quantum field theory with masses $m_i^2$ for each of the propagators,
and massless external on-shell states, as depicted in Fig. 5.
Couplings are identical for
all $m_i$, and are non-derivative $\phi^3$. The box diagram (Euclidian)
Feynman integral can be performed as usual after introducing
Feynman parameters $u_i$ and exponentiating the denominator (see
Appendix D). We obtain
$$
\int d^dk\prod_{i=1}^4{1\over (k+p_i)^2+m_i^2}
=\int_0^{\infty}{d\tau\over\tau^{{d\over 2}-3}}
\int_0^1\prod_{i=1}^4du_i\delta(1-\sum_{i=1}^4u_i)
e^{2\pi\tau(su_1u_3+tu_2u_4-\sum_{i=1}^4u_im_i^2)}
\eqno(3.22)
$$

In space-time dimension $d=10$ and with the masses $m_i^2$ positive even
 integers
(as dictated by the superstring spectrum), the $\phi^3$-like box diagram
is essentially the same as the partial superstring amplitude of (3.21).
A well-understood difference is that the $d\tau_2$ integral in the superstring
amplitude is truncated away from 0. This is a remnant of duality [9].
Duality implies modular invariance, which restricts the domain of
integration in $A_1(s,t)$ to a fundamental domain for $SL(2,{\bf Z})$,
and insures ultra-violet finiteness. Of interest to us is rather
the issue of how the poles on top of branch cuts expected in string theory
(and absent in an infinite superposition of $\phi^3$ like theories) can emerge.
The fact that the summands are of the same form and have no poles shows that
 there must
be difficulties in resumming the series resulting from (3.19).
These convergence issues are important manifestations of the difference between
string theory and mere infinite superpositions of field theories.

\bigskip

\noindent{\it Singularities of the Green's function}

We turn now to a precise analysis of the problem of convergence, leading
 ultimately
to the proof of Theorem 1. First the product expansion for the Green's function
can be expressed in terms of the real coordinates $(x,y)$ as
$$
exp(-G(z,0))
={1\over 4\pi^2}
|q|^{y^2-y}|1-q^ye^{2\pi ix}|^2|1-q^{1-y}e^{-2\pi ix}|^2R(z)\eqno(3.23)
$$
with
$$
R(z)=\prod_{n=1}^{\infty}|1-q^{n+y}e^{2\pi ix}|^2|1-q^{n+1-y}e^{-2\pi ix}|^2
|1-q^n|^{-4}
\eqno(3.24)
$$
Since $|q|<e^{-2\pi\sqrt 3}$ in the fundamental domain $F$, the function
$R(z)$ is uniformly bounded from above and below in all variables. Thus the
singularities of $G(z,w)$ as $z\rightarrow w$ and/or $\tau\rightarrow\infty$
are described precisely by the first three factors in (3.23). The expansion
(3.23) leads to a further splitting in the factorization
(3.10) for the integrand of $A_1(s,t)$
$$
{\cal R}(|q|^{u_i},\alpha_i;s,t)
={\cal I}(|q|^{u_i},\alpha_i;s,t){\cal R}^{*}(|q|^{u_i},\alpha_i;s,t)
\eqno(3.25)
$$
with
$$
\eqalignno{
{\cal I}(|q|^{u_i},\alpha_i;s,t)=
& \prod_{i=1}^4 \bigl |1-e^{i\alpha_i} |q|^{u_i}  \bigr |^{-s_i}
\bigl |1-\prod_{j\not=i}e^{i\alpha_j} |q|^{u_j}  \bigr |^{-s_i}\cr
&\times\prod_{i=2k}\prod_{j=2m+1} \bigl |1-e^{i(\alpha_i+\alpha_j)}
|q|^{u_i+u_j}\bigr |^{s_i+s_j}\cr
{\cal R}^{*}=&\bigg ({R(z_2-z_1)R(z_3)\over R(z_3-z_1)R(z_2)}\bigg )^{-s}
\bigg ({R(z_3-z_2)R(z_1)\over R(z_3-z_1)R(z_2)}\bigg )^{-t}&(3.26)\cr}
$$
As in the case of (3.24) for the Green's function, ${\cal R}^{*}$
is uniformly bounded from above and below in all variables, and the 12 factors
in ${\cal I}$ are the only ones incorporating singularities in ${\cal R}$.
The expansion (3.25) shows immediately that the integrations over the vertex
operator locations in $A_1(s,t)$ converges only in the infinite strip
$$
\Re (s),\Re (t)\leq 0,\ \Re (s+t)>-2
\eqno(3.27)
$$
Conversely, in this region, the last factor $|q|^{-(su_1u_3+tu_2u_4)}$
remains bounded as $\tau_2\rightarrow\infty$, so that the $d^2\tau$
integral is finite as well. Thus we have established that (3.27)
is the domain of convergence of $A_1(s,t)$. Note, however, that
all three expressions on the right hand side of
(3.1) - and hence the full amplitude
$A_1(s,t,u)$ - are simultaneously convergent only when $\Re(s)=\Re(t)=\Re(u)=0$
when the constraint $s+t+u=0$ is enforced.
This confirms what we foresaw in \S II from a mere qualitative analysis.

\bigskip

\noindent{\it Proof of Theorem 1}

The factorization (3.25)-(3.26) shows that the full factor ${\cal R}$
cannot be expanded as a uniformly convergent series in the variables
$|q|^{u_i}e^{i\alpha_i}$, as we had anticipated on physical grounds in our
 earlier comparison
of string theory with field theory. The only factor which can be expanded
uniformly is the factor ${\cal R}^{*}$, which leaves us with
the seemingly intractible 12 factors of ${\cal I}$. The key observation behind
Theorem 1 is that actually the eight factors in ${\cal I}$ involving
$composites$ of $|q|^{u_i}e^{i\alpha_i}$ can also be expanded, at the harmless
expense of discarding purely meromorphic terms which do not
affect the behavior of the amplitude near the branch cuts.

To see this,
we begin by observing that for fixed $\tau$,
the integrals $du_id\alpha_i$ over the location of the vertex operators always
produce globally meromorphic functions of $s$ and $t$, with
at most double poles in $s$, $t$, and $u=-(s+t)$ at integers $\geq2$.
Indeed,
for fixed $\tau$, the factor $|q|^{-(su_1u_3+tu_2u_4)}{\cal R}^{*}$
is a smooth, bounded function of the $u_i,\alpha_i$, and only ${\cal I}$
plays a role in determining convergence of the $du_id\alpha_i$ integrals.
Setting $w_i=|q|^{u_i}e^{i\alpha_i}$, these integrals
can be viewed as integrals over a subdomain of $|w_i|\leq 1$.
In view of the constraint $|q|^{u_1+u_2+u_3+u_4}=|q|<e^{-\pi\sqrt 3}$, at most
6 of the 12 factors in ${\cal I}$ can approach 0. Thus
the $du_i\alpha_i$ integrals reduce to integrals in
the complex variables $w_i$'s of the form
\medskip
\item{(i)}
$$
\int_{\half \leq |w_1|\leq 1} d^2w_1
|1-w_1|^{-s} E(w_1)
$$
\item{(ii)}
$$
\int_{\half \leq |w_2|\leq|w_1|\leq 1}
d^2w_1 d^2w_2|1-w_1|^{-t}|1-w_2|^{-s}
|1-w_1w_2|^{s+t} E(w_1,w_2)
$$
\item{(iii)}
$$
\eqalign{\int_{\half \leq |w_i| \leq 1\atop
|w_{i+1}|\geq|w_i|}
\prod_{i=1}^3 d^2w_i&|1-w_1|^{-t}|1-w_2|^{-s}
|1-w_3|^{-t}|1-w_1w_2|^{s+t}\cr
&\times |1-w_2w_3|^{s+t} |1-w_1w_2w_3|^{-s}
E(w_1,w_2,w_3)\cr}
$$

Here $E(w_i)$ are smooth bounded functions which incorporate
$|q|^{-su_1u_2-tu_2u_4}$, ${\cal R}^{*}$ as well as well-behaved factors in
${\cal I}$. Our claim follows from
the analytic continuations of integrals of the form (i)-(iii)
derived in Appendix A.
\bigskip
The meromorphicity of the contribution of each fixed $\tau$ implies the
meromorphicity of the contribution of any bounded region in $\tau$. In
 particular
the region $\{|\tau_1|\leq1/2, |\tau|\geq1, \tau_2\leq1\}$ contributes
only a meromorphic term which can be absorbed in the expression $M_N(s,t)$
in Theorem 1. We may restrict then the $\tau$ domain of integration to
the semi-infinite strip
$$
\{|\tau_1|\leq1/2, \tau_2\geq1\}
\eqno(3.28)
$$
in which the variable $\alpha_4=2\pi\tau_1-\alpha_1-\alpha_2-\alpha_3$ can be
treated as an angle on the same footing as the other three angle variables
 $\alpha_1$,
$\alpha_2$, $\alpha_3$. The contribution of (3.28)
is then
invariant under each interchange $u_1\leftrightarrow u_2$
and $u_3\leftrightarrow u_4$. The region of integration can thus be restricted
to $u_1\leq u_2$ and $u_3\leq u_4$ upon including an overall factor
of 4. We need to consider separately the contributions of
the following two regions
$$
\eqalignno{({\rm I})&=\{(u_1+u_2)\tau_2<1\}\cr
({\rm II})&=\{(u_1+u_2)\tau_2\geq 1\}&(3.29)\cr}
$$

\medskip

In Region I, the expression
$$
|q|^{-(su_1u_2+tu_3u_4)}
$$
remains bounded. This implies that the contributions of each $\tau$
in this region can be summed up in a convergent integral and produce no
new singularities. The careful mathematical arguments are given in Appendix A.
The net outcome is that Region I just produces a meromorphic function
of $s$ and $t$ which can be absorbed into the expression $M_N(s,t)$ allowed
in Theorem 1.

\medskip

In Region (II), we observe that $\tau_2(u_i+u_j)\leq 1$ whenever one subindex
 $i$
or $j$ is even and the other one is odd. Under these conditions
 $|q|^{u_i+u_j}\leq
e^{-2\pi}<1$ and, except for the first four factors
$$
\prod_{i=1}^4 \bigl |1-e^{i\alpha_i}|q|^{u_i} \bigr |^{-s_i}
\eqno(3.30)
$$
in the infinite product expression for ${\cal R}$, all the other
factors (i.e. ${\cal R}^{*}$ as well as 8 of the factors
in ${\cal I}$) are bounded away from 0. The expansion (3.16), which is
just the expansion of these non-vanishing factors in a series
in $|q|^{u_i}$ and $e^{i\alpha_i}$, is then uniformly convergent. We
assert now that for each $N$ it suffices to consider a finite number of terms
in
 this
expansion in order to obtain an analytic continuation
in $s$ and $t$ to the half-space $\Re ( s)<N$, $\Re ( t)<N$. Since $N$
is arbitrary, we actually obtain in this way an analytic continuation to
the whole $s$ and $t$ planes. Fix then $N$. The uniform convergence
of the series (3.16) for ${\cal R}$ allows us to rewite it under the
form of a {\it limited} Taylor expansion
$$
\eqalignno{
{\cal R} (|q|^{u_i},\alpha_i ;s,t)
=
\prod_{i=1}^4 \bigl |1-e^{i\alpha_i} &|q|^{u_i} \bigr |^{-s_i}
\bigg ( \sum_{n_1+\cdots+n_4\leq4N}\sum_{|\nu_i|\leq n_i}
P_{n_i\nu_i}(s,t)\prod_{i=1}^4|q|^{n_iu_i}e^{i\nu_i\alpha _i}\cr
&+\sum
|q|^{n_1u_1+\cdots+n_4u_4} E_{n_i}(|q|^{u_i},\alpha_i;s,t)\bigg )&(3.31)\cr}
$$
where the second sum on the right hand side is over
indices $n_i$ satisfying
$$
n_1+n_3= n_2+n_4=N
$$
and $E_{n_i}$ are suitable smooth functions. To see (3.31) we begin by
observing
that any term $w_{ij}q^n$ in the factorization (3.7) for ${\cal R}$ satisfies
$$|(n_1+n_3)-(n_2+n_4)|\leq1$$
when written under the form $\prod_{i=1}^4|q|^{n_iu_i}
e^{i\nu_i\alpha_i}$. Consider next any product
 $\prod_{\ell=1}^Lw_{i_{\ell}j_{\ell}}q^{n_{\ell}}$
where each individual factor corresponds to exponents
$\{n_i^{\ell},\nu_i^{\ell}\}$.
The exponents $\{n_i\nu_i\}$ corresponding to the product will satisfy
$$|(n_1+n_3)-(n_2+n_4)|\leq L\eqno(3.32)$$
Now except for the first four factors (3.30) which are not expanded, all other
factors involve $w_{i_{\ell}j_{\ell}}q^{n_{\ell}}$ with
$${\rm min}(n_1^{\ell}+n_3^{\ell},n_2^{\ell}+n_4^{\ell})\geq1$$
This implies $L\leq{\rm min}(n_1+n_3,n_2+n_4)$, which combined with (3.32),
implies in turn
$$
{1\over 2}(n_1+n_3)\leq n_2+n_4\leq 2(n_1+n_3)
\eqno(3.33)
$$
Returning now to the Taylor expansion of ${\cal R}$, we note that any term
not included in the first sum on the right hand side of (3.31) must
satisfy either $n_1+n_3>2N$ or $n_2+n_4>2N$. In either case, the other
pair of indices must add up to more than $N$, in view of (3.33). Thus such
terms
can be absorbed into the second sum on the right hand side of (3.31).

Evidently the above truncation
respects the $u_1\leftrightarrow u_3$, $u_2\leftrightarrow u_4$ symmetry. We
observe now that the last term gives rise to only a
meromorphic function in the given half-planes $\Re(s),~\Re(t)<N$. Indeed
$$
 |q|^{-su_1u_3 +n_1u_1+n_3u_3}|q|^{-tu_2u_4+n_2u_2+n_4u_4}
\eqno (3.34)
$$
is bounded then as $\tau_2\rightarrow\infty$.
As in the case of Region I, the meromorphic contributions corresponding to
these terms for each $\tau$ again add up to
just a meromorphic function. Thus it suffices
to treat the terms in the first sum in (3.31). These are precisely
the terms stated in Theorem 1, except that the integration region has been
truncated to Region II. Since Region I with the
simplified integrand still produces only a meromorphic function, we
may reattach it. Furthermore, if we proceed
in groups of 4 terms respecting the $1\leftrightarrow 3$, $2\leftrightarrow 4$
symmetry, we can rewrite the integral over $u_1\leq u_3$, $u_2\leq u_4$ in
terms of the original full region $0\leq u_i\leq 1, i=1,\cdots,4$.
This establishes Theorem 1.

\bigskip
\bigskip
\bigskip

\centerline
{\bf IV. DOUBLE DISPERSION RELATIONS}

\bigskip

With Theorem 1, the problem of analytically continuing
the amplitude $A_1(s,t)$ reduces to the problem of analytically continuing
$A_{\{n_i\nu_i\}}$ for each fixed $n_i$, $\nu_i$. In this section, we
carry this out by showing that $A_{\{n_i\nu_i\}}$ can be expressed as
a double dispersion relation.
Such a relation gives immediately an analytic
continuation to the whole plane in $s$ and $t$ cut along the positive real
axis.
The corresponding (double) spectral density has a natural
interpretation as a density of intermediate states,
and we exhibit it explicitly. In particular its support can be parametrized
by conics.

\bigskip

\noindent
{\it The minimal amplitudes} $A_{\{n_i\nu_i\}}$ {\it in terms of hypergeometric
functions}

The first step in our derivation of the double dispersion relations
is to express the amplitudes $A_{\{n_i\nu_i\}}$ in terms of
Laplace transforms of Gauss' hypergeometric functions $F(a,b;c;x)$.
Recall that $A_{\{n_i\nu_i\}}$ is given by (3.18). Thus it can be rewritten as
$$
\eqalignno{A_{\{n_i\nu_i\}}(s,t)
=\int_1^{\infty}{d\tau_2\over\tau_2^2}\int_0^1
du_i&\delta(1-\sum_{i=1}^4u_i)|q|^{-su_1u_3-tu_2u_4+\sum_{i=1}^4u_i
(n_i+|\nu_i|)
 }\cr
&\times
\prod_{i=1}^4C_{|\nu_i|}(s_i)F({s_i\over 2},{s_i\over
2}+|\nu_i|;|\nu_i|+1;|q|^{2u_i})&(4.1)\cr}
$$
since each $d\alpha$ integral produces a hypergeometric function
$$
\int_0^{2\pi}{d\alpha\over 2\pi}e^{i\alpha\nu}|1-xe^{i\alpha}|^{-s}=
C_{|\nu|}(s)x^{|\nu|}F({s\over 2},{s\over 2}+|\nu|;|\nu|+1;x^2)\eqno(4.2)
$$
If we introduce the (inverse) Laplace transform $\varphi_{n\nu}(s;\beta)$
of $F({s\over 2},{s\over
2}+|\nu|;|\nu|+1;x^2)$ by
$$
\eqalignno{C_{|\nu|}(s)x^{n+|\nu|}F({s\over 2},{s\over
2}+|\nu|;|\nu|+1;x^2)&=
\int_0^{\infty}d\beta\, x^{\beta}\varphi_{n\nu}(s;\beta)\cr
\varphi_{n\nu}(s;\beta)&=0\ {\rm for}\ \beta<0&(4.3)\cr}
$$
we immediately obtain the desired formula.
\bigskip
\noindent{\bf Lemma 1}. {\it The amplitude} $A_{\{n_i\nu_i\}}$
{\it can be written as}
$$
\eqalignno{
A_{\{n_i\nu_i\}}(s,t)  =\int\nolimits_0^\infty  \prod_{i=1}^4 d\beta_i~ &
\Psi_{\cn} (s,t;\beta_i)
\int\nolimits_1^\infty {d\tau_2\over\tau_2^2}
\int\nolimits_0^1 \prod_{i=1}^4 du_i~\delta\bigg(1-\sum_i u_i
\biggr )\cr
&\times \exp \biggl \{-2\pi\tau_2 \sum_i u_i\beta_i + 2\pi\tau_2
\bigl ( su_1 u_3+ tu_2u_4\bigr ) \biggr \}&(4.4)\cr}
$$
{\it with}
$$
\Psi_{n_i\nu_i}(s,t;\beta_i)=\prod_{i=1}^4\varphi_{n_i\nu_i}
(s_i;\beta_i)\eqno(4.5)
$$
{\it In particular the support of} $\Psi_{n_i\nu_i}$ {\it is contained in}
$\{\beta_i\geq 0,\ i=1,\cdots,4\}$.
\bigskip
The analyticity properties of $\varphi_{n\nu}(s;\beta)$ as well as those
of the closely related Mellin transform $f_{n\nu}(s,\alpha)$ of
$F({s\over 2},{s\over 2}+|\nu|;|\nu|+1;x^2)$
are given in Appendix B. For the moment, we note only
that $\varphi_{n\nu}(s;\beta)$ is an infinite superposition of Dirac point
masses or delta functions
$$
\varphi _{n\nu}(s;\beta) = \sum_{k=0}^\infty
C_k(s) C_{k+|\nu|} (s) \delta (2k +n+|\nu| -\beta)
\eqno(4.6)
$$
In particular, its integrals in $\beta$ over any finite interval are entire
functions of $s$, although they can have poles if the interval is infinitely
extended, as shown in Appendix B.

\bigskip
\noindent{\it Derivation of the Double Dispersion Relation}

To recast $A_{\{n_i\nu_i\}}$ in the form of a
double dispersion relation, we start from the formula (4.4)
provided by Lemma 1. The variables $u_3$ and $u_4$ are changed to
$$
\sigma _3 = 2\pi\tau_2u_3 {}\qquad \ \sigma_4 = 2\pi \tau_2 u_4
$$
Without loss of generality, we may extend the integration region for
$u_3$ and $u_4$, and hence for $\sigma_3$ and $\sigma_4$ to the
half-line $[0,\infty]$, in view of the presence of the $\delta$-function
on the $u$'s in (4.4).  Having done so, we may view the
$\delta$-function now as a $\delta$-function for the modulus $\tau_2$,
and we can easily integrate out $\tau_2$.  The result is
$$
\eqalignno{
A_{\{n_i\nu_i\}}(s,t) =  2\pi
\int_0^\infty\!\!\!\prod_{i=1}^4d\beta_i& \Psi_{\{n_i\nu_i\}}
(s,t;\beta_i) \int\nolimits_0^1\!\!\! du_1
\int\nolimits_0^{1-u_1} \!\!\!\!\!\!du_2
\int\nolimits_0^\infty \!\!d\sigma_3
\int\nolimits_0^\infty \!\!d\sigma_4 {(1-u_1-u_2)^2 \over
(\sigma_3+\sigma_4)^3}\cr
&\times \theta\big (\sigma _3 + \sigma _4 - 2\pi (1-u_1-u_2)   \big )\cr
&\times \exp
\biggl \{ (su_1-\beta_3)\sigma_3 + (tu_2-\beta_4)\sigma_4\cr
&\qquad\qquad\qquad\qquad - {\sigma_3+\sigma_4\over
1-u_1-u_2}(u_1\beta_1 + u_2\beta_2)\biggr \}&(4.7)\cr}
$$
We make a further change of variables from $\sigma_3, \sigma_4$ to $\mu ~\in
[0,\infty ]$
and $\alpha~\in [0,1]$ by
$$
\eqalign{
& \sigma_3 = 2\pi (1-u_1-u_2)\mu \alpha\cr
& \sigma_4 = 2\pi (1-u_1-u_2)\mu(1-\alpha)\cr}
$$
and arrive at
$$
\eqalignno{
A_{\{n_i\nu_i\}}(s,t) =
\int\nolimits_0^\infty\!\!\!   \prod_{i=1}^4d\beta_i~& \Psi_{\{n_i\nu_i\}}
(s,t;\beta_i)\! \int\nolimits_0^1\!\!\! du_1
\int\nolimits_0^{1-u_1}\!\!\! du_2
(1-u_1-u_2)\cr
&\times\int\nolimits_0^1 \!\!\!d\alpha ~ \int\nolimits_1^\infty
 {d\mu\over\mu^{2}}
e^{-\mu \kappa}&(4.8)\cr}
$$
Here we have defined the function $\kappa$, linear in $s$ and $t$ by
$$
\eqalign{
\kappa = 2\pi \biggl \{ u_1\beta_1 + u_2\beta_2
& + (1-u_1-u_2) (\alpha\beta_3 + (1-\alpha)\beta_4 )  \cr
& - s(1-u_1-u_2) u_1 \alpha - t (1-u_1-u_2)u_2(1-\alpha)\biggr \}\cr}
\eqno(4.9)
$$
Now we come to the actual analytic continuation.  We rewrite the
integral over $\mu$ in terms of the following expression
$$
\int\nolimits_1^\infty  {d\mu\over \mu^2}\ e^{-\mu \kappa} =
\half
\int\nolimits_0^\infty d\rho \biggl [ {\rho^2\over (\rho+\kappa)^2} - 1
+{2 \kappa\over\rho+1} \biggr ]+E(\kappa)
\eqno(4.9)
$$
where $E(\kappa)$ is the entire function of $\kappa$ given by
$$
E(\kappa)=-\int\nolimits_0^1  {d\mu\over \mu^2}
(e^{-\mu \kappa} - 1 + \kappa\mu ) + (c_2-{3\over 2})\kappa+1
\eqno(4.10)
$$
Here $c_2$ is a constant. The following Lemma shows that
$E(\kappa)$ does not affect the branch cuts:
\bigskip
\noindent{\bf Lemma 2}. {\it The contribution of
$E(\kappa)$ to the amplitude} $A_{\{n_i\nu_i\}}$
$$
\eqalign{
M_{\{n_i\nu_i\}}(s,t) =  \prod_{i=1}^4
\int\nolimits_0^\infty\!\!\! d\beta_i~ \Psi_{\{n_i\nu_i\}}
(s,t;\beta_i)\! \int\nolimits_0^1\!\!\! du_1
\int\nolimits_0^{1-u_1}\!\!\! du_2
\int\nolimits_0^1 \!\!\!d\alpha ~ (1-u_1-u_2)
E(\kappa)\cr}
\eqno(4.11)
$$
{\it is a globally meromorphic function of both} $s$ {\it and} $t$.
{\it Furthermore, its poles are at integers on the real axes in $s$ and $t$,
with real coefficients.}
\bigskip
Indeed, all the expressions entering (4.11) are actually moments of
 $\varphi_{n\nu}$
and hence are all given by special values of the hypergeometric function,
its derivatives, and its integrals at $z=1$. Such values are meromorphic
 functions
of $s$. The details are in Appendix B.
\medskip
Henceforth we
can ignore this part of $A_{\{n_i\nu_i\}}$, as it does not enter the
double dispersion representation. We turn now to the first expression
on the right hand side of (4.9). The argument is based
on a few successive changes of variables. For greater clarity we
work them out only on the main contribution,
which is formally that of the term $\rho^2(\rho+\kappa)^{-2}$
$$
\int_0^{\infty}\prod_{i=1}^4 d\beta_i
\Psi_{\{n_i\nu_i\}}(s,t;\beta_i)
\int\nolimits_0^{1-u_1}\!\!du_1 (1-u_1-u_2) \int\nolimits_0^1 \!\!d\alpha
\int\nolimits_0^{\infty} d\rho  {\rho^2\over(\rho+\kappa)^2}
\eqno(4.12)
$$
Since $\kappa$ is a linear function of $\alpha$, the
$\alpha$-integral may be carried out explicitly, and (4.12) becomes
$$
\eqalign{
\int\nolimits_0^\infty \prod_{i=1}^4d\beta_i &
\int\nolimits_0^1 du_1
\int\nolimits_0^{1-u_1} du_2 (1-u_1-u_2)^2\Psi_{\{n_i\nu_i\}}(s,t;\beta_i)\cr
& \times \int_0^{\infty}dx
{x^2\over (x+x_0+\beta_3-su_1) (x+x_0+\beta_4-tu_2)}\cr}
\eqno(4.13)
$$
where we made the change of variables $\rho = 2\pi (1-u_1-u_2)x$ and
set
$$
x_0 \equiv \q {u_1\beta_1 + u_2\beta_2\over 1-u_1-u_2}
\eqno(4.14)
$$
Upon performing a last change of variables from $\beta_3$ and $\beta_4$
to $\sigma$ and $\tau$
$$
\beta_3 = - x -x_0 + u_1\sigma \qqq \beta_4 = - x-x_0 + u_2\tau
\eqno(4.15)
$$
the main term (4.12) takes the form of a double dispersion
relation
$$
\int\nolimits_0^\infty d\sigma \int\nolimits_0^\infty d\tau
{\rho ( s,t;\sigma,\tau)\over (s-\sigma)(t-\tau)}
\eqno(4.16)
$$
The double spectral density $\rho (s,t;\sigma,\tau)$ in (4.16) for each
minimal amplitude is given by
$$
\eqalign{
\rho_{\{n_i\nu_i\}}(s,t;\sigma,\tau) = \int\nolimits_0^\infty d\beta_1&
\int\nolimits_0^\infty d\beta_2
\int\nolimits_0^1 du_1
\int\nolimits_0^{1-u_1}  du_2~
(1-u_1-u_2)^2 \int\nolimits_{x_{0}}^\infty dx (x-x_0)^2\cr
& \cr
&\times \Psi_{\{n_i\nu_i\}}  (s,t;\beta_1, \beta_2, -x+ u_1\sigma, -x+u_2\tau
 )\cr}
\eqno(4.17)
$$
For given $s$ and $t$, and fixed $\sigma$ and $\tau$, all the integrals
in the definition of $\rho (s,t;\sigma ,\tau)$ run over finite ranges.
To see this, we note that from the structure of $\Psi_{\{n_i\nu_i\}}$,
its first four arguments must
be positive, which requires that $0 \leq x \leq \sigma,\tau$ and in
view of $x_o \leq x$, also $0 \leq \beta_1 \leq \sigma$ and $0\leq
\beta_2 \leq \tau$.  As a result, $\rho (s,t;\sigma,\tau)$ is a
completely finite function of $s,t$ and $\sigma$ and $\tau$.  In fact
for fixed $\sigma$ and fixed $\tau$ it is polynomial in $s$
and $t$, and the degree of this polynomial increases with
$\sigma$ and $\tau$.

Next, the same calculation leads to the following formula
for the contribution of the remaining term $1-2\kappa(\rho+1)^{-1}$ in (4.9)
$$
\int_0^{\infty}\int_0^{\infty}d\sigma d\tau\Lambda_{\{n_i\nu_i\}}
(s,t;\sigma,\tau)
$$
with $\Lambda_{\{n_i\nu_i\}}(s,t;\sigma,\tau)$ given by
$$
\eqalignno{\Lambda_{\{n_i\nu_i\}}(s,t;\sigma,\tau)
=\int_0^{\infty}& d\beta_1\int_0^{\infty} d\beta_2\int_0^{\infty}dx
\Psi(s,t;\beta_1,\beta_2,\sigma u_1-x_0-x,\tau u_2-x_0-x)\cr
&\times\bigg(1-2\pi(1-u_1-u_2){u_1(\sigma-s)+u_2(\tau-t)-2x\over
2\pi(1-u_1-u_2)+1}\bigg)
&(4.18)\cr}
$$
In the same way as the term $1-2\kappa(\rho+1)^{-1}$ guaranteed
the convergence of the $d\rho$ integral in (4.9), the role of
$\Lambda_{\{n_i\nu_i\}}$
in (4.18) is as a subtraction term guaranteeing the convergence of the
$d\sigma d\tau$ integrals in the final expression (4.16)+(4.18) for
$A_{\{n_i\nu_i\}}$. In practice, however, it produces only meromorphic
functions of $s$ and $t$. Thus the behabior of $A_{\{n_i\nu_i\}}$ at the
branch cuts is completely described by the double dispersion relation
(4.16), and for all practical purposes, we may safely drop
$\Lambda_{\{n_i\nu_i\}}$ from our considerations.
\bigskip
We have thus established all the statements except the last one in
the following theorem:

\bigskip

\noindent
{\bf Theorem 2}. {\it The minimal amplitudes}
$A_{\{n_i\nu_i\}}(s,t)$ {\it can be expressed as}
$$
A_{\{n_i\nu_i\}}(s,t)=M_{\{n_i\nu_i\}} (s,t) +
\int_0^{\infty}\int_0^{\infty}d\tau d\sigma
\bigg ({\rho_{\{n_i\nu_i\}}(s,t;\sigma,\tau)\over
(s-\sigma) (t-\tau)}-\Lambda_{\{n_i\nu_i\}}\bigg )
\eqno(4.19)
$$
{\it where the density} $\rho_{\{n_i\nu_i\}}$ {\it and the
meromorphic subtraction term} $\Lambda_{\{n_i\nu_i\}}$
{\it are given by (4.17) and (4.18).}
{\it The term} $M_{\{n_i\nu_i\}}$ {\it is another globally meromorphic
function of $s$ and $t$, whose properties are described in Lemma 2}.
{\it The integral on the right hand side in (4.19) defines a holomorphic
function of} $s,t$ {\it in the cut plane} $s,t\in{\bf C}\setminus{\bf R}_+$.
{\it More precisely, its domain of holomorphy is given by}
$$
s\in{\bf C}\setminus[(m_1+m_3)^2,+\infty),\ t\in{\bf C}\setminus[(m_2+m_4)^2,
\infty)
\eqno(4.20)
$$
{\it where we have set}
$$
m_i^2=n_i+|\nu_i|,\ i=1,\cdots,4\eqno(4.21)
$$
\bigskip
The last statement will be established in the next section.
The factorized expression (4.5) for $\Psi_{\{n_i\nu_i\}}$ in terms of four
$\varphi_{n\nu}$-functions shows that
the double spectral density can also be expressed as
$$
\eqalign{
\rho_{\{n_i\nu _i\}} (s,t;\sigma,\tau) = \int\nolimits_0^\infty \!\! d\beta_1&
\int\nolimits_0^\infty \!\! d\beta_2
\varphi _{n_1 \nu _1}(t;\beta _1 ) \varphi _{n _2 \nu _2} (s;\beta _2)
\int\nolimits_0^1 \!\! du_1
\int\nolimits_0^{1 -u_1} \!\!\!\! du_2(1-u_1-u_2)^2\cr
&\times \int_{x_{0}}^\infty dx (x-x_o)^2
\varphi _{n_3 \nu _3} (t; u_1 \sigma-x )
\varphi _{n_4 \nu _4} (s; u_2 \tau -x )\cr}
\eqno(4.22)
$$
This generalizes the result presented in [7] for the case $n_i=0$, $\nu_i=0$.

\bigskip
\noindent
{\it Location of cuts.}

The expression (4.20) for the amplitude $A_{\{n_i\nu_i\}}$ shows that
the branch cuts in both $s$ and $t$ lie within the positive real axis. Where
they begin depends however on $n_i~\nu_i$. To address this issue, we need a
 closer
inspection of the support of the density
$\rho_{\{n_i\nu_i\}}(s,t;\sigma,\tau)$.
Since the Laplace transform $\varphi_{n\nu}(s;\beta)$
of the hypergeometric function $F$ can be expanded into a series of Dirac point
masses (see (4.6)), we may also expand $\rho_{\{n_i\nu_i\}}$
into a series of ``monomial" densities
$$
\rho_{\{n_i\nu_i\}}(s,t;\sigma,\tau)
=\sum_{k_i=0}^{\infty}\prod_{i=1}^4C_{k_i}(s_i)C_{k_i+|\nu_i|}(s_i)
{}~\rho^{\{2k_i+n_i+|\nu_i|\}}(s,t;\sigma,\tau)
\eqno(4.23)
$$
where the monomial density $\rho^{M_i^2}$ is defined by the same
integral (4.19), with however the following choice for $\Psi(\beta_i)$
$$
\Psi_{\{M_i^2\}}=\prod_{i=1}^4\delta(\beta_i-M_i^2)
\eqno(4.24)
$$
Thus it suffices to determine the support of each density
$\rho^{\{M_i^2\}}$, for each $M_i^2=2k_i+n_i+|\nu_i|$. The $d\beta_i$ integrals
can now be carried out, and we find
$$
\eqalignno{\rho^{\{M_i^2\}}(s,t;\sigma,\tau)
=\int_0^1du_1\int_0^{1-u_1}&du_2(1-u_1-u_2)^2\int_{x_0}^{\infty}
dx(x-x_0)^2\cr
&\times\delta(x+M_3^2-\sigma u_1)\delta(x+M_4^2-\tau u_2)
&(4.25)\cr}
$$
with
$$
x_0={u_1M_1^2+u_2M_2^2\over 1-u_1-u_2}\eqno(4.26)
$$
These expressions coincide exactly with those for the spectral density
of the box diagram for $\phi^3$ field theory (c.f. (D.5) in Appendix D).
Thus the parametrization (D.8) derived in Appendix D for the $\phi^3$
spectral density holds in the string case as well. In particular,
the support of $\rho^{\{M_i^2\}}$ as a function of $\sigma$ and $\tau$
is contained in the region $A\geq B\geq 0$, where $A$ and $B$ are given by
(D.7), with $m_i^2$ replaced by $M_i^2$.

Let $A_{\{n_i\nu_i\}}^{\{M_i^2\}}$ denote the contribution
of the monomial density $\rho^{\{M_i^2\}}$ to the amplitude $A_{\{n_i\nu_i\}}$.
We begin by identifying the domain of
holomorphy of  $A_{\{n_i\nu_i\}}^{\{M_i^2\}}$. It is given by
$$
\{s\notin\pi_{\sigma}({\rm support}\,\rho^{\{M_i^2\}})
\}
\cup \{ t\notin\pi_{ \tau}({\rm support}\rho^{\{M_i^2\}})\}\eqno(4.27)
$$
where $\pi_{\sigma}$ and $\pi_{\tau}$ denote respectively the projections of
$(\sigma,\tau)$ on the first and the second variable. Now
the support of $\rho^{\{M_i^2\}}$ is determined by
the condition $A\geq B\geq 0$, which can be rewritten as
$$
(M_1^2+M^2_3){1\over\sigma}
+(M^2_2+M^2_4){1\over \tau}+2({1\over \sigma}+{1\over\tau})
(M^2_1M^2_3{1\over\sigma}+M^2_2M^2_4{1\over\tau})^{1/2}\leq 1
\eqno(4.28)
$$
The left hand side defines a function $f$ of $({1\over\sigma},{1\over\tau})$
which is increasing, in the sense that
$$
f(\alpha,\beta)<f(\alpha',\beta')
$$
for all $0\leq\alpha\leq\alpha'$, $0\leq\beta\leq\beta'$,
$\alpha+\beta<\alpha'+\beta'$. Thus its contour must have the form
depicted in Fig. 6.
In particular for any level set $f(\alpha,\beta)=constant$, the point with the
greatest $\alpha$ value is the point on the $\alpha$ axis. In the case at
hand, this point is given by
$$\alpha_{M_i^2}=(M_1+M_3)^{-2}\eqno(4.29)$$ Similarly, the extreme
point on the $\beta$ axis is given by
$$\beta_{M_i^2}=(M_2+M_4)^{-2}\eqno(4.30)$$
The inverses of these values are then the beginnings of the branch cuts of
$A_{\{n_i\nu_i\}}^{\{M_i^2\}}$ in $s$ and $t$ respectively.

Returning to the amplitude $A_{\{n_i\nu_i\}}$ proper, we observe that the
branch
cuts in $s$ and $t$ move out to the right as any $k_i$ increases. Thus all the
cuts are contained in the leading one at $k_i=0$, $i=1,\cdots, 4$,
in which case $M_i^2\equiv m_i^2=n_i+|\nu_i|$. and we find
that the domain of holomorphy of $A_{\cn}(s,t)$ is given by Fig. 6(b)
where
$$
(s,t)\in\big({\bf
 C}\setminus[(m_1+m_3)^2,+\infty)\big)
\times
\big({\bf C}\setminus[(m_2+m_4)^2,+\infty)\big)
$$
This completes the proof of Theorem 2.

\bigskip
\bigskip
\bigskip

\centerline
{\bf V. POLES FROM DOUBLE DISPERSION RELATIONS}
\centerline{\bf  AND STRING DUALITY}

\bigskip

Formally, the partial amplitude $A_1(s,t)$ is an infinite sum of
$\phi ^3$-like box diagrams and as such exhibits double branch cut
singularities in both $s$ and $t$.
These branch cuts are clearly displayed by the double dispersion
relation derived in the previous section.
However, in \S II, we saw that additional pole singularities
on top of branch cuts are expected as well, and their appearance
must be elucidated.

In the present section, we show that these poles in $s$ and $t$,
lying on top of branch cuts are indeed included in the double
dispersion representation, and can be directly extracted from it.
In local quantum field theory -- with a finite number of fields --
poles never arise from dispersion relations (in perturbation
theory), and we shall link their appearance to the property of
string duality.

\medskip
\noindent
{\it Exhibiting simple and double poles op top of branch cuts}

We show how the integration in one of the dispersion
parameters (say $\tau$), produces a simple spectral density which
exhibits double and simple poles (in $s$) at positive even integers.  We
represent the amplitude by a simple dispersion relation
$$
A_{\{n_i \nu _i\}} (s,t) =
\int\nolimits_0^\infty d\sigma  ~{\rho _{\{n_i \nu _i\}}(s,t;\sigma)\over
\sigma-s}
\eqno(5.1)
$$
where the simple spectral density is defined by
$$
\rho_{\{n_i\nu_i\}}(s,t;\sigma) = \int\nolimits_0^\infty d\tau~
{\rho_{\{n_i \nu _i\}} (s,t;\sigma,\tau)\over \tau-t}
\eqno(5.2)
$$
The double spectral density $\rho _\cn (s,t;\sigma, \tau)$ was defined
and constructed explicitly in \S IV (4.17).
The $\tau$-integration may be carried out because $\rho _\cn$
depends on $\tau$ only through a single $\varphi$-function.
We make use of a standard relation between the Mellin transform
$f$ and the inverse Laplace transform $\varphi$ (see Appendix \S B, in (B.25))
:
$$
\int\limits_0^\infty d\tau {1\over \tau -t}
\varphi _{n_4 \nu _4} (s, u_2\tau - x) = f_{n_4\nu_4} (s, u_2t -x)
\eqno (5.3)
$$
and evaluate $\rho _\cn$ :
$$
\eqalign{
\rho _{\cn}(s,t;\sigma) =
\int\nolimits_0^\infty \!\! d\beta_1
\int\nolimits_0^\infty \!\! d\beta_2~ &
\varphi _{n_1\nu _1} (t, \beta_1)
\varphi _{n_2\nu _2} (s, \beta_2)
\int\nolimits_0^1 \!\! du_1
\int\nolimits_0^{1-u_1} \!\! du_2 (1-u_1-u_2)^2\cr
&
\times \int\nolimits_{x_{o}}^\infty dx (x-x_o)^2
\varphi _{n_3\nu_3} (t, u_1\sigma-x ) f_{n_4\nu_4}  (s,  u_2 t -x)\cr}
\eqno(5.4)
$$
{}From the appearance of $f$ as the last factor in the integrand,
it is clear that simple poles in $s$ occur at positive integers.
Moreover, there are further simple poles in $s$ appearing from the fact that
the $\beta _2$ integration now extends to $\infty$.
We shall now identify these additional double poles more
explicitly.

For fixed $\sigma$ and any values of $n_i$
and $\nu _i$, we have $0 \leq x \leq \sigma$,
and $ 0\leq\beta_1 \leq \sigma$, so that in an expansion in
coefficients $C_k(t)$, both functions $\varphi(t,\cdot)$ produce only a
finite number of terms, indeed, of order $\sigma$.  The only possible
origin then of poles as a function of $s$ is from the range $\beta_2
\rightarrow \infty$, which occurs when $u_2 \rightarrow 0$.  We see that
for fixed $\sigma$, $\rho (s,t;\sigma)$ is an entire function of $t$.
We make use of this analytic behavior in $t$ to expand both
$\varphi(t,\cdot)$ functions in a series, as given by (B.18). As
a result, we obtain
$$
\rho _{\cn}(s,t;\sigma) = \sum_{k_1,k_3 = 0}^\infty
C_{k_1}(t) C_{k_1 +|\nu _1|}(t)
C_{k_3}(t) C_{k_3 +|\nu _3|}(t)  \rho _{\cn,k_1,k_3}
(s,t;\sigma)
\eqno(5.5)
$$
In view of the preceding remarks, the functions
$\rho _{\cn,k_1,k_3}(s,t;\sigma)$ vanish whenever $k_1$ or $k_3 > \sigma$,
and the above sum is effectively finite.
Furthermore, we have
$$
\eqalign{
\rho _{\cn,k_1,k_3} (s,t;\sigma) =
\int\nolimits_0^\infty \!\! d\beta_2 & ~\varphi _{n_2 \nu _2} (s,\beta _2)
\int\nolimits_0^1 \!\! du_1
\int\nolimits_0^{1-u_1} \!\! du_2 f _{n_4\nu_4} (s, M^2_3 - u_1\sigma  + u_2 t)
\cr
&\times\theta(u_1\sigma - M^2_3 - x_0)
(u_1\sigma -M^2_3 - x_0)^2 (1-u_1-u_2)^2
 \cr}
\eqno(5.6)
$$
where we have defined $M_i^2 = 2k_i +n_i +|\nu _i|$ for $i=1,3$, and
$x_0 = (M^2_1 u_1+\beta_2 u_2) (1-u_1-u_2)^{-1}$.
A useful integral representation is obtained by also expanding
$f_{n_4 \nu _4}$ in an infinite Taylor series in powers of $t$.
For each term, of order $t^p$, we must perform the following integral over
$u_2$, and then over $\beta_2$:
$$
\eqalignno{
\int\nolimits_0^\infty \!\! d\beta_2&
\int\nolimits_0^{1-u_1} \!\! du_2 (1-u_1-u_2)^2 u_2^p
(u_1\sigma - m^2_3 - x_0)^2\cr
& = {2p!\over (p+3)!}
\biggl ( u_1(1-u_1)\sigma - m^2_3 (1-u_1)
- m^2_1  u_1 \biggr )^{p+3} \theta (u_1 \sigma - m^2_3 ) \cr
& \qqq \qqq \qqq
\times\int\nolimits_0^\infty \!\! d\beta_2~\varphi _{n_2 \nu _2} (s,\beta_2)
(u_1\sigma -m^2_3 + \beta_2)^{-1-p}\cr
& = {2\over (p+3)!}
\bigl (u_1 (1-u_1) \sigma -(1-u_1) m^2_3 - u_1 m^2_1 \bigr ) ^{p+3}
f_{n_2\nu _2} ^{(p)} (s,m^2_3 - u_1\sigma )&(5.7)\cr}
$$
Assembling all other factors as well in (5.6), we finally obtain
$$
\eqalign{
\rho _{\cn,k_1,k_3} (s,t;\sigma) = \sum_{p=0}^\infty ~ &
{2 t^p \over p!(p+3)!}
\int\nolimits_0^1 du ~\theta\biggl (\sigma -
{M^2_3 \over u} - {M^2_1 \over 1-u}\biggr )\cr
& \qqq
\times\bigl (\sigma u(1-u) - (1-u) M^2_3 -u M^2_1 \bigr ) ^{p+3}\cr
& \qqq
\times f_{n_2 \nu _2} ^{(p)} \bigl (s,M^2_3 - u\sigma\bigr )
f_{n_4 \nu _4} ^{(p)} \bigl (s,M^2_3 - u\sigma\bigr )
\cr}
\eqno(5.8)
$$
Here, we use the notation that $f_{n\nu}^{(p)} (s,\alpha) = \partial_\alpha^p
f_{n\nu}(s,\alpha)$ and this function has simple poles in $s$ at positive
integers, strictly greater than $p$, as can be seen from (B.13)
and (B.14).
As a result, $\rho$ contains terms holomorphic in $s$ but also
terms with single and double poles in $s$ at positive integers.
These poles occur in the single spectral density so that after
integration over $\sigma$, they will be present in the full
amplitude on top of a branch cut in $s$ on the positive real axis.

An instructive special case is when $t=0$, which forces $k_1=k_3=0$
in the sum (5.5) and $p=0$ in (5.8), so that
$$
\eqalignno{\rho _{\{n_i \nu _i\}} (s,0;\sigma) = {1 \over 3} \delta _{\nu _1,0}
\delta _{\nu _3,0}
\int\nolimits_0^1 du ~&
\theta (\sigma -{n_3\over u}-{n_1 \over 1-u})
\{ (u(1-u) \sigma -n_3(1-u) - un_1\}^3\cr
&\times f_{n_2\nu_2} (s,n_3 - u\sigma) f_{n_4\nu _4} (s,n_3-u\sigma)
&(5.9)\cr}
$$
Here double and simple poles in $s$ are clearly exposed.

\bigskip
\noindent
{\it Connection with String Duality}

We have seen in Sections III and IV that, formally, the superstring
amplitude is a sum of $\phi^3$-like box diagrams.
However, this sum -- which in particular goes over all
masses of internal propagators -- is not uniformly convergent
and actually produces poles.
Physically, the appearance of poles is related to string duality,
as was already the case to tree level. To see this, we write
the tree level amplitude as a sum of poles in the $s$-channel
$$
A_o(s,t,u) = \sum _{n=0} ^\infty
{\Gamma (-u/2) \Gamma (t/2 +n+1) \over u/2 \Gamma (1+s/2)
\Gamma (1+t/2)^2 n! } ~ {1 \over -s/2 +n}
\eqno (5.10)
$$
Each term also has poles in $u$, but not in $t$.
The series is absolutely convergent for $\Re (t) <0$, and thus we
have only poles in $s$ in this region, as well as poles in $u$,
but the series is holomorphic in $t$.
To define the series for $\Re (t) \geq 0$, we must analytically
continue it from $\Re (t) <0$.
The analytic continued expression has poles in $t$ this time,
which can be seen by simply interchanging $s$ and $t$ in (5.10).
Pictorially, this result is represented in Fig. 8.
What happens to one loop level is completely analogous.
Summations over intermediate masses in the string box diagram
are only convergent for $\Re (s), ~\Re (t) <0$, and must be analytically
continued. In doing so, one produces poles in the crossed channels.
This is precisely the meaning of expressions such as (5.1,2)
and (5.8) which contain poles. Pictorially, what we have shown is that the
sum over box graphs must be analytically continued and
that poles are found in the cross channels in the way depicted in Fig. 9.
To state that the box diagrams and the triangle graph or the
polarization graphs are equal to one another -- as is customarily
done in the old dual model formulation -- is misleading;
instead the one loop amplitude contains all contributions at once,
after proper analytic continuation.

\medskip

\noindent
{\it Cancellation of poles at odd integers}

The Mellin transform $f_{n\nu} (s,\alpha)$ is meromorphic in $s$
with simple poles for positive integers.
The poles at even positive integers correspond to physical
on-shell superstring states, and are expected on physical
grounds, as explained in \S II.
The poles at odd integer values of $s$ are spurious and do not
correspond to the propagation of any physical degrees of freedom.
In our construction of the analytically continued
amplitude, the existence of these poles results from splitting
up the full amplitude $A_1(s,t,u)$ into partial amplitudes
$A(s,t)$, $A(t,u)$ and $A(u,s)$.
This splitting up of the domain is performed in such a way that
pole type singularities arising from complex integrals are integrated
over only 180 degrees around the singularity, instead of 360 degrees.
In fact this phenomenon is identical to and results from the same
splitting up problem that arises at tree level.

For example the tree level amplitude $A_0(s,t,u)$ is an integral
over the full complex plane as in (2.5), and poles only arise
at positive even integers in $s,t,u$.
Now let us split the integration into the one over the unit disk
$|z|\leq 1$ and the complementary region $|z|\geq 1$.
Each of these integrals is just precisely the Mellin transform $f$
$$
\eqalign{
A_0(s,t,u) = & -{2 \over s^2} \int _{{\bf C}} d^2 z |z|^{-2-t}
|1-z|^{-s} \cr
	          = & -{4\pi \over s^2} \bigl [ f_{00} (s,t) + f_{00} (s,u) \bigr ]
\cr }
\eqno (5.14)
$$
It is shown in Appendix B that $f_{00}(s,t)$ has simple poles
in $t$ at positive even integers only. Thus the pole structure
in $t$ and $u$ is not affected by the splitting up of the $z$-domain
of integration.
On the other hand, the splitting into $|z|\leq 1$ and $|z|\geq 1$
precisely cuts the pole region around $z=1$ in half, and as
a result, each half now has additional poles at odd positive integers
as well. But in the sum these poles cancel out.

The same Mellin transforms occur in the one-loop analytically
continued amplitude, and the poles at odd integers cancel
as well once all partial amplitudes have been put together.
To see this, one has to make heavy use of the detailed
properties of the coefficient functions $P_{n_i \nu _i} (s,t)$,
and we shall not provide the check of this cancellation here.

\bigskip
\bigskip
\bigskip

\centerline
{\bf VI.  HETEROTIC STRING AMPLITUDES}

\bigskip

We consider the one-loop scattering amplitudes for four massless
bosonic string states in the heterotic $E_8 \times E_8$ or ${\rm Spin}(32)/Z_2$
superstrings [20].  Whereas for Type II superstrings, there was a single
massless supermultiplet, for heterotic strings there is a supergravity
multiplet as well as a super Yang Mills multiplet.  So each $S$-matrix
state can be either one of these.  To exemplify the analytic
continuation procedure, we shall specialize to a simple subcase:  that
of the scattering of four massless charged gauge bosons.

The scattering amplitude for four gauge bosons with space-time momenta
$k_i^\mu$ polarization vectors $\epsilon_i^\mu $ and root (weight)
lattice momenta $K_i^I$ $(i = 1,\cdots, 4$; $\mu=1,\cdots, 10;$
$I =1,\cdots,16)$ is given by
$$
\eqalign{
A_\ell ^H  (k_i, \epsilon_i, K_i ) =
(2\pi)^{10} \delta(k)g^4
& \epsilon_1^{\mu_1} (k_1)
\epsilon_2^{\mu_2} (k_2)
\epsilon_3^{\mu_3} (k_3)
 \epsilon_4^{\mu_4} (k_4)\cr
& K_{\mu_1\mu_2 \mu_3 \mu_4} (k_i) A_\ell ^H (s,t,u; S,T,U)\cr}
\eqno(6.1)
$$
The kinematical factor $K$ is identical to the one encountered in (2.3),
and the root (weight) lattice Mandelstam variables $S,T$ and $U$ as
defined by
$$
\eqalign{
& S = - (K_1 + K_2)^2 = - 4 - 2K_1\cdot K_2\cr
& T = - (K_2 + K_3)^2 = -4 -2 K_2\cdot K_3\cr
& U = - (K_1 + K_3)^2 = -4 -2 K_1\cdot K_3\cr}
\eqno (6.2)
$$
Lattice momentum conservation implies
$$
\sum_i K_i = 0 ; \q S+T+U = -8
\eqno (6.3)
$$
Since the lattice is self-dual and even, the allowed values for
$K_i\cdot K_j$ are 2, 1, 0, -1 or -2, or equivalently $S,T,U$ take on
the values -8, -6, -4, -2 or 0.
The tree level answer is given by [20]
$$
A_0 ^H(s,t,u;S,T,U) = \pi {\Gamma (-{s+S\over 2 }  - 1)
                        \Gamma (-{t+T\over 2 }  - 1)
                        \Gamma (-{u+U\over 2 }  - 1)\over
\Gamma(1+{s\over 2}) \Gamma(1+{t\over 2}) \Gamma(1+{u\over 2} ) }
\eqno (6.4)
$$
with obvious and physically acceptable analytic continuation, analogous
to the Type II case.

The one-loop level reduced scattering amplitude is given by [20]
$$
A_1 ^H(s,t,u; S,T,U) = \int\nolimits_F {d^2\tau\over \tau_2^2}
\prod_{i=1}^4 \int {d^2z_i\over \tau_2} \prod_{i>j} e^{\half s_{ij}
G(z_i, z_j)}\overline{{\cal L}(K_i,z_i,\tau)}
\eqno (6.5)
$$
Here the Green function $G$ and the Mandelstam variables are those of (2.12)
and (2.3) respectively.
The function ${\cal L}$ is the modification required in going from the Type
II to the heterotic case, and is given in terms of the following lattice
sum
$$
\eqalign{
{\cal L}(K_i,z_i,\tau) = & \eta(\tau)^{-24} \prod_{i<j} \biggl
({\vartheta_1(z_i-z_j|\tau )\over
\vartheta'_1(0|\tau)} \biggl )^{K_i\cdot K_j}
\vartheta _\Lambda(\sum _i K_i z_i|\tau ) \cr
&\cr
& \vartheta_\Lambda(x^I|\tau)
= \sum_{p \in \Lambda} e^{i\pi\tau p\cdot p + 2\pi i
p \cdot  x }\cr}
\eqno(6.6)
$$
where $\vartheta_\Lambda$ is the $\Lambda$ - lattice $\vartheta$-function,
for $\Lambda$ the lattices corresponding to $E_8 \times E_8$ or
${\rm Spin}(32)/Z_2$.  Here the Dedekind function is defined by
$$
\eta(\tau) = e^{i\pi\tau/12} \prod_{n=1}^\infty (1-e^{2\pi i \tau n} )
\eqno(6.7)
$$
The above expression may be obtained by straightforwardly translating
the expressions in [17], or by direct calculation using chiral splitting
techniques, as explained in [2,22].  The latter easily generalizes to the
case of other scattering amplitudes as well, and also to higher point
functions.
Notice that ${\cal L}$ is a function only of $K_i^2 = 2$ and $K_i\cdot K_j$,
which may all be expressed in terms of $S,T$ and $U$ only.

As in Type II superstring case, we set $z_4=0$ by translation
invariance, and the remaining integration region may be naturally
decomposed according to the orderings of $0 \leq \Im ( z_1), \Im ( z_2),
\Im ( z_3) \leq \tau _2 $
$$
A_1 ^H(s,t,u;S,T,U) = 2A_1 ^H(s,t;S,T) + 2 A_1 ^H(t,u;T,U) + 2 A_1 ^H(u,s;U,S)
\eqno(6.8)
$$
where $A_1 ^H(s,t; S,T)$ is given by (6.5) with the ordering (2.12), and we
express $u = -s-t$, $U = - S-T-8$.  We study the function
$A_1 ^H(s,t;S,T)$ which is given by equation (6.5) with the definite
ordering
$$
0 \leq \Im (z_1) \leq \Im ( z_2) \leq \Im ( z_3 ) \leq \tau_2
\eqno(6.9)
$$
The one loop amplitude for scattering of four gauge bosons
takes on a form completely analogous to that of Type II
superstring in (3.11) and we have
$$
\eqalign{
A_1 ^H(s,t;S,T) = \int\nolimits_F {d^2\tau\over \tau_2^2}
\int\nolimits_0^ {2\pi}&
\prod_{i=1}^3 {d\alpha_i\over 2\pi} \int\nolimits_0^1 \prod_{i=1}^4 du_i \cr
&\times|q|^{-(su_1u_3 + tu_2u_4)} {\cal R} ^H (|q|^{u_i},\alpha _i;s,t;S,T)\cr}
\eqno(6.10)
$$
where we have as a result of (6.5)
$$
{\cal R} ^H (|q|^{u_i},\alpha _i;s,t;S,T) = {\cal R} (|q|^{u_i},\alpha _i;s,t)
\overline { {\cal R} ^H (|q|^{u _i},\alpha _i;S,T) }
\eqno (6.11)
$$
For the Type II string in (3.16), we employed an expansion in a uniformly
and absolutely convergent power series
in which we treated exactly four of the $\vartheta$ - function factors.
The analogous expansion in the heterotic case is given as follows :
$$
\eqalign{
{\cal R} ^H (|q|^{u_i}, \alpha _i;s,t;S,T)
= \prod _{i=1} ^4 & ( 1 - e^{i \alpha _i} |q| ^{u_i} )^{-s_i/2}
(1 - e^{-i \alpha _i} |q|^{u_i}) ^{-s_i/2 -S_i/2 -2} \cr
& \times \sum _{n_i =0} ^\infty \sum _{\nu _i = -n_i} ^{n_i}
P^{(4)H} _{n_i \nu _i} (s,t;S,T) \prod _{i=1} ^4 |q|^{u_in_i}
e^{i \alpha _i \nu _i} \cr }
\eqno (6.12)
$$
The fundamental hypergeometric functions for the heterotic string case are
given
 by
$$
\Phi (s,t;S,T; |q|^{u_i})
= \sum _{n_i=0} ^\infty \sum _{\nu _i =-n_i} ^{n_i}
P_{\{n_i \nu _i\}} ^{(4)H} (s,t;S,T) ^H \Phi ^H_{\{n_i \nu _i\}} (s,t;S,T;
 |q|^{u_i})
\eqno (6.13)
$$
A formula for the expansion coefficients $P^{(4)H} _{n_i \nu _i}$ is
contained in Appendix C.
The functions $\Phi ^H_{\{n_i \nu _i\}}$ are defined by
$$
\Phi ^H_{\{n_i \nu _i\}} (s,t;S,T;|q|^{u_i} )
= \prod _{i=1} ^4 \Phi_{n_i\nu_i}(s_i,s_i+S_i+4;|q|^{u_i})
$$
with
$$
\Phi ^H_{n \nu} (s,S;|q|^{u} )
=\int _0 ^{2 \pi} {d \alpha  \over 2 \pi}
|q|^{(n -1)u} e^{i (\nu  +1)\alpha _i}
(1-e^{i \alpha } |q|^{u} ) ^{-s/2}
(1-e^{-i \alpha } |q|^{u} ) ^{-S/2}
\eqno (6.14)
$$
As before, it is convenient to introduce the Laplace transform
 $\vp_{n\nu}^H(s,S;\beta)$
by
$$
\Phi_{n\nu}^H(s,S;x^2)=\int_0^{\infty}d\beta\,x^{\beta}\vp_{n\nu}^H(s,S;\beta)
\eqno (6.15a)
$$
and the Mellin transform $f_{n\nu}^H(s,S;\alpha)$ by
$$
f_{n\nu}^H(s,S;\alpha)
=\int_0^1dx\,x^{-\alpha-1}\Phi_{n\nu}^H(s,S;x^2)
\eqno (6.15b)
$$
In terms of hypergeometric functions, the functions $\Phi_{n\nu}^H(s,S;\beta)$
can be expressed as
$$
\Phi_{n\nu}^H(s,S;|q|^{u})=
C_{|\nu+1|}(s_{\nu})F({s\over 2}+(\nu+1)_-,{S\over 2};
|\nu+1|+1;|q|^{2u})|q|^{(n+|\nu+1|-1)u}
\eqno (6.16)
$$
where $s_{\nu}$ is $S$ or $s$, and $(\nu+1)_-$ is 0 or $|\nu+1|$,
depending respectively on whether $\nu+1$ is non-negative or non-positive.
This implies in particular that $f_{n\nu}^H(s,S;\alpha)$
is a meromorphic function with poles
in $s+S$ at positive integers. Furthermore, $\varphi_{n\nu}^H(s,S;\beta)$ can
be
 expanded
in a series of the form (B.18)
$$
\varphi_{n\nu}^H(s,S;\beta)
=\sum_{k=0}^{\infty}
C_{k,\nu}^H(s,S)\delta(2k+n+|\nu+1|-1-\beta)
\eqno (6.17)
$$
with
$$
C_{k,\nu}^H(s,S)=C_k(s+2(\nu+1)_-)C_k(S)C_{k+|\nu+1|}(s_{\nu})\big(C_k(s_{\nu}
+2|\nu+1|)\big)^{-1}
\eqno (6.18)
$$
Analytic continuations can now be carried out along precisely the same lines as
 in
the case of Type II superstrings. Thus if we set
$$
\Psi_{\cn}(s_i,S_i;\beta_i)
=\prod_{i=1}^4\vp_{n_i\nu_i}^H(s_i,s_i+S_i+4;\beta)
$$
we arrive at
\bigskip
\no{\bf Theorem 4}. {\it a) For any fixed integer $N$, the scattering amplitude
 of four massless charged
gauge bosons in the heterotic string $A_1^H(s,t)$ can be expressed as
$$
A^H(s,t;S,T)=\sum_{n_1+\cdots+n_4\leq4N}
\sum_{|\nu_i|\leq n_i}P_{\cn}^{(4)H}(s,t;S,T)A_{\cn}^H(s,t)+M_N^H(s,t)
\eqno (6.19)
$$
where $M_N^H(s,t;S,T)$ is a meromorphic function of both
$s$ and $t$ in the half-space $Re\,s, Re\,t<N$, and the partial amplitudes
$A_{\cn}^H(s,t)$ are given by double dispersion relations
$$
\eqalign{A_{\cn}^H(s,t)=\int_0^{\infty}d\sigma\int_0^{\infty}d\tau~ &
\bigg ( {\rho_{\cn}^H(s,t;S,T;\sigma,\tau)
\over(\sigma-s)(t-\tau)}-\Lambda_{\cn}^H(s,t;S,T;\sigma,\tau)\bigg )\cr
&\qquad+M_{\cn}^H(s,t;S,T)\cr}
\eqno(6.20)
$$
The double spectral
density $\rho _{\cn}^H(s,t;S,T;\sigma,\tau)$ is given by
the formula (4.17), with $\Psi_{\cn}$ there replaced by the
corresponding $\Psi_{\cn}^H$ for the heterotic string.
The subtraction
terms $\Lambda_{n_i\nu_i}^H$ are all globally meromorphic functions.
For each ${\cn}$, the double dispersion relation in (6.20)
defines a holomorphic function of $s$ and $t$ in the complex plane
cut along the positive real axis, beginning at
$(m_1+m_3)^2$ and $(m_2+m_4)^2$ respectively, where
$$
m_i^2=n_i+|\nu_i+1|-1
\eqno(6.21)
$$
b) The above double dispersions can be recast as simple double dispersion
relations of the form (5.1), with simple spectral
densities $\rho _{\cn}(s,t;S,T;\sigma)$ given by (5.5-5.8),
where the expansion coefficients $C_k(s)C_{k+|\nu|}(s)$
and the Mellin transforms $f_{n\nu}(s,\alpha)$ have been replaced
respectively by their heterotic analogues $C_{k,\nu}^H(s,S)$ and
$f_{n\nu}^H(s,S;\alpha)$. In particular, the simple spectral density
$\rho _{\cn}^H(s,t;S,T;\sigma)$ has double poles in
$2s+S+4$ at positive even integers.

c) As illustration, for the forward scattering amplitude $t=0$
for the heterotic string we have with $m_i^2$ as in (6.21)}
$$
\eqalign{\rho_{\cn}^H(s,t=0;S,T;\sigma)=&
\theta(\nu_1+1)\theta(\nu_3+1)C_{\nu_1+1}(T+4)C_{\nu_3+1}(T+4)\cr
&\times\int_0^1du\,\theta(\sigma-{m_3^2\over u}-{m_1^2\over 1-u})
\times\big(u(1-u)\sigma-m_3^2(1-u)-m_1^2u)\cr
&\times f_{n_1\nu_2}^H(s,s+S+4;m_3^2-u\sigma)
f_{n_4\nu_4}^H(s,s+S+4;m_3^2-u\sigma)\cr}
\eqno (6.22)
$$

\bigskip
\bigskip
\bigskip

\centerline
{\bf VII.  APPLICATIONS }

\bigskip

We shall complement the abstract treatment of the analytic continuation
of the one-loop amplitudes in the preceding sections with a discussion
of three simple, but very important applications.

\bigskip
\no
{\it The $i\epsilon$ Prescription}

The first application follows directly from the existence and explicit
forms of the analytic continuation  constructed before.
It concerns the correct specification in string loop amplitudes of an
$i\epsilon$ prescription, familiar from quantum field theory.  The rules
of quantum field perturbation theory naturally provide each propagator
with an $i\epsilon$ prescription, and this prescription uniquely
indicates the analytic structure of amplitudes as a function of external
momenta.  In Appendix D, it is shown how this prescription gives rise to
dispersion relations to one loop order in quantum field theory.

In string perturbation theory however, the formulation is not in terms
of all the propagators for a given string diagram.  In a most standard
fashion, no internal momenta are retained, and the amplitude is written
directly as an integral over moduli.  In this formulation, the
introduction of an $i\epsilon$ prescription if obscure.  Even in a
formulation where internal loop momenta are retained, all internal
propagators are not exhibited, and again the correct introduction of an
$i\epsilon$ prescription is obscure.

With the above analytic
continuation results however, and especially with their expression in
terms of double dispersion relations, the problem
of introducing a consistent $i\epsilon$ prescription
becomes completely
transparent.  Indeed, the correct $i\epsilon$ prescription in the
amplitude is obtained by defining the following partial amplitudes
first:
$$
A_{i\epsilon}(s,t) = A_m (s+i\epsilon,t+i\epsilon) +
\int\nolimits_0^\infty d\sigma \int\nolimits_0^\infty d\tau \q
{\rho (s,t;\sigma,\tau)\over (s-\sigma+i\epsilon)(t-\tau+i\epsilon)}
\eqno(7.1)
$$
The full amplitude is now gotten by adding the three partial amplitudes:
$$
A(s,t,u) = 2A_{i\epsilon} (s,t) + 2A_{i\epsilon}(t,u) + 2A_{i\epsilon }(u,s)~.
\eqno(7.2)
$$
In view of the above construction, these amplitudes are finite, with
appropriate poles and cuts on the positive real $s,t$ and $u$ axes.

\bigskip
\no
{\it Decay Rates of Massive Strings}

The second application concerns special sub parts of the full amplitude.
Of particular interest are the residues of the amplitude at double poles
in a given channel, say the $s$-channel.  In string theory, as in
quantum field theory, the residue at the double pole gives the
mass-shift of the respective string states as well as the decay width of
the incoming massive string.  Massless string states exhibit no
mass shifts in superstring models because of the space-time supersymmetry
non-renormalization theorems [23].

We note that the mass shift receives contributions from both the
meromorphic part of the amplitude as well as from the double dispersion
part.  The decay width on the other hand receives contributions only
from the double dispersion part.  This situation arises from the fact
that the mass-shift is affected by all (virtual) string intermediate
states, whereas the decay width arises from those string states whose
mass is less than the original state.  As a result, decay widths are
easier to calculate from the above construction than the mass shifts.
Below we shall present a complete treatment of the partial and total
decay widths and we shall present an example of a mass-shift calculation
for the lowest massive string state.

We have established in \S V that decay widths essentially arise from a
sum of quantum field theory square diagrams with all string states
moving through the loop.  To compute the partial decay width of a string
state of mass$^2 = 2N$, into 2 string states of masses squared $m_1^2$
and $m_3^2$, we use the factorization of the superstring 4 point
amplitude onto a double pole at $s=2N$.  The $t$-dependence of the
residue may be expanded into Legendre functions indexed by the spin of
the massive string states.

The basic formula for the total decay width of a massive string state of
mass squared $2N$ is given by
$$
\half \Gamma(2N,t) = \Im \lim_{s\rightarrow N} (s-2N)^2
\int\nolimits_0^\infty d\sigma \int\nolimits_0^\infty d\tau
{\rho (s,t;\sigma,\tau) \over (s-\sigma+i\epsilon)(t-\tau) }
\eqno(7.3)
$$
The partial width into string states of mass squared $m_1^2$ and $m_3^2$
can easily be deduced from this amplitude by restricting to the relevant
terms in $\rho (s,t;\sigma,\tau)$.
Note that we included the three point function square on the left hand
side, as this factor must be truncated away from the residue to obtain
the decay rate.

We begin by computing the right hand side of (7.3).  As we are
interested only in the double poles in the $s$-channel, we do not quite
need the full four factor approximation and expansion.  In fact we must
only retain two factors, namely those responsible for producing the
double poles in $s$.  This is valid for $|t|<2$. Thus, we begin by expanding as
 follows:
$$
\eqalign{
\int\nolimits_0^{2\pi}  {d\alpha_1\over 2\pi}
\int\nolimits_0^{2\pi} {d\alpha_3\over 2\pi}
{\cal I}(|q|^{u_i},\alpha _i;s,t)& {\cal R}(|q|^{u _i},\alpha _i;s,t)
 =  \bigl |1-e^{i\alpha_2}|q|^{u_2} \bigr |^{-s}
\bigl |1-e^{i\alpha_3}|q|^{u_3} \bigr |^{-s} \cr
& \times \sum _{n_i =0} ^\infty
\sum_{ {\scriptstyle | \nu_2| \leq n_2}\atop
\scriptstyle | \nu_4 | \leq n_4}
P_{n_i;\nu_2,\nu_4} ^{(2)} (s,t)
|q|^{n_iu_i} e^{i\nu _2\alpha_2+i\nu _4\alpha _4}\cr}
\eqno(7.4)
$$
The coefficient functions $P_{n_i; \nu_2, \nu_4} ^{(2)}(s,t)$
are polynomials in
$s$ and $t$, and formulas are given in Appendix C for their evaluation.
{}From the $\alpha_2$ and $\alpha_4$ integrals of expansion (7.4) and its
subsequent inverse Laplace transform, we derive the {\it  density
function}
$$
\Psi(s,t;\beta_i) = \sum_{n_i=0}^\infty \sum_{\nu_2,\nu_4} \!\!
P_{n_i; \nu_2, \nu_4} ^{(2)} (s,t)
\delta (\beta_1-n_1)\delta(\beta_3-n_3)
\varphi_{n_2\nu_2}(s;\beta_2)\varphi_{n_4\nu_4}(s;\beta_4)
\eqno(7.5)
$$
The single spectral density, as discussed in \S V for this function
$\Psi$, can be readily deduced along the lines of the calculations in
that section.  To obtain the partial width into $m_1^2$ and $m_2^2$ mass
squared string states, we specialize to $n_1 = m_1^2$ and $n_3 = m_3^2$,
dropping the sum over $n_1$ and $n_3$.  We find, for this partial single
spectral density:
$$
\eqalign{
\int_0^\infty d\tau {\rho (s,t;\sigma,\tau)\over \tau-t}\bigg |_{m_1^2,
m_3^2} = & \sum_{\scriptstyle n_2,n_4=0\atop\scriptstyle
\nu_2,\nu_4}^\infty \!\!
P_{n_i; \nu_2, \nu_4} ^{(2)} (s,t)\sum_{p=0}^\infty
{2t^p\over p!(p+3)!}\cr
&\times \int\nolimits_0^1 du
f_{n_2\nu_2}^{(p)} (s;m_3^2 - u\sigma)
f_{n_4\nu_4}^{(p)} (s;m_3^2 - u\sigma) \cr
& \times \theta \biggl (\sigma - {m_3^2\over u} - {m_1^2\over 1-u}\biggr )
\bigl ( \sigma u(1-u) - (1-u) m_3^2 - um_1^2 \bigr ) ^{p+3}\cr}
\eqno(7.6)
$$
The double pole part at $s=2N$ is easily extracted by using the
following residue formula derived in Appendix B :
$$
\lim_{s\rightarrow 2N} (s-2N) f_{n\nu}(s,\alpha) = F_\nu(2N,\alpha-n)
\eqno(7.7)
$$
where the residue is a polynomial in $\alpha$ of degree $2N-2$
$$
F_\nu(2N,\alpha) = - {1\over 2\Gamma(N)^2} \prod_{k=1}^{N-1}
\bigl \{ k+{\alpha +\nu \over 2} \bigr \} \bigl \{ k +
{\alpha - \nu \over 2} \bigr \}
\eqno(7.8)
$$
Furthermore, the 3-point function in (7.3) can be read off from (2.6),
so we obtain

\bigskip

\noindent
{\bf Theorem 5}. {\it The partial decay width of a string state
of mass $2N$ into two string states of masses $m_1^2$ and $m_3^2$ is given by
$$
\eqalign{
\half \Gamma(2N,t) \bigg |_{ {m_{1}^2},{m_{3}^2}} {8\pi\over t^2}
C_N(t)^2 = &
\sum_{\scriptstyle {n_2,n_4 = 0 \atop\scriptstyle \nu_2,\nu_4} }^\infty
P_{m_1^2, n_2, m_3^2, n_4; \nu_2,\nu_4} ^{(2)} (2N,t)
\sum_{p=0}^\infty {2t^p \over p!(p+3)!}
\cr
&\times\int_0^1 \!\!\! du
\times ~\theta \biggl (2N-{m_3^2\over u}
 - {m_1^2\over 1-u} \biggr )\cr
&\times
(2Nu(1-u) - (1-u)m_3^2-um_1^2) ^{ p+3} \cr
&\times F_{\nu_{2}}^{(p)} (2N,m_3^2-2Nu-n_2)
F_{\nu_{4}}^{(p)} (2N,m_3^2-2Nu-n_4)\cr}
\eqno(7.9)
$$
As $F_\nu(2N,\alpha)$ is a polynomial of degree $2N-2$, only terms with
$p \leq 2N-2$ contribute, and the $p$ sum is finite for fixed $N$.}

\bigskip
\no{\it Mass Shifts to One Loop}

The mass shifts to one-loop are given by the real parts of the residues of
$A_1(s,0)$ at the double poles at $s=2N$. Unlike the decay rates, the real
part incorporates also the contributions of the terms we have set aside as
globally meromorphic. In principle, it is possible, although exceedingly
 tedious, to work
out the residues of the globally meromorphic terms at any double pole and sum
them up. Thus it is more efficient to proceed in the manner we describe
next.

For $t=0$ the amplitude $A(s,t,u)$ is given by the expression
$$
A(s,t,u)|_{t=0}
={1\over 2}\int_F
{d^2\tau\over\tau_2^2}\int\prod_{i=1}^4 d^2z_i
\exp {s\over 2}\big ( G(z_1,z_2)+G(z_3,z_4)
-G(z_1,z_3)-G(z_2,z_4)\big )
\eqno(7.10)
$$
The double poles arise from the integral
$$
\int{d^2z_2\over\tau_2}exp\big({s\over 2}G(z_1,z_2)\big)
exp(-{s\over 2}G(z_2,z_4)\big)\times
\int{d^2z_3\over\tau_2}exp\big({s\over 2}G(z_3,z_4)\big)
exp\big(-{s\over 2}G(z_1,z_3)\big)
\eqno(7.11)
$$
over the region when $z_2$ and $z_3$ are close respectively to $z_1$ and
$z_4$. In this region, $exp(-G(z_1,z_2))\sim|z_1-z_2|^2$ and
$exp(-G(z_3,z_4))\sim|z_3-z_4|^2$.
In view of the arguments leading to Lemma B.2 in Appendix B, each of
the two factors in (7.8) is a
globally meromorphic function of $s$, with poles at $s$ integer $\geq 2$.
The residues at $s=2$ are given respectively by $exp(-G(z_2,z_4))|_{z_2=z_1}$
and
$exp(-G(z_1,z_3))|_{z_3=z_4}$ and thus are both equal to $exp(-G(z_1,z_4))$.
Integrating now with respect to $z_1$ and $z_4$, we find in view of
translation invariance that the residue (i.e. the coefficient
of the most singular term in the Laurent expansion) of the forward scattering
amplitude $A_1(s,t)|_{t=0}$ at the double pole $s=2$ is given by
$$
Res \big |_{s=2}=A(-4)
$$
where the function $A(s)$ has been defined by
$$
A(s)=\int_{F}{d^2\tau\over\tau_2^2}\int{d^2z \over\tau_2}
exp\big({s\over 2}G(z,0)\big)
\eqno(7.12)
$$
The integral (7.12) converges only when $0\leq Re\,s<2$, and thus $A(-4)$ has
to
be understood in the sense of analytic continuation. The methods we developed
earlier readily produce
this analytic continuation. In fact, we have
\bigskip
\no{\bf Theorem 6}. {\it a) The function $A(s)$ can be extended to a
meromorphic
 function
of $s$ in the complex plane with a cut along the negative real axis.
All the poles of $A(s)$ are simple, and can occur only at
positive integers starting from $s=2$;

\no b) More precisely, for any fixed positive even integer $N$, $A(s)$ can be
written as
$$
A(s)=s\,{\rm ln}\,s+
\sum_{k=1}^{N/2-1}a_k(s)\int_{2k}^{\infty}{d\alpha\over\alpha^2}{\rm ln}
(s+\alpha)+A_N(s)
\eqno(7.13)
$$
where $A_N(s)$ are holomorphic in the domain $-N<\Re\,s<2$, and the
coefficients
$a_k(s)$ are polynomials in $s$.

\no c) The holomorphic terms $A_N(s)$ in (7.13) can be expressed in terms of
 convergent
integrals. In particular
$$
\eqalignno{16\pi^4A(-4)
=&\int_0^1duL_2(-2u(1-u))
+\int_{-1/2}^{1/2}d\tau_1\int_{\sqrt{1-\tau_1^2}}^1
\int_0^1dye^{4\pi\tau_2y(1-y)}\cr
&\qquad+\int_F{d^2\tau\over\tau_2^2}\int{d^2z\over\tau_2}\big(exp(-2G(z,0))-
e^{4\pi\tau_2y(1-y)}\big)&(7.14)\cr}
$$
where the function $L_2(s)$ is defined by}
$$
L_2(s)=s{\rm ln}s+\int_0^s{dx\over
 x}(e^{-x}-1)+e^{-s}-\big(\int_0^{\infty}{dx\over x}
e^{-x}+\int_0^1{dx\over x}(e^{-x}-1)\big)s
\eqno (7.15)
$$
\bigskip
Since (a) and (b) of Theorem 6 can be established along the same lines as
Theorems 1-4, we discuss only (c). We consider separately the contributions of
 the two
regions $\Im \tau_2\leq 1$ and $\Im \tau_2\leq 1$ of moduli space.
The first gives
an absolutely convergent integral. The second may be reexpressed as
$$
({1\over 4\pi^2})^s\int_1^{\infty}{d\tau_2\over\tau_2^2}
\int_0^1du|q|^{-su(1-u)}\Phi(s,|q|^{2u},|q|^{2(1-u)})|_{s=2}
\eqno(7.16)
$$
where the function $\Phi(s,|q|^{2u},|q|^{2(1-u)})$ is the following
version of hypergeometric functions
$$
\Phi(s,|q|^{2u},|q|^{2(1-u)})
=\int_{-1/2}^{1/2}d\tau_1\int_0^1dx \bigg | \prod_{n=0}^{\infty}
(1-q^ne^{2\pi iz})(1-q^{n+1}e^{-2\pi iz})\bigg |^s
\eqno (7.17)
$$
Comparing with (B.4), it is easy to see that the poles of $\Phi$ now occur
in $s$ at $negative$ integer values. In particular
$\Phi(s,|q|^{2u},|q|^{2(1-u)})$ is holomorphic at $s=2$. Its radius of
convergence in each of the variables $|q|^{2u}$ and $|q|^{2(1-u)}$ is 1. By
symmetry we may restrict our considerations to $u\leq 1-u$. We observe now that
the integral (7.16) over the region $\tau_2u\leq 1$ is convergent, since there
the factor $|q|^{-2u(1-u)}$ remains bounded. In the complementary region
$\tau_2u\geq1$, we have $|q|^{2(1-u)}\leq |q|^{2u}\leq e^{-2\pi}$. Since these
values lie well within the radius of convergence of $\Phi$, it follows that
the expression
$$
|q|^{-su(1-u)}\big[\Phi(s,|q|^{2u},|q|^{2(1-u)})-1\big]|_{s=2}
$$
remains bounded. This reduces the analytic continuation of (7.13) to that of
the simple expression
$$
({1\over 4\pi^2})\int_1^{\infty} {d\tau _2 \over \tau _2^2}
\int_0^1du|q|^{-su(1-u)} \big |_{s=2}
$$
This can be done explicitly in terms of the function $L_2(s)$, (see Appendix
F) giving
$$
({1\over 4\pi^2})^2\int_0^1du\,L_2 (-2u(1-u))
$$
Collecting all the terms gives the expression (7.14).

\vfill\eject

\centerline
{\bf Appendix A: MEROMORPHIC ANALYTIC CONTINUATIONS}

\bigskip

In this appendix, we gather the analytic continuations leading only
to meromorphic functions. The simplest is provided by the following
lemma, the proof of which was given in the discussion of tree-level string
amplitudes in \S II, see (2.9)

\bigskip

\no{\bf Lemma A.1.}
{\it Let $f(x)$ be any smooth function of $x$ on the interval
[0,1]. Then the integral
$$
\int_0^1dx\,x^{-1-s}f(x)
$$
can be extended as a meromorphic function of $s$ in the complex plane, with
simple poles at $s=n$, $n=0,1,2,\cdots$ and residues $-f^{(n)}(0)/n!$.}

\bigskip

\no {\bf Lemma A.2}
{\it Let $\phi(x,y)$ be a smooth function of $(x,y)$ which
does not vanish in the unit square $0\leq x,y\leq 1$.
Then for any function $E(x,y,s)$
smooth in $x$ and $y$ and holomorphic in $s$, the integral
$$
\int\nolimits_0^1 \int\nolimits_0^1 dx dy
(x^2+y^2\phi^2)^{-s/2}E(x,y,s)
$$
can be extended as a meromorphic function in the whole $s$-plane, with poles at
$s = 2,3,4,\cdots$}.
\medskip
\no {\it Proof}.  The basic idea, applying here and elsewhere in
presence of several different scales, here $x$ and $y$, is to order them in
increasing order, and integrate them successively, beginning with the smallest.
More precisely, we write
$$
\int\nolimits_0^1 \int\nolimits_0^1 dx dy =
\int\nolimits_0^1 dx \int\nolimits_0^{x}dy +
\int\nolimits_0^1 dy \int\nolimits_0^{y}dx
$$
In the first region on the right hand side, we change variables
$y\equiv x z$, obtaining
$$
\int\nolimits_0^1 dx~x^{1-s} \biggl \{
\int\nolimits_0^1 dz (1+z^2 \phi^2)^{-s/2} E(x, z x,s)\biggr
\}
$$
The integral between brackets produces a $C^\infty$ function of $x$.
Lemma A.1 thus applies, showing that the $dx$ integral produces in
turn a meromorphic function of $s$ in the whole plane, with poles at
$s=2,3,4,\cdots$.  In the second region, we integrate first with respect
to $x$, using the change of variables $x\equiv y z$, $0 \leq z
\leq 1$.  The proof of Lemma A.2 is complete.
\bigskip
\no{\bf Lemma A.3}. {\it Let $E(w_i)$ be smooth functions
of $w_i$ in the disks $|w_i|\leq1$, $i=1,2,3$. Then the integrals
\medskip
\item{(i)}
$$
\int_{\half \leq |w_1|\leq 1} d^2w_1
|1-w_1|^{-s} E(w_1)
$$
\item{(ii)}
$$
\int_{\half \leq |w_2|\leq |w_1|\leq 1}
d^2w_1 d^2w_2|1-w_1|^{-t}|1-w_2|^{-s}
|1-w_1w_2|^{s+t} E(w_1,w_2)
$$
\item{(iii)}
$$
\eqalign{\int_{\half \leq |w_i| \leq 1\atop
|w_{i+1}|\geq|w_i|}
\prod_{i=1}^3 d^2w_i&|1-w_1|^{-t}|1-w_2|^{-s}
|1-w_3|^{-t}|1-w_1w_2|^{s+t}\cr
&\times|1-w_2w_3|^{s+t} |1-w_1w_2w_3|^{-s}
E(w_1,w_2,w_3)\cr}
$$
can be analytically continued as meromorphic functions of $s$ and $t$
in the whole complex plane. They admit poles of order at most two,
in $s$, $t$, and $u=-(s+t)$ at integers greater or equal to 2.}
\medskip
\no{\it Proof}. For (i), we introduce the variables $x_1\equiv 2
 (1-|w_1|)$ and
$y_1\equiv {\alpha_1\over \pi}$ and observe that $|1-w_1|$ can be written
as
$$
|1-w_1| = \half (x_1^2 + y_1^2 \phi^2(x_1,y_1))^{-1/2} \qqq {\rm
for}~~~0\leq y_1\leq 1
\eqno(A.1)
$$
with a smooth function $\phi(x_1,y_1)$ bounded away
from $0$ for $\half\leq|w_1|\leq 1$.  In this range, the factor
$E(w_1)$ is also a smooth function of $x_1$ and $y_1$.  Thus the
integral in (i) can be rewritten as
$$
\int\nolimits_0^1\!\!\!\int\nolimits_0^1 dx_1dy_1 (x_1^2+y_1^2
\phi^2)^{-s/2}E_1(x_1,y_1, s)
$$
for another function $E_1$ still smooth in $x_1,y_1$ and holomorphic in $s$.
Part (i) follows at once from Lemma A.2.

To deal with (ii) and (iii), there are more scales, and
we refine the arguments of Lemma A.2 as follows.

First by exploiting periodicity in the $\alpha_i$'s $(w_i \equiv
e^{i\alpha_i}|w_i|)$, we can reduce the study of (ii) to that of
an expression of the form
$$
\int\nolimits_{1/2}^1 d|w_i|
\int\nolimits_0^{\pi} d\alpha_i
\big |1-e^{i\alpha_1}|w_1| \big |^{-s}
\big |1-e^{i\alpha_2}|w_2| \big |^{-t}
\big |1-e^{i(\alpha_1\pm\alpha_2)}|
w_1||w_2| \big |^{(s+t)/2} E(w_1,w_2)\eqno (A.2)
$$
Consider first the simpler case where it is
$(\alpha_1-\alpha_2)$ which appears in (A.2).
We again order the
angles as either $\alpha_1\leq\alpha_2$ or $\alpha_2\leq\alpha_1$, and consider
for example the region with
$$
0 \leq \alpha_1 \leq \alpha_2 \leq \pi
$$
We set then
$$
\eqalign{
&0\leq y_1 \equiv \alpha_1/\pi\leq 1\cr
&0\leq y_2 \equiv (\alpha_2-\alpha_1)/\pi\leq1\cr
& x_i\equiv 1-|w_i|\cr}
$$
and exploit the formula (A.1) in the range where all angles $\pi y_i$ are
less than $\pi$ to write (A.2) as
$$
\eqalign{\int_0^{1/2}dx_1\int_0^{x_1}dx_2
\int_0^1dy_1\int_0^{1-y_1}dy_2 &(x_1^2+y_1^2\phi_1^2)^{-s/2}
(x_2^2+(y_1+y_2)^2\phi_2^2)^{-t/2}\cr
&(x_1^2+x_2^2 + y_2^2 \phi_3^2)^{(s+t)/2} E_1(x_1,x_2,y_1,y_2)\cr}
\eqno(A.3)
$$
Here $E_1,\phi_1,\phi_2,\phi_3$ are all
smooth functions, with $\phi_1$, $\phi_2$, $\phi_3$
bounded away from $0$.
The integral (A.3) can now be treated by integrating successive scales.
Thus the scaling $x_2\rightarrow x_1 x_2$ transforms (A.3) into
$$
\eqalign{
\int_0^{1/2}dx_1x_1 \int_0^1 dx_2
\int_0^1dy_1int_0^{1-y_1}
&dy_2  (x_1^2+y_1^2\varphi_1^2)^{-s/2}
(x_1^2x_2^2 + (y_1+y_2)^2\phi_2^2)^{-t/2}\cr
&\times (x_1^2 + y_2^2\phi_4^2)^{(s+t)/2}
E_2(x_1,x_2,y_1,y_2)\cr}
$$
where $E_2$, $\phi_4$ are still smooth functions, with $\phi_4$ bounded away
from 0. If we set $y_i=yk_i$, $y=y_1+y_2$, and replace the $dy_1dy_2$ integral
by a $dydk_1dk_2$ integral, the integrand becomes
$$
\delta(1-k_1-k_2)
(x_1^2+y^2k_1^2\phi_1^2)^{-s/2}
(x_1^2x_2^2+y^2\phi_2^2)^{-t/2}
(x_1^2+y^2k_2^2\phi_4^2)^{(s+t)/2}
E_2
$$
This can be now finished off by comparing the $x_1$ and $y_1$ scales
$$
\int\nolimits_0^{1/2} dx_1 \int\nolimits_0^1 dy =
\int\nolimits_0^{1/2}dx_1 \int\nolimits_0^{2x_1} dy +
\int\nolimits_0^1 dy \int\nolimits_0^{\half y} dx_1
$$
These two terms lead respectively to
$$
\eqalignno{
\int_0^{1/2}dx_1& x_1 \int_0^1 dx_2
\int\nolimits_0^1 dy
\int_0^1 dk_1 \int_0^1 dk_2~\delta(1-k_1-k_2)\cr
&\times (1+y^2k_1^2\varphi_1^2)^{-s/2}
(x_2^2+y^2\phi_2^2)^{-t/2}
(1+y^2k_2^2\phi_4^2)^{s+t}
E_2&(A.4)\cr}
$$
and
$$
\eqalignno{
\int_0^1 dy~y^2
&\int_0^{1/2}dx_1x_1
\int_0^1 dx_2 \int_0^1 dk_1
\int_0^1dk_2\delta(1-k_1-k_2)\cr
&\times (x_1^2+k_1^2\phi_1^2)^{-s/2}
(x_1^2 x_2^2 + \phi_2^2)^{-t/2}
(x_1^2+k_2^2\phi_4^2)^{s+t}
E_2&(A.5)\cr}
$$
In (A.4) only the factor $(x_2^2+y^2 \varphi_2^2)^{-t/2}$ can create
singularities.  In view of Lemma A.2, it produces a meromorphic
function of $t$, with simple poles at most at $t = 2,3,4,\cdots$.

In (A.5) we view the $dk_i$ integrals as
$$
\int_0^1 \int\nolimits_0^1 dk_1 dk_2~\delta(1-k_1-k_2) =
\int_0^{1/2} dk_1\bigg |_{k_2=1-k_1} +
\int_0^{1/2}dk_2\bigg |_{k_1=1-k_2}
$$
Thus of the three factors in (A.5), only one can contribute
singularities. In the range $0\leq k_1\leq 1/2$ it is the factor
$(x_1^2+k_1^2\varphi_1^2)^{-s/2}$, which produces a
meromorphic function
of $s$ with simple poles in $s$ at $s=2,3,4,\cdots$.  In the range
$0\leq k_2 \leq 1/2$, the factor $(x_1^2+k_2^2\varphi_1^2)^{(s+t)}$
produces simple poles in $u$ ($=-(s+t)$) at $2,3,4,\cdots$.  This
completes the treatment of (A.2) when the expression
$\alpha_1-\alpha_2$ appears.
\medskip
We turn now to the case with $\alpha_1+\alpha_2$, and break the
$d\alpha_1 d\alpha_2$ region of integration further into four regions,
defined by $\alpha_i \leq
\pi/2$ or $\alpha_i \geq \pi/2$.  In three regions, the above
arguments apply essentially unmodified.  In the last, ${\pi\over 2}
\leq\alpha_i \leq \pi$, we write the factor with $\alpha_1+\alpha_2$ as
$$
\big|1-e^{i(\alpha_1+\alpha_2)} |\lambda_1||\lambda_2|\big|^{s+t} =
\big(x_1^2 + x_2^2 + (1-{\alpha_1+\alpha_2\over 2\pi})^2
\phi_3^2\big)^{(s+t)/2}
$$
Obviously the other factors are bounded away from $0$.  Letting
$\beta_i = \pi - \alpha_i$ we obtain
$$
\big|1-e^{i(\alpha_1+\alpha_2)}|\lambda_1||\lambda_2|\big|^{s+t} =
(x_1^2 +x_2^2 + (\beta_1 +\beta_2)^2 \phi_4^2)^{(s+t)/2}
$$
with $0 \leq \beta_i \leq {\pi\over 2}$ and $\phi_4$ bounded away from 0.
An easy adaptation of Lemma
A.2 shows at once that this term is meromorphic in $u$, with simple
poles at $u=4,5,6,\dots$.  This establishes (ii).
\medskip
Although more tedious because of the presence of a greater number of scales,
the proof of (iii) is exactly along the same lines. This completes our
discussion of Lemma A.2.
\bigskip
The above scaling arguments also yield easily the following version, which
we need to handle the error terms in the derivation of the double dispersion
relations:
\bigskip
\no{\bf Lemma A.4}. {\it Let $E(\lambda,y)$ and $\phi(\lambda,y)$ be smooth
functions of $\lambda$ and $y$ in the range $0\leq\lambda_i,y\leq1$. Then the
integral
$$
\int_0^1\prod_{i=0}^M d\lambda_i~\lambda_i^{\ell_i}
\int_0^1 dy
\biggl (\prod_{i=0}^M \lambda_i^2 + y^2\varphi^2\biggr )^{-s/2}
y^q E(\lambda,y)\eqno(A.6)
$$
can be analytically continued as a meromorphic function in the $s$-plane.
In fact, for any fixed positive integer $N$, there exist entire functions
$c_{n_0n_1\cdots n_M}(s)$ so that it can be expressed as
$$
\sum_{n_i=q+2+\ell_i}^N
c_{n_{0} n_{1}\cdots n_{M}}(s)
\prod\limits_{i=0}^M(s-n_i)^{-1}
$$
up to a holomorphic function in $Re\, s < N$.}
\bigskip
We can discuss now meromorphic continuations as they appear in the reduction
process stated in Theorem 1.
\bigskip
\no {\bf Lemma A.5}.  {\it Let
${\cal I}(u,\alpha, q)$ be as in (3.26), and let $E(u,\alpha,q)$ be any smooth
 function of $u,\alpha$, and $q$. Dependence on $s$ and $t$ is implicit
in these functions and is assumed to be holomorphic.  Set
$$ {\cal E}(s,t,q) \equiv \int\nolimits_0^{2\pi}
\prod_{i=1}^3{d\alpha_i\over 2\pi} \int\nolimits_0^1 \prod_{i=1}^4 du_i
\delta(1-\sum_{i=1}^4u_i){\cal I}(|q|^{u_i},\alpha _i,q)
E(|q|^{u_i},\alpha _i ,q)
\eqno(A.7)$$
Then
${\cal E}$ can be analytically continued as a meromorphic
function in the whole plane in both $s$ and $t$, with at most poles in
$s,t$ and $-(s+t)$ at integers greater or equal to 2.  The
coefficients of a pole in one variable are entire in the other variable.
Furthermore, ${\cal E}$ is uniformly bounded as $\tau_2\rightarrow\infty$
on any compact set $K$ in $s$ and $t$ away from the poles. Near the poles,
the coefficients of the Laurent expansion as well as the holomorphic parts
are also uniformly bounded as $\tau_2\rightarrow\infty$. The bounds depend
only on $K$ and on bounds for a finite number of derivatives of
$E(|q|^{u _i},\alpha _i,q)$ of the form}
$$
{\rm sup}\tau_2^{-a}|\partial_u^a \partial_\alpha^b E(|q|^{u_i},\alpha _i,q)|
$$
\medskip
\no{\it Proof}.
The region of integration in the integral (A.7) can
be divided into $2^4$ subregions by ordering the $u_i$'s.  Consider for
example the region
$$
0 \leq u_1 \leq u_2 \leq u_3\leq u_4
$$
We choose then $u_1,u_2,u_3$ as independent variables, and enforce the
$\delta(1-\sum\limits_{i=1}^4 u_i)$ constraint by setting $u_4 = 1 -
(u_1+u_2+u_3)$.  Since $u_4 \geq {1\over 4}$ in this region, all factors
in ${\cal I}(|q|^{u_i},\alpha _i,q)$ with $|q|^{2u_4}$ are $C^\infty$
and bounded away from $0$.
They can all be absorbed in a single factor $E(|q|^{u _i},\alpha _i,q)$
which is $C^\infty$ and bounded away from $0$.  Thus ${\cal E}(s,t,q)$ can be
expressed as
$$
\eqalign{
\int\nolimits_0^{2\pi} \prod_{i=1}^3 {d\alpha_i\over 2\pi}
\int \prod_{i=1}^3  &du_i
|1-e^{i\alpha_1}|q|^{u_1}|^{-t}
|1-e^{i(\alpha_1+\alpha_2}|q|^{u_1+u_2}|^{s+t} \cr
&\times
|1-e^{i\alpha_2}|q|^{u_2}|^{-s}
|1-e^{i\alpha_3}|q|^{u_3}|^{-t}
|1-e^{i(\alpha_2+\alpha_3)}|q|^{u_2+u_3}|^{s+t}\cr
&\times
|1-e^{i(\alpha_1+\alpha_2+\alpha_3)}|q|^{u_1+u_2+u_3}|^{-s}
E(|q|^{u_i},\alpha _i,q)\cr}
$$
with the $\Pi du_i$ integral over the region
$0\leq u_1\leq u_2\leq u_3,~~u_1+u_2+2u_3\leq 1$.  If we again introduce
$w_i \equiv e^{i\alpha_i}|q|^{u_i}$, the range spanned by
$|w_1|$ is $|q|\leq |w_i|\leq 1$. We note however that
analytic continuation is
only needed when $|w_i|$ is close to 1, say
$\half \leq |w_i|\leq 1$.  It is necessary to separate out the regions
where $|q| \leq |w_i| \leq \half$, and not change to the new
variable $w_i$ there, since the Jacobian of that change introduces
negative powers of $|q|$ which blow up as $|q|\rightarrow 0$.  Thus we
break up the region $\{0 \leq u_1\leq u_2\leq u_3, u_1+u_2+2u_3 \leq
1\}$ further into
$$
\eqalign{
\{|q| & \leq |w_3|\leq |w_2|\leq |w_1|\leq 1;
|w_1||w_2||w_3|^2\geq |q|\}\cr
& = \{|q|\leq|w_3|\leq|w_2|\leq|w_1|\leq \half;
|w_1||w_2||w_3|^2 \geq |q|\}\cr
&\qquad \cup\{|q|\leq|w_3|\leq|w_2|\leq\half\leq|w_1|;
|w_1||w_2||w_3|^2 \geq |q|\}\cr
&\qquad \cup\{|q|\leq|w_3|\leq\half\leq|w_2|\leq|w_1|;
|w_1||w_2||w_3|^2 \geq |q|\}\cr
&\qquad\cup\{|q|\leq\half\leq|w_3|\leq|w_2|\leq|w_1|;
|w_1||w_2||w_3|^2 \geq |q|\}\cr}
$$
The first region obviously produces an entire function of $s$ and $t$.
The contribution of the next region can be expressed as
$$
\eqalign{
\int\nolimits_0^{2\pi} &\prod_{i=1}^3 {d\alpha_i\over 2\pi}
\int \prod_{i=1}^3 du_i |1-e^{i\alpha_1}|q|^{u_1}|^{-t}
E_1(|q|^{u_i},\alpha _i,q)\cr
& = \int_{\half\leq|w_1|\leq 1}
d^2w_1|1-w_1|^{-1}
\bigg[\int\limits
\prod\limits_{i=1}^2 {d\alpha_idu_i\over
2\pi}|w_1|^{-1}
E_1(w_1,|q|^{u_2}, |q|^{u_3},\alpha_2,\alpha_3,q)\bigg]
\cr}
$$
The integral between brackets is over the region
$$
u_i\geq (2\pi\tau_2)^{-1}{\rm ln}2,\ u_2+2u_3\leq 1+\tau_2^{-1}{\rm ln}|w_1|
$$
It produces a
smooth function of $w_1$, bounded together with its derivatives in
the range $\half \leq |w_1|\leq 1$, uniformly as
$\tau_2\rightarrow\infty$.  Part (i) of Lemma A.3 gives then the
desired statement.

In the same way, the last two regions give rise to integrals of the
form
$$
\eqalign{
\int\int_{\half\leq|w_i|\leq 1}
& d^2w_1d^2w_2
|1-w_1|^{-t}|1-w_2|^{-s}
|1-w_1w_2|^{s+t}\cr
&\times
\bigg[\int\int
{d\alpha_3du_3\over 2\pi}
|w_1|^{-1}|w_2|^{-1}
E_2(w_1,w_2,|q|^{u_3},\alpha_3,q)\bigg]
\cr}
$$
where the integral between brackets is over the region
$$(\pi\tau_2)^{-1}{\rm ln}2\leq 2u_3\leq 1+\tau_2^{-1}({\rm ln}|w_1|
+{\rm ln}|w_2|$$
and of the form
$$
\eqalign{
\int\int\int_
{\half\leq|w_i|\leq 1}\!\!\!\!\!\!\!\!
d^2w_1 d^2w_2 d^2w_3
& |1-w_1|^{-t}|1-w_2|^{-s}|1-w_3|^{-t}|1-w_1w_2
|^{s+t}\cr
&\times |1-w_2w_3|^{s+t}|1-w_1w_2w_3|^{-s}
E_3(w,q)\cr}
$$
Applying respectively (ii) and (iii) of Lemma A.3 gives Lemma A.5.
\bigskip
The bounds formulated in Lemma A.5 allow us to determine when we can integrate
in $\tau$ and still have meromorphic functions. The case of interest to us
is covered by the following:
\bigskip

\no
{\bf Lemma A.6}.  {\it Let }${\cal I}$ {\it be as in Lemma A.5 and let
$E(|q|^{u_i},\alpha _i,q)$
be a smooth function of $u_i$, $\alpha_i$, and $q$ which
is bounded together with all its derivatives as $q\rightarrow0$.
 Then the integral
$$
\eqalign{\int {d^2\tau\over \tau_2^2} \int \prod_{i=1}^4 {d\alpha_i\over 2\pi}
\delta(2\pi\tau_1-\sum_{i=1}^4\alpha_i)\int& \prod_{i=1}^4 du_i~\delta \biggl
 (1-\sum_{i=1}^4 u_i\biggr )
|q|^{-(su_1u_2+tu_2u_4)}\cr
&\times {\cal I}(|q|^{u_i},\alpha _i,q) E(|q|^{u_i},\alpha _i,q)\cr}
$$
over the region $D\cap\{\tau _2(u_1+u_2)\leq 1\}$ can be analytically
continued as a meromorphic function of both $s$ and $t$ in the whole
plane.  It has poles at most in $s$, $t$, and $u=-(s+t)$ at integers, with
residues entire in the remaining variables.}
\bigskip
\no{\it Proof}.
First we note that
Lemma A.5 still holds when the region of integration in $\prod_{i=1}^4du_i$
is modified to $\tau_2 (u_1+u_2) < 1$ in the integral (A.7),
with a few simple modifications of the same proof. Next, we need to show that
the integral $d\tau^2/\tau_2^2$ over $D\cap\{\tau _2(u_1+u_2) < 1\}$ of the
analytic continuations obtained through Lemma A.5 for each fixed $\tau$ is
finite. For each $\tau$, Lemma A.5 shows that the integral is a meromorphic
function in the whole plane with the indicated poles.  Furthermore, the
statement (c) reduces estimating the size of the integral to estimating the
size
of
$$
\prod_{i=1}^4 \tau_2^{-a_i}\biggl ( {\partial\over\partial u_i}\biggr )^{a_i}
|q|^{-su_1u_2 - tu_2u_4}
$$
in the region where $\tau_2 u_i \leq 1$, and of
$
|q|^{-(su_1u_2 + tu_2u_4)}
$
in the domain of integration $\tau_2(u_1+u_2) \leq 1$.  These are
uniformly bounded as $\tau_2 \to \infty$, and thus the
integral over $D\cap\{\tau_2(u_1+u_2) < 1\}$ can be carried out,
giving the desired result.

\bigskip
\bigskip
\bigskip

\centerline
{\bf Appendix B : HYPERGEOMETRIC FUNCTIONS}

\bigskip

We define binomial coefficients by the relation
$$
(1-z)^{-s/2} =
\sum _{k=0} ^\infty C_k (s) z^k \qqq
C_k(s) = {\Gamma (k+s/2) \over \Gamma (s/2) \Gamma (k+1)}
\eqno (B.1)
$$
The hypergeometric function ${}_2F_1 = F$ may be defined by the
absolutely convergent series for $|z|<1$:
$$
F(a, b;c;z) = \sum_{k=0}^\infty \q {\Gamma(a+k) \Gamma(b+k) \Gamma(c)\over
\Gamma(a)\Gamma(b)\Gamma(c+k)~k!} ~~z^k
\eqno(B.2)
$$
For $|z|\geq 1$, it can be defined by analytic continuation, with a
branch cut along $[1,\infty]$.  As a function of $a$, $b$ and $c$, the
function is meromorphic.  $F$ satisfies the famous Gauss hypergeometric
differential equation, which we shall not need here.  $F$ also satisfies
an important ``reciprocity'' relation
$$
\eqalign{
F(a, b;c;z)& = {\Gamma(c)\Gamma(c-a-b)\over \Gamma(c-a)\Gamma(c-b)}~~
F(a, b;a+b+1-c;1 -z)\cr
& + (1-z)^{c-a-b}
{\Gamma(c)\Gamma(a+b-c)\over \Gamma(a)\Gamma(b)}~~
F(c-a, c-b;c-a-b+1;1-z)\cr}
\eqno(B.3)
$$
In particular $F$ and its derivatives at $z=1$ are meromorphic functions
of $a$, $b$, and $c$.
\noindent
{\it Mellin transformation of hypergeometric functions}

In our discussion of the Type II superstring,
we encountered a special case of the hypergeometric function
$$
z^{n+|\nu |} C_{|\nu|} (s)
F \bigl ({s\over 2}, {s\over 2} + |\nu |; |\nu |+ 1; z^2\bigr ) =
\int\nolimits_0^{2\pi} {d\alpha \over 2\pi}
\bigg | 1-z~e^{i\alpha }\bigg |^{-s} e^{i \nu \alpha} z^n
\eqno(B.4)
$$
For the heterotic string, the corresponding integral was given in \S VII.
This expression further generalizes (B.4), but in fact since
$S,\ T,\ U$ are negative even integers, the expressions for the
heterotic string are simply related to (B.4).
The Mellin transform of the special hypergeometric function of (B.4) is
defined by
$$
f_{n\nu} (s,\alpha) = C_{|\nu|} (s)
\int\nolimits_0^1 dx ~~x^{-\alpha-1 + n + |\nu |}
F({s \over 2}, { s \over 2} + |\nu|; |\nu | +1; x^2)
\eqno(B.5)
$$
Note that we have $f_{n\nu} (s,\alpha ) = f_{0\nu} (s,\alpha -n)$,
however, it is convenient to exhibit the $n$-dependence explicitly
in our string formulas.
For $\Re (\alpha) < 0$, and $\Re (s) < 2$, we may write an absolutely
convergent  series expansion
$$
f_{n\nu}(s,\alpha) = \sum_{k=0}^\infty ~~ C_k(s) C_{k+|\nu|}(s)
{1 \over 2k+n+|\nu| -\alpha}
\eqno(B.6)
$$
which allows us to identify this functions with a higher hypergeometric
function evaluated at unity:
$$
f _{n\nu} (s,\alpha) =  {C_{|\nu|} (s) \over
n+|\nu| - \alpha  }
{}~{}_3 F_2 ~\biggl (
 \matrix{{s\over2}, {s\over 2} +|\nu|,
{n +|\nu| -\alpha \over 2};1\cr
1+|\nu |, 1 + {n +|\nu|-\alpha \over 2} \cr} \biggr )
\eqno(B.7)
$$
In string theory, this function arises from the following integral
representation
$$
f_{n\nu} (s, t ) = {1\over 2\pi}
\int\nolimits_{|z|\leq 1} d^2z~~
|z|^{-t-2} |1-z |^{-s} z^{\half (n+\nu )} \bar z ^{\half (n - \nu)}
\eqno(B.8)
$$
The integral extended over the full complex plane equals a ratio
of $\Gamma$-functions, related to the Virasoro-Shapiro amplitude
in string theory.
Splitting the complex plane into the regions inside and outside
the unit circle results in two integrals of the type (B.8),
whose sum is again a ratio of $\Gamma$ functions :
$$
f_{0\nu}(s,t ) + f_{0\nu} (s,u ) = -
{s^2\over 4} ~ {\Gamma(-s/2)\Gamma((-t+\nu)/2)\Gamma((-u+\nu)/2)\over
\Gamma(1+s/2)\Gamma(1+(t +\nu)/2)\Gamma(1+(u+\nu) /2) }
\eqno (B.9a)
$$
This expression may be rewritten in terms of the generalized
hypergeometric functions yielding an interesting identity :
$$
\eqalign{
{1 \over  |\nu | - t}
{}~{}_3 F _2 \biggl (\matrix{ {s\over 2}, {s\over 2} +|\nu| , {|\nu | -t \over
2};1\cr
																			1+|\nu |, 1 + {|\nu |-t \over 2}  \cr} \biggr )
+&
{1 \over  |\nu |+s+t}
{}~{}_3 F _2 \biggl (\matrix{ {s\over 2}, {s\over 2} +|\nu| , {|\nu |+s+t \over
2};1\cr
																			1+| \nu | , 1 + {|\nu | + s +t \over 2} \cr} \biggr ) \cr
& \cr & =
 { \Gamma (1-{s \over 2}) \Gamma (|\nu |+1 ) \Gamma (-{t \over 2}
+{|\nu |\over 2}) \Gamma ({s+t+|\nu |\over 2})
\over
\Gamma (|\nu |+{s \over 2}) \Gamma (1 + { |\nu |+ t \over 2})
\Gamma (1 +{ |\nu |-s -t \over 2}) } \cr }
\eqno (B.9b)
$$

\medskip

\noindent
{\it Analytic Continuation of the Mellin Transform}

We shall now discuss the analytic structure of $f _{n\nu}(s,\alpha)$ and show
that $f$ extends to a meromorphic function throughout the
complex plane with simple poles in
$\alpha$ and $s$ at all positive integers.
We shall derive also the residues at these poles.
We begin by splitting $f = f^+ + f^-$, with $0 < \delta < 1$:
$$
\eqalign{
& f^+ _{n\nu} (s,\alpha) = C_{|\nu|} (s)
\int\nolimits_\delta^1 dx ~~x^{-\alpha +n+|\nu|-1}
F ({s\over 2}, {s\over 2}+|\nu|; |\nu|+1; x^2 )\cr
& f^- _{n\nu} (s,\alpha) = C_{|\nu|} (s)
\int\nolimits ^\delta _0 dx ~~x^{-\alpha +n+|\nu|-1}
F ({s\over 2}, {s\over 2}+|\nu|; |\nu|+1; x^2 )\cr}
\eqno(B.10)
$$
Clearly, $f^+$ is an {\it entire} function of $\alpha$, whereas $f^-$ is
an {\it entire} function of $s$.  Analytic continuation of $f^-$ in
$\alpha$ is standard, and proceeds along the lines of Lemma I of \S A.
In this way we obtain an analytic continuation of $f^-$
throughout the half-plane $\Re (\alpha) < N$
$$
\eqalign{
f^- _{n \nu} (s,\alpha)  = &{\Gamma({\alpha -n -|\nu| \over 2} -N+1)
\over \Gamma({\alpha -n -|\nu| \over 2} +1)} ~~f^{-~(N)}_{n\nu}
(s,\alpha-2N)\cr
& - \sum_{k=0}^{N-1} ~~{\Gamma({\alpha -n-|\nu| \over 2} -k)
\over \Gamma({\alpha -n -|\nu| \over 2} +1)}~~
\delta^{-2\alpha +2k+2n+2|\nu| }~~F^{(k)} \bigl ({s\over 2}, {s\over 2}+|\nu|
;|\nu|+ 1; \delta ^2 \bigr )\cr }
\eqno(B.11)
$$
Analytic continuation of $f^+ _{n\nu}(s,\alpha)$ in $s$ may be
achieved by using the reflection formula (A.2), and we obtain
$$
\eqalign{
f^ + _{n \nu} (s,\alpha) = {1\over 2\cos ~{\pi s\over 2}} &
\sum_{k=0}^\infty
\biggl [
C_k(s) C_{k+s-1} (2|\nu| +2-s)
\int\nolimits_\delta^1 dx~x^{-\alpha-1+n+|\nu|} (1-x^2)^k\cr
& \cr
&- C_k (2|\nu| +2 -s) C_{k-s+1} (s)
\int\nolimits_\delta^1 dx~x^{-\alpha-1+n+|\nu|} (1-x^2)^{k+1-s}
\biggr ]\cr}
\eqno(B.12)
$$
This series is absolutely convergent, since the integrals produce an
exponential suppression factor of order $(1-\delta ^2)^k$ for large $k$; of
course, coefficients for small values of $k$ may have poles themselves,
and these must be isolated first.

\bigskip

\noindent
{\it Pole Structure and Residues}

It is easy to see that poles in $s$ occur only at positive
integers.
The structure of the residue depends upon whether $s$ is even
or odd.
The poles at positive even integers $s$ are similar in nature
to the poles in $\alpha$ of the full function $f_{n\nu}(s,\alpha)$
and contribute to the poles in $s$ in (B.9).
The poles in $s$ at positive odd integers have no analogues
in terms of poles in $\alpha$, and arise because the
region around the singularity at $z=1$ is integrated over only
partially.
Such poles cancel out of (B.9) between the two $f$ terms.

The residues are defined as follows
$$
F_\nu (N,\alpha) = \lim _{s \mapsto N} ~ (s-N) f_{0\nu} (s,\alpha)
\eqno (B.13)
$$
and are given by the following expressions
$$
\eqalign{
F_\nu (2N,\alpha) &= - \half \sum _{k=0} ^{N-1}
{ \Gamma (N-|\nu|) \Gamma ({|\nu| - \alpha \over 2})
\over
\Gamma (N) \Gamma (N-k-|\nu|) \Gamma (N-k) \Gamma (k+1)
\Gamma ({|\nu| - \alpha \over 2} +2 +k -2N) } \cr
&\cr
F_\nu (2N+1,\alpha) &= {1 \over 2 \pi ^2} \sum _{k=0} ^{2N-1}
{ \Gamma (N+|\nu|-\half -k) \Gamma ({|\nu| - \alpha \over 2})
\Gamma (N-\half -k) \Gamma (\half -N)
\over
\Gamma (2N-k) \Gamma (-N+\half +|\nu|)
\Gamma ({|\nu| - \alpha \over 2} -k ) } \cr }
\eqno (B.14)
$$
Using the binomial resummation formula
$$
{\Gamma (A+B-1) \over \Gamma (A+B-N) \Gamma (N)}
=
\sum _{k=0} ^{N-1}
{\Gamma (A) \Gamma (B)
\over
\Gamma (A-k) \Gamma (k+1) \Gamma (B-N+1+k) \Gamma (N-k) }
\eqno (B.15)
$$
we may rewrite the residue expression at even integers
as follows, in terms of a factorized expression
$$
\eqalign{
F_\nu (2N,\alpha) =& -
{\Gamma ( {|\nu| - \alpha \over 2}) \Gamma ( {-|\nu| -\alpha \over 2})
\over
2 \Gamma (N)^2 \Gamma ({|\nu|-\alpha \over 2} +1 -N)
\Gamma ({-|\nu|-\alpha \over 2} +1 -N) } \cr
= & -{1 \over 2 \Gamma (N)^2}
\prod _{k=1} ^{N-1} \bigl \{ k +{\alpha +|\nu| \over 2} \bigr \}
                    \bigl \{ k +{\alpha -|\nu| \over 2} \bigr \} \cr}
\eqno (B.16)
$$
It does not seem that a similar resummation is possible for the
residues at odd positive integers.
Of course, as pointed out previously, the r\^ole of these poles is
very different in string theory and we shall not make
use of their expressions.

\vfill\break

\noindent
{\it Inverse Laplace Transform}

The inverse Laplace transform of (B.4) is defined as follows
$$
e^{(n+|\nu |)\omega } C_{|\nu|} (s)
F ({s\over 2}, {s\over 2}+|\nu |; |\nu |+ 1;
e^{-2\omega} ) \theta(\omega) =
\int\nolimits_0^\infty d\beta ~e^{-\omega\beta} \varphi _{n\nu}(s,\beta)~.
\eqno(B.17)
$$
Formally, we have
$$
\varphi _{n\nu}(s,\beta) = \sum_{k=0}^\infty
C_k (s) C_{k+|\nu|} (s) ~\delta(2k+n+|\nu |-\beta )~.
\eqno(B.18)
$$
and from this expression, it is clear that one may obtain $\varphi$
directly as a discontinuity of the Mellin transform:
$$
\varphi _{n\nu}(s,\beta) \theta(\beta) = {1\over 2\pi i}
\biggl ( f_{n\nu}(s,\beta+i\epsilon) - f_{n\nu}(s,\beta-i\epsilon)\biggr )
\eqno(B.19)
$$
This expression inplies that $\varphi_{n\nu}(s,\beta)$ is entire as a function
of $s$, since the poles of $f_{n\nu}$ cancel out. More precisely, since
$\varphi_{n\nu}(s,\beta)$ should be interpreted as a distribution in $\beta$,
the integrals of $\varphi_{n\nu}$ over any finite $\beta$ interval are entire
in $s$. However, integrals over $[0,\infty)$ produce poles in $s$ at positive
integers. We can see this explicitly in the following formulas which
we require in Section III.2 and which
can be obtained by differentiating (B.17) with respect to
$\omega$ at $\omega=0$
$$
\eqalignno{
\int_0^{\infty}d\beta\varphi_{n\nu}(s,\beta)
=&C_{|\nu|}(s)F({s\over 2},{s\over 2}+|\nu|,|\nu|+1,1)\cr
\int_0^{\infty}d\beta\,\beta\varphi_{n\nu}(s,\beta)
=&C_{|\nu|}(s)\big((n+|\nu|)F({s\over 2},{s\over 2}+|\nu|,|\nu|+1,1)\cr
&\qquad+F'({s\over 2},{s\over 2}+|\nu|;|\nu|+1,1)\big)&(B.20)\cr}
$$
In Section III.2, we actually also need to establish the meromorphicity
in $s$ of expressions of the form (B.18), with however
additional integrations in the $\omega$ variable. The simplest such expression
involves a single integration in $\omega$, and can be rewritten as
$$
\int_0^1d\omega\int_0^{\infty}d\beta
\varphi_{n\nu}(s,\beta)e^{-\omega\beta}
=
\int_0^1d\omega\int_0^{2\pi}{d\alpha\over 2\pi}
|1-e^{i\alpha-\omega}|^{-s}e^{i\nu\alpha}e^{-n\omega}\eqno(B.21)
$$
in view of (B.4). In terms of the complex variable $w=exp(i\alpha-\omega)$, the
 integral
(B.21) reduces to an integral of the form (i) in Lemma A.3. The only difference
is the change of domain of integration from $e^{-1}\leq |w|\leq 1$
to
$1/2\leq |w|\leq 1$, which gives rise only to a manifestly entire correction
term. Thus (B.21) is again a meromorphic function of $s$, with poles
at most at positive integers. A more complicated integral which may arise
is one of the form
$$
\int_0^1{d\omega\over\omega}\int_0^{\infty}d\beta\varphi_{n\nu}(s,\beta)
(e^{-\omega\beta}-1)
=\int_0^1dt\int_0^1d\omega\int_0^{\infty}\varphi_{n\nu}(s,\beta)
e^{-t\omega\beta
 }\eqno(B.22)
$$
The identity (B.4) leads now to integrals with the following type of
singularities
$$
\int_0^1dt\int_0^1d\omega(t^2\omega^2+\alpha^2U^2)^{-s}E(t,\omega,\alpha)
\eqno (B.23)
$$
where $E$, $U$ are smooth functions of all variables, and $U$ is non-vanishing.
The scaling methods of Appendix A apply (c.f. Lemma A.4), and show that (B.23)
is a meromorphic function of $s$. Evidently the above arguments will also
establish the meromorphicity in $s$ of any moments of $\varphi_{n\nu}$
of the form
$$
\int_0^1{d\omega\over\omega^k}\int_0^{\infty}d\beta\,\varphi_{n\nu}(s,\beta)
\big(e^{-\omega\beta}-\sum_{m=0}^{k-1}{(-1)^m(\omega\beta)^m\over m!}\big)
\eqno (B.24)
$$
since higher order Taylor expansions simply lead to the insertion of powers
of $(1-t)$ in (B.23). Finally, the Mellin transform $f$ is given by a Hilbert
transform of the inverse Laplace transform
$$
\int _0 ^\infty { \varphi _{n\nu} (s;\tau - x) \over \tau -t} =
f_{n \nu} (s,t-x)
\eqno (B.25)
$$
This formula clearly exhibits the poles of $f_{n \nu}$ as a function of $t$.

\bigskip
\bigskip
\bigskip

\centerline
{\bf Appendix C.  EXPANSION POLYNOMIAL COEFFICIENTS}

\bigskip

We need to determine three types of expansion coefficients for truncated
$\vartheta $ functions both for the Type II and for the heterotic
strings.

\bigskip

\noindent
{\it Type II Superstrings}

First the monomial expansion
$$
\prod _{i=1} ^4 \int _0 ^{2 \pi} { d\alpha _i \over 2 \pi}
{\cal R} (|q|^{u_i},\alpha _i;s,t) = \sum _{n_i =0} ^\infty
 P^{(0)} _{n_i } (s,t) |q|^{\sum _i n _i u _i}
\eqno (C.1)
$$
Second the expansion with two factors isolated
$$
\eqalignno{\int _0 ^{2 \pi} {d \alpha _1 \over 2 \pi}
\int _0 ^{2 \pi} {d \alpha _3 \over 2 \pi}
{\cal R} (|q|^{u_i},\alpha _i;s,t) = \prod _{i=2,4}
\bigl | 1 - e^{i \alpha _i} |q|^{u_i} \bigr | ^{-s}&
\times\sum _{n_i =0} ^\infty \sum _{|\nu _i | \leq n_i}
P ^{(2)} _{n_i;\nu _2 ,\nu _4 } (s,t) \cr & |q|^{\sum _i n _i u _i}
e^{i\nu _2 \alpha _2 + i \nu _4 \alpha _4}&(C.2)\cr}
$$
Third, the expansion coefficients with four factors isolated
$$
{\cal R} (|q|^{u_i},\alpha _i;s,t) =
\prod _{i=1} ^4 \bigl | 1 - e ^{i \alpha _i} |q|^{u_i} \bigr |^{-s_i}
 \sum _{n_i =0} ^\infty \sum _{|\nu _i | \leq n_i}
P ^{(4)} _{n_i;\nu _i } (s,t) |q|^{\sum _i n _i u _i} e^{i \sum _i
\nu _i \alpha _i}
\eqno (C.3)
$$
To do so, we recast the function ${\cal R}$ of (6.5) in terms
of a product of simple functions
$$
\eqalign{
T\big (z,q;s\big ) & = \prod_{n=0}^\infty \big (1-q^n z\big)^{-s/2}
= \sum_{k=0}^\infty \sum_{l=0}^\infty T_{k,l} (s) q^k z^l\cr
&\cr
T^\ast \big (z,q;s\big ) & = \prod_{n=1}^\infty \big (1-q^n z\big
)^{-s/2} = \sum_{k=0}^\infty \sum_{\ell=0}^k T_{k,l}^\ast (s) q^k
z^{l}\cr}
\eqno(C.4)
$$
We recall that the modulus $q$ is not an independent variable,
but is given by (3.12).
As a result, we have the simple expression
$$
{\cal R} (|q|^{u_i},\alpha _i;s,t ) = \prod _{i=1}^{12}
\bigl | T(w_i,q; s_i) \bigr |^2
\eqno(C.5)
$$
Here, we have defined $w_i$ as in (3.6) and (3.13) $s_i$ as in (3.15).
Upon using the expansion of (C.4) in (C.1-3), we shall
have to collect powers of $w_i$, which are given by
the following sums
$$
\eqalign{
L_1 & = l_1 + l_5 + l_8 + l_9 + l_{10} + l_{12} + \sum _i k_i \cr
L_2 & = l_2 + l_5 + l_6 + l_7 + l_9    + l_{10} + \sum _i k_i \cr
L_3 & = l_3 + l_6 + l_7 + l_8 + l_{10} + l_{11} + \sum _i k_i \cr
L_4 & = l_4 + l_7 + l_8 + l_9 + l_{11} + l_{12} + \sum _i k_i \cr}
\eqno (C.6)
$$
and powers $\bar L_i$ of $\bar w_i$ are given by barred
quantities in sums (C.6).

We may now easily write down the expansions of (C.1-4) in terms of
the expansion coefficients $T$ and $T^\ast$.
First, in terms of monomials only
$$
P ^{(0)} _{n_i } (s,t) =
\sum _ {k_i, \bar k_i \geq 0} \sum _{l_i, \bar l _i \geq 0}
\prod _{j=1} ^{12} T_{k_j l_j} (s_j) T_{\bar k_j \bar l_j} (s_j)
\eqno (C.7a)
$$
with the following restrictions on the sums
$$
n_i = 2 L_i = 2 \bar L_i \qqq i=1,2,3,4
\eqno (C.7b)
$$
Second, the expansion retaining two factors is given by
$$
P ^{(2)} _{n_i;\nu _2 \nu _4 } (s,t) =
\sum _ {k_i, \bar k_i \geq 0} ~\sum _{l_i, \bar l _i \geq 0}~
\prod _{\alpha =2,4} T^\ast _{k_\alpha l _\alpha} (s)
T^\ast _{\bar k_\alpha \bar l _\alpha} (s)
\prod _{j\not=2,4} ^{12} T_{k_j l_j} (s_j) T_{\bar k_j \bar l_j} (s_j)
\eqno (C.8a)
$$
with the following restrictions on the sums
$$
\eqalign{
i= & 1,3 \qqq n_i = 2 L_i = 2 \bar L_i \cr
i= & 2,4 \qqq n_i +\nu _i = 2 L_i \qq n_i - \nu _i = 2 \bar L_i \cr}
\eqno (C.8b)
$$
Third, the expansion coefficients of the expansion with four
factors isolated are given by
$$
P ^{(4)} _{n_i;\nu _i } (s,t) =
\sum _ {k_i, \bar k_i \geq 0} ~\sum _{l_i, \bar l _i \geq 0}~
\prod _{\alpha =1} ^4 T^\ast _{k_\alpha l _\alpha} (s _\alpha)
T^\ast _{\bar k_\alpha \bar l _\alpha} (s _\alpha )
\prod _{j=5} ^{12} T_{k_j l_j} (s_j) T_{\bar k_j \bar l_j} (s_j)
\eqno (C.9a)
$$
with the restrictions on the sums
$$
n_i +\nu _i = 2 L_i \qq n_i - \nu _i = 2 \bar L_i \qqq i=1,2,3,4
\eqno (C.9b)
$$

\bigskip

\noindent
{\it Heterotic Superstrings}

We shall furthermore provide analogous formulas for the
expansion coefficients that occur in the heterotic
four point amplitudes for the scattering of gauge bosons.
These coefficients are defined by the following expressions
First the monomial expansion
$$
\prod _{i=1} ^4 \int _0 ^{2 \pi} { d\alpha _i \over 2 \pi}
{\cal R} ^H (|q|^{u_i},\alpha _i;s,t;S,T) = \sum _{n_i =0} ^\infty
 P^{(0)H} _{n_i } (s,t;S,T) |q|^{\sum _i n _i u _i}
\eqno (C.10)
$$
Second the expansion with two factors isolated
$$
\eqalign{
\int _0 ^{2 \pi} {d \alpha _1 \over 2 \pi}
\int _0 ^{2 \pi} {d \alpha _3 \over 2 \pi}
& ~{\cal R} ^H (|q|^{u_i},\alpha _i;s,t;S,T) \cr
& = \prod _{i=2,4}
\bigl ( 1 - e^{i \alpha _i} |q|^{u_i} \bigr ) ^{-s/2}
\bigl ( 1 - e^{-i \alpha _i} |q|^{u_i} \bigr ) ^{-s/2-S/2-2} \cr
& \times
\sum _{n_i =0} ^\infty \sum _{|\nu _i | \leq n_i}
P ^{(2)H}  _{n_i;\nu _2 ,\nu _4 } (s,t;S,T) |q|^{\sum _i n _i u _i}
e^{i\nu _2 \alpha _2 + i \nu _4 \alpha _4} \cr }
\eqno (C.11)
$$
Third, the expansion coefficients with four factors isolated
$$
\eqalign{
{\cal R} ^H (|q|^{u_i},\alpha _i;s,t;S,T) =
\prod _{i=1} ^4 & \bigl ( 1 - e ^{i \alpha _i} |q|^{u_i} \bigr )^{-s_i/2}
\bigl ( 1 - e ^{-i \alpha _i} |q|^{u_i} \bigr )^{-s_i/2-S_i/2-2} \cr
& \times  \sum _{n_i =0} ^\infty \sum _{|\nu _i | \leq n_i}
P ^{(4)H} _{n_i;\nu _i } (s,t) |q|^{\sum _i n _i u _i} e^{i \sum _i
\nu _i \alpha _i} \cr}
\eqno (C.12)
$$
We may now write down the expansions of (C.8-11) in terms
of the expansion coefficients $T$ and $T^*$.
First for the monomial terms only
$$
P ^{(0)H} _{n_i } (s,t;S,T) =
\sum _ {k_i, \bar k_i \geq 0} \sum _{l_i, \bar l _i \geq 0}
\sum _{p\in \Lambda} \sum _{n=0} ^\infty f_n
\prod _{j=1} ^{12} T_{k_j l_j} (s_j) T_{\bar k_j \bar l_j} (s_j +S_j+4)
\eqno (C.13a)
$$
with the following restrictions on the sums
$$
\eqalign{
n_1 = 2L_1  & = 2 \bar L _1 +  p            ^2 +2n-2 \cr
n_2 = 2L_2  & = 2 \bar L _2 + (p -K_1 )     ^2 +2n-2 \cr
n_3 = 2L_3  & = 2 \bar L _3 + (p +K_3 +K_4 )^2 +2n-2 \cr
n_4 = 2L_4  & = 2 \bar L _4 + (p +K_4 )     ^2 +2n-2 \cr}
\eqno (C.13b)
$$
Second, the expansion retaining two factors is given by
$$
\eqalign{
P ^{(2)H} _{n_i;\nu _2 \nu _4 } (s,t;S,T) =
\sum _ {k_i, \bar k_i \geq 0} ~\sum _{l_i, \bar l _i \geq 0}~
& \sum _{p\in \Lambda} \sum _{n=0} ^\infty f_n
\prod _{\alpha =2,4} T^\ast _{k_\alpha l _\alpha} (s)
T^\ast _{\bar k_\alpha \bar l _\alpha} (s+S+4)\cr
& \prod _{j\not=2,4} ^{12} T_{k_j l_j} (s_j) T_{\bar k_j \bar l_j}
(s_j+S_j+4)\cr}
\eqno (C.14a)
$$
with the following restrictions on the sums
$$
\eqalign{
n_1 =2 L_1  & = 2 \bar L _1 +  p            ^2 +2n-2 \cr
n_3 =2 L_3  & = 2 \bar L _3 + (p +K_3 +K_4 )^2 +2n-2 \cr
n_2+\nu _2 = 2L_2 \qqq n_2 - \nu _2 & = 2 \bar L _2 + (p -K_1 ) ^2 +2n-2 \cr
n_4+\nu_4  = 2L_4 \qqq n_4 - \nu _4 & = 2 \bar L _4 + (p +K_4 ) ^2 +2n-2 \cr }
\eqno (C.14b)
$$
Third, the expansion coefficients of the expansion with four
factors isolated are given by
$$
\eqalign{
P ^{(4)H} _{n_i;\nu _i } (s,t;S,T) =
\sum _ {k_i, \bar k_i \geq 0} ~\sum _{l_i, \bar l _i \geq 0}~
& \sum _{p \in \Lambda } \sum _{n=0} ^\infty f_n
\prod _{\alpha =1} ^4 T^\ast _{k_\alpha l _\alpha} (s _\alpha)
T^\ast _{\bar k_\alpha \bar l _\alpha} (s _\alpha +S_\alpha +4 ) \cr
& \prod _{j=5} ^{12} T_{k_j l_j} (s_j) T_{\bar k_j \bar l_j}
(s_j+S_j+4)\cr}
\eqno (C.15a)
$$
with the restrictions on the sums
$$
\eqalign{
n_1 - \nu _1 & = 2 \bar L _1 +  p            ^2 +2n-2 \cr
n_2 - \nu _2 & = 2 \bar L _2 + (p -K_1 )     ^2 +2n-2 \cr
n_3 - \nu _3 & = 2 \bar L _3 + (p +K_3 +K_4 )^2 +2n-2 \cr
n_4 - \nu _4 & = 2 \bar L _4 + (p +K_4 )     ^2 +2n-2 \cr
&\cr
n_i + \nu _i & = 2      L _i  \qqq i=1,2,3,4 \cr}
\eqno (C.15b)
$$

\vfill\break

\noindent
{\it Recursion Relations for Expansion Coefficients $T$ and $T^*$}

It remains to determine the coefficient functions $T_{kl} (s)$
and $T^\ast _{kl} (s)$.
This may simply be achieved by obtaining a functional relation
$$
\eqalign{
T      \big (qz,q;s\big ) &= \big (1-z \big )^{s/2} T  \big ( z, q; s\big )\cr
T^\ast \big (qz,q;s\big ) &= \big (1-zq\big )^{s/2} T^\ast\big (z,q;s\big )\cr}
\eqno(C.16)
$$
These relations translate into recursion relations on the coefficients
$T$ and $T^\ast$ as follows:
$$
\eqalign{
& T_{k,l}      (s) = \sum_{p=0}^l  C_{l-p} (s) T_{k-p,p} (s)\cr
& \cr
& T_{k,l}^\ast (s) = \sum_{p=0}^{k-l} C_{l-p} (s) T_{k-l,p}^\ast (s)\cr}
\eqno(C.17)
$$
supplemented with the following boundary conditions
$$
T_{k,0} (s) = \delta_{k,0} \qqq T_{0,0}^\ast (s) = 1
\eqno(C.18a)
$$
Furthermore, we should note that by construction, we have
$$
\eqalign{
T_{k,l} & =0 \q T_{k,l} ^\ast =0 \q {\rm for}  ~k<0~ {\rm  or}~ l<0 \cr
T_{k,l} & =0 \qqq \q {\rm for }~ k<l \cr}
\eqno (C.18b)
$$
It is easy to produce explicit solutions for the first few
values of $l$ and all values of $k$, say for the coefficients
$T$.
It is convenient to express these in terms of the binary
function
$$
\theta _k (n) =\left \{
\eqalign{
1 & \q {\rm if}~ k ~{\rm divides }~ n \cr
0 & \q {\rm if } ~k ~ {\rm does~ not~ divide } ~n \cr} \right \}
\eqno (C.19)
$$
We find
$$
\eqalign{
T_{k,1} (s) = & C_1 (s) \cr
T_{k,2} (s) = & \theta _2 (k) C_2 (s) +\half (k+1 -\theta _2 (k)) C_1 (s)^2 \cr
T_{k,3} (s) = & \theta _3 (k) C_3 (s) + \half (3\theta _3 (k-2)
-\theta _2 (k-1) +2 \theta _3 (k-1) +k )C_1(s)C_2(s) \cr
& \qqq +{1 \over 12} (k^2-1-3\theta _2(k) +4 \theta _3(k) ) C_1(s) ^3 \cr }
\eqno (C.20)
$$

\vfill\break

\centerline{\bf Appendix D: ANALYTIC STRUCTURE OF THE BOX DIAGRAM}

\bigskip

We consider $\phi^3$ scalar field theory in $d$ space-time dimensions,
and evaluate the box diagram with arbitrary mass parameters.  The four
external particles are taken to be on-shell and assumed to be massless.
It is easy to write the amplitude using Feynman parameters with the
conventions of Fig. 5.
$$
T(s,t) = 3! \int {d^dk\over (2\pi)^d} \int\nolimits_0^1 du_i
{\delta (1-u_1-u_2-u_3-u_4) \over [k^2+\sum\limits_i m_i^2
u_i-u_1u_3s-u_2u_4t]^4}
\eqno(D.1)
$$
With the help of the exponential representation of the denominator, and
upon carrying out the $k$-integration, we get
$$
T(s,t) = {(2\pi)^4\over (8\pi^2)^{d/2} } \int\nolimits_0^\infty\!\!\!\!\!
{ d\tau \over \tau^{d/2-3}}
\int\nolimits_0^1\!\!\!\!\! du_i ~\delta(1-\sum _i u_i)
\exp \bigl \{ -2\pi\tau  [\sum\limits_i u_i m_i^2-u_1u_3s-u_2u_4t] \bigr \}
\eqno(D.2)
$$
Remarkably, this amplitude has the same structure as the string
amplitude of (4.4), but with fixed value of $\beta_i = m_i^2$ and full
range for $\tau$ from 0 to $\infty$ and space-time dimension $d=10$.
Despite these differences, we may use the
steps explained in \S IV, (4.7) -- (4.17) to derive the double spectral
density of the box diagram.  We find
$$
\eqalign{
T(s,t)  = { (4\pi)^{-d/2} \over \Gamma({d\over 2}-2)}
\int\nolimits_0^1 du_1 & \int\nolimits_0^{1-u_1} du_2 (1-u_1-u_2)^{-3+d/2}
\int\nolimits_0^\infty dx ~x^{{d\over 2}-3}\cr
& \times \big [ (x+x_0+m_3^2 - su_1)(x+x_0+m_4^2-tu_2)\big]^{-1}\cr}
\eqno(D.3a)
$$
with
$$
x_0 = { u_1 m_1 ^2 + u_2 m_2 ^2 \over 1 -u_1 -u_2}
\eqno (D.3b)
$$
The four point amplitude may be represented by a double dispersion
relation in the following way
$$
T(s,t) = \int\nolimits_0^\infty d\sigma \int\nolimits_0^\infty d\tau
{\rho_{\{ m_i^2\} }(\sigma ,\tau)\over (\sigma-s)(\tau-t) }
\eqno(D.4)
$$
and the double spectral density is given by
$$
\eqalign{
\rho_{\{ m_i^2\} }(\sigma,\tau)  = {(4 \pi )^{2 -d/2} \over
4 \Gamma ({d\over 2}-2)} &
\int\nolimits_0^1 du_1  \int\nolimits_0^1 du_2 ~\theta(1-u_1-u_2)
(1-u_1-u_2)^{-3+d/2}\cr
& \times \int\nolimits_{x_{0}}^\infty dx(x-x_0)^{-3 + {d\over 2}} \delta
(x+m_3^2-\sigma u_1) \delta(x+m_4^2 -\tau u_2)\cr}
\eqno(D.5)
$$
It is easy to carry out the $u_1$ and $u_2$ integrations:
$$
\rho_{\{ m_i^2\} }(\sigma,\tau)  = { (4 \pi ) ^{ 2 - d/2} \over
4 \Gamma({d\over 2}-2) ~\sigma\tau}
\int\nolimits_0^\infty dx ~\theta
\biggl (1-{x+m_3^2\over \sigma} - {x+m_4^2\over\tau} \biggr )
\eqno(D.6)
$$
where $M$ is a quadratic function of $x$:
$$
M \equiv - {\sigma+\tau\over\sigma\tau} (x^2-2xA + B^2)\qq
\eqno(D.7a)
$$
with
$$
A = {\sigma\tau - (m_1^2+m_3^2)\tau - (m_2^2+m_4^2)\sigma \over
2(\sigma+\tau) }
\eqno(D.7b)
$$
$$
B^2 = {m_1^2 m_3^2 \tau  + m_2^2 m_4^2\sigma \over \sigma+\tau}\qqq\qqq
\eqno(D.7c)
$$
It is clear that $M>0$ implies that the $\theta$-function in (D.6)
equals 1 as long as $x,~\sigma,~\tau >0$.
Hence the first $\theta$
function under the integral is redundant.  Now the $x$-integral is
easily performed, and we find
$$
\rho_{\{ m_i^2\} }(\sigma,\tau)  = { 4 \pi^{5/2}\over
(4\pi)^{d/2}\Gamma({d-3\over 2}) }
{(\sigma+\tau)^{ {d\over 2}-3}\over (\sigma\tau)^{{d\over 2}-2} }
(A^2-B^2)^{d-5\over 2} \theta (A^2-B^2) \theta (A)
\eqno(D.8)
$$
The spectral representation (D.4) strictly speaking only holds for
$d < 8$, at which point ultraviolet divergences occur and subtractions
are required.  In dimension $d$,
there are $[ {d-6\over 2} ]$ subtractions necessary, which may be handled
by dimensional regularization.
For the case $d=4$, this result coincides
with the one given by Itzykson and Zuber [21] for massless
external particles.

\bigskip

\no
{\it   Support of the Spectral Density for the Box Diagram.}

The support of the spectral density $\rho_{m_i ^2}$ is determined by the
inequalities resulting from the $\theta$ functions in (D.8):
$$
A\geq 0 \qquad A^2-B^2 \geq 0 \qquad \sigma,\tau \geq 0
\eqno(D.9)
$$
This support region is the same in all dimensions of space-time $d\geq
4$, and a non-zero $\rho_{st}$ results in a branch cut in the $s,t$
complex planes.  We shall now solve the above inequalities and
parametrize the domain explicitly.  We begin by solving $A^2-B^2 \geq
0$, temporality ignoring the remaining 3 constraints.  Though the
inequality appears to be quartic in $\sigma$ and $\tau$, it is in
fact quadratic in  $1/\sigma$ and $1/\tau$.  Introducing an
arbitrary mass scale
$m^2$, we can recast $A^2-B^2 \geq 0$ in the form
$$
\eqalign{
\biggl (1- {m_1^2+m_3^2  \over \sigma } - {m_2^2 + m_4^2
\over \tau }\biggr )^2 & +
\biggl ( {m_1^2m_3^2 - m^4\over \sigma m^4} + {m_2^2  m_4^2 - m^2 \over
\tau m^2} \biggr )^2 \cr
& \geq \biggl ( {m_1^2m_3^2 + m^4\over \sigma m^2} + {m_2^2  m_4^2 + m^4
\over \tau m^2} \biggr )^2 \cr}
\eqno(D.10)
$$
This region can be parametrized by two real numbers $x,y$, satisfying
$x^2+y^2 \geq 1$:
$$
\eqalign{
{m_1^2 m_3^2 - m^4\over \sigma } + {m_2^2  m_4^2 - m^4
\over \tau} & =
x \biggl ( {m_1^2 m_3^2 + m^4\over \sigma } + {m_2^2  m_4^2 + m^4
\over \tau} \biggr ) \cr
1 - {m_1^2+m_3^2 \over \sigma } -
{m_2^2+m_4^2 \over \tau } & =
y \biggl ( {m_1^2 m_3^2 + m^4\over \sigma m^2 } + {m_2^2  m_4^2 + m^4
\over \tau m^2} \biggr ) \cr}
\eqno(D.11)
$$
The additional requirements that $\sigma , \tau \geq 0$ imply the
obvious restriction that $x_- \leq x \leq x_+$ with
$$
x_\pm =  {\max \atop \min}
\biggl [
{m_1^2 m_3^2 - m^4\over m_1^2 m_3^2 + m^4} ~~,~~
{m_2^2  m_4^2 - m^4\over m_2^2 m_4^2 + m^4}
\biggr ]
\eqno (D.12)
$$
The restriction that $A \geq 0$ implies $y\geq 0$ in view of
$\sigma,\tau \geq 0$.  It is now straightforward to solve for $\sigma$ and
$\tau$ and we find
$$
\cases{\sigma = + \lambda A_2^{-1}\cr
        \tau  = - \lambda A_1^{-1}\cr} \qquad\qquad
\cases{A_1  =&  $m_1^2 m_3^2 - m^4 - x(m_1^2 m_3^2 + m^4)$\cr
        A_2  =& $m_2^2 m_4^2 - m^4 - x (m_2^2 m_4^2 + m^4)$\cr
        \lambda =& $(m_1^2+m_3^2)A_2 - (m_2^2+m_4^2)A_1-2y Dm^2$\cr
        D = &   $m_1^2 m_3^2 - m_2^2 m_4^2$\cr}
\eqno(D.13)
$$
The scale $m^2$ is arbitrary in this parametrization and may be
picked in the most convenient way; for example if $m_1^2 m_3^2 > m_2^2
m_4^2$, we may choose $m^4 = m_2^2 m_4^2$, so that $x_- = 0$.
Notice that when $D > 0$, we have $A_1 \geq 0, A_2 \leq 0$ and hence
$\lambda \leq 0$; on the other hand when $D < 0$, then $A_1 \leq 0, A_2
\geq 0$ and hence $\lambda \geq 0$.

When $D=0$, the above parametrization is degenerate and this case
may be treated separately.
The function $B^2$ is independent of $\sigma$ and $\tau$
and we get
$$
\rho_{\{ m_i^2\} }(\sigma,\tau)  = { 2^{7-d} \pi^{5/2}\over
(4\pi)^{d/2}\Gamma({d-3\over 2}) }
{(\sigma \tau)^{{d\over 2}-3}\over (\sigma + \tau)^{{d\over 2}-2} }
\prod _{\epsilon = \pm }
\biggl ( 1 -{ (m_1 + \epsilon m_3)^2 \over \sigma } -
           { (m_2 + \epsilon m_4)^2 \over \tau } \biggr )
^{{d-5 \over 2}}
\eqno (D.14)
$$
and the region of (D.9) corresponds to a hyperbola in $\sigma$,
$\tau$ space
$$
1 -{ (m_1 +  m_3)^2 \over \sigma } -
           { (m_2 +  m_4)^2 \over \tau }  \geq 0
\eqno (D.15)
$$
which is represented in Fig. 7.
This region is easily parametrized by standard methods, which
we shall not elaborate on here.
It is easy to see that all these results on the box diagram
agree with those given in [21].

\bigskip

\no
{\it  Parametrization of the Box Graph Amplitude.}

The amplitude for the box graph may now be recovered by using the
double spectral integral (D.4). For the case $D\not=0$, the
parametrization of (D.13) yields a natural solution to the
$\theta$-function inequalities of (D.8), and we get
$$
T(s,t) = 4 m^{2d-4}|D| ^{-1+{d\over 2} } \int\nolimits_{x_{-}}^
{x_+} \!\! dx \int\nolimits_0^\infty \!\! dy~
{\theta(x^2+y^2-1) (x^2+y^2-1)^{d-5\over 2} \over
|\lambda |^{ {d\over 2}-2} (\lambda-s A_2)(\lambda+tA_1) }
(1-x)^{2-{d\over 2} }
\eqno (D.16)
$$
The $x$ and $y$ integrations may be decoupled by introducing the
variable $ y = (1-x^2)^\half z$, and making the definitions:
$$
\eqalign{
& c = {(m_2^2 + m_4^2) A_1-(m_1^2+m_3^2)A_2\over
2D(1-x^2)^\half} \geq 0\cr
& a = c+ {sA_2\over 2D(1-x^2)^\half } \qquad b = c - {tA_1\over
2D(1-x^2)^\half}\cr}
\eqno (D.17)
$$
Assuming that the dimension of space-time is even $d/2 -2 = k$, and $k$
integer $k\geq 0$, we may rewrite $T$ in the following form:
$$
T(s,t) = 2^{-d/2} m^d |D| \int\nolimits_{x_{-} }^{x_+} dx
(1-x)^{{-\half}-{d\over 4} } (1+x)^{-{5\over 2}+ {d\over 4} }I_k (a, b, c)
\eqno (D.18)
$$
where the integral $I_k$ is defined as
$$
I_k(a, b, c)\equiv \int\nolimits_1^\infty dz {(z^2-1)^{k-\half}\over
(z+a)(z+b)(z+c)^k}
\eqno (D.19)
$$
The latter integral is elementary, and we have
$$
I_k(a, b, c) = {1\over (k-1)!} \biggl ( - {\partial\over \partial
c} \biggr )^{k-1} ~
\biggl \{ {J_k(a)-J_k(c)\over (a-b)(a-c)} +
{J_k(b) - J_k(c)\over (b-a)(b-c) } \biggr \}
\eqno (D.20)
$$
with $J_k(a) = 2(a^2-1)^{k-\half}$ Argch($a$).  We have omitted in
$I_k(a,b,c)$ large $z$ subtractions that render the integral convergent.
There are $k$ subtractions to be made which are polynomial in $a,b$
and $c$, which means polynomial in $s,t$ and $u$, and hence correspond
to the usual ultraviolet subtractions in dispersion relations.

For the case $D=0$, this answer may be considerably simplified,
and may be obtained by taking the limit $D\rightarrow 0^+$ in (D.16).
It is convenient to set $m_1m_3=m^2$, so that $x_-=0$
and $x_+ \sim D \rightarrow 0$.
We find
$$
T(s,t) = 2 (m_1m_3)^{d-4} \int _0^1 d\alpha  \int _1 ^\infty  dy~
{ (y^2-1)^{d-5\over 2} \over
|\mu |^{ {d\over 2}-2} (\mu -\alpha s )(\mu -(1-\alpha ) t) }
\eqno (D.21)
$$
where we have used the abbreviation
$$
\mu = (m_1^2 + m_3^3)\alpha + (m_2^2 + m_4^2) (1-\alpha) + 2ym_1m_3
\eqno     (D.22)
$$
This integral is easily evaluated using the formulas (D.19)
and (D.20).

\bigskip
\bigskip
\bigskip

\centerline{\bf Appendix E: FORWARD-LIKE SCATTERING AMPLITUDES}

\bigskip

The forward scattering amplitude is given by $A_{\cn }(s,t)$ with $t=0$.
Its analytic continuation is considerably simpler than that of the full
amplitude $A_{\cn}(s,t)$. In this Appendix we show how to analytically
continue forward-like scattering amplitudes, and how to actually obtain
$A_{\cn}(s,t)$ from them. In effect, this provides another
method of analytic continuation, which is more general than
that of dispersion relations and is perhaps easier to generalize to the case
of higher point functions. For simplicity, we restrict ourselves
to the case $n_i=\nu_i=0$, and denote the amplitude $A_{\{n_i\nu_i\}}(s,t)$
by $A(s,t)$.
\medskip
The forward-like scattering amplitudes $A_{kl}(s;S,T_0)$ we need are defined as
 follows$$
\eqalignno{
A_{kl}(s;S,T_0)=\int_1^{\infty}{d\tau_2\over\tau_2^2}\int_0^1\prod_{i=1}^4
du_i&\delta(1-\sum_{i=1}^4u_i)
|q|^{-(Su_1u_3+T_0u_2u_4+ku_2+lu_4)}\cr
&\prod_{m=0}^1F({s\over 2},{s\over 2};1;|q|^{2u_{2m}})
&(E.1)\cr}
$$
for some fixed value of the parameter $T_0$ between 0 and -1, and $k,l$
positive
integers. We also require
$$
\eqalign{
A(s,t;S,T) = \int\nolimits_1^\infty {d\tau_2\over \tau_2^2}
\int \prod_{i=1}^4 du_i~&\delta \bigl (1 - \sum_{i=1}^4 u_i\bigr )
|q|^{-(Su_1u_3+Tu_2u_4)}\cr
&\times \prod_{i=1}^4F({s_i\over 2},{s_i\over 2};1;|q|^{2u_i})\cr}
\eqno(E.2)
$$
Evidently
$$A(s,t;S,T)|_{S=s,T=t}=A(s,t)$$
so our problem is to analytically continue $A(s,t;S,T)$ in all variables.
For this we require the following properties of $A(s,t;S,T)$
\medskip
\item{(i)} ${\partial^{k+l}A\over\partial S^k\partial T^l}(s,t;S,T)\vert_{
S=S_0,T=T_0}$ is a
globally meromorphic function of $s,t$ for all $k,l$;
\item{(ii)} For each $n$,
the derivatives ${\partial^{2n}A\over\partial^nS\partial^nT}(s,t;S.T)$ can be
analytically continued to the half-space $Re\, S,Re\, T< 2+3n/4$ cut along the
positive real axis and contain no poles in this region;
\item{(iii)} For any fixed $S_0,T_0$ between 0 and -1, and for $s,t$ varying
only in an arbitrary fixed half-space $Re\,s,\ Re\,t<2(N+1)$, we have
$$
A(s,t;S,T_0)
=\sum_{k,l=0}^{\infty}C_k(s)C_k(t)A_{kl}(s;S,T_0)+M_N(s,t)$$
where $M_N(s,t)$ is a meromorphic function of both $s$ and $t$ in the
half-space
$Re\,s,\ Re\,t<2(N+1)$. Similar relations hold between the derivatives
${\partial^mA\over\partial T^m}(s,t;S,T)\vert_{T=T_0}$ and expressions of the
 form
(C.1) with an additional insertion $(u_1u_3)^k$ in the integrand.

\bigskip
Assuming (i)-(iii) for the moment, we make use of
the following version of Taylor's formula, applied to $A(s,t;S,T)$ as a
function of $S$ and $T$:
$$
\eqalign{
A(s,t;S,T) = &- \sum_{k,l=0}^{N-1} {S^kT^l\over k!l!}
{\partial^{k+l} A\over \partial  S^k\partial T^l} (s,t;S_0,T_0)\cr
&+ \sum_{k=0}^{N-1} {S^k\over k!} {\partial^kA\over \partial S^k} (s,t;S_0,T)
+  \sum_{k=0}^{N-1}{T^k\over k!} {\partial^kA\over \partial T^k} (s,t;S,T_0)\cr
&+\biggl [{1\over (n-1)!}\biggr ]^2 \int\nolimits_{S_0}^S \!\! dS_1
\int\nolimits_{T_0}^T \!\! dT_1 (S-S_1)^{n-1} (T-T_1)^{n-1}
{\partial^{2n}A\over \partial S^n\partial T^n }(s,t;S_1,T_1)\cr}
\eqno(E.3)
$$

The first term on the right hand side of (E.3) is
meromorphic by (i), while the last term admits a holomorphic
continuation in the $S$ and $T$ planes cut along the real positive axis
by (ii).
Thus it suffices to analytically continue the third term in (E.3),
the second term being similar with the roles of $s,S$ and $t,T$ interchanged.
In view of (iii), this reduces to the analytic continuation of the
forward-like scattering amplitudes $A_{kl}(s;S,T_0)$, as we had stated earlier.
\medskip
We sketch now the arguments for (i)-(iii) and an analytic
continuation for $A_{kl}(s,t)$.
\bigskip
\no{\it Verification of (i)-(iii).}
\medskip
For $S_0$ and $T_0$ between $-1$ and $0$, the factor
 $|q|^{-S_0u_2u_4-T_0u_1u_3}$
remains bounded. The $du_i$ integrals of hypergeometric functions just produce
poles, as we saw in Appendix B. This implies (i). For (ii) we need an
intuitive understanding of how poles on top of cuts emerge. The cuts arise
from the analytic continuation of the $d\tau_2$ integral. To perform this
integral, we need to expand the hypergeometric functions in
series in $|q|^{2u_i}$. But the series after the integral
is performed converge only when $\Re s,\Re t<2$. This is how the poles at $s=2$
and $t=2$ manifest themselves. To obtain a better radius of convergence, we
consider instead the derivatives $\partial^{2n}/\partial S^n\partial T^n$.
Each derivative $\partial^2/\partial S\partial T$ brings down a factor
$\prod_{i=1}^4u_i$, which extends simultaneously the radii of convergence of
all the factors involved by 3/4. Our general statement for any $n$
follows at once.
\medskip
To see (iii), we exploit the $u_1\leftrightarrow u_3$ symmetry to restrict the
region of integration to $0\leq u_1\leq u_3$. When $|q|^{2u_1}\geq e^{-2\pi}$,
the integrand of $A(s,t;S,T_0)$ is bounded and hence the contribution of this
region is a global meromorphic function. When $|q|^{2u_1}\leq e^{-2\pi}$, we
also have $|q|^{2u_3}\leq e^{-2\pi}$. Both hypergeometric functions can now be
 expanded
into series which are convergent for all $t$
$$
\eqalign{\prod_{m=1}^2F({t\over 2},{t\over 2};1;|q|^{2u_{2m+1}})
&=\sum_{k,l=0}^{N}C_k(t)C_l(t)|q|^{2ku_1+2lu_3}\cr
&\qquad+|q|^{2(N+1)u_1}e_N(|q|^{2u_1})F({t\over 2},{t\over 2};1;|q|^{2u_3})\cr
&\qquad+|q|^{2(N+1)u_3}e_N(|q|^{2u_3})F({t\over 2},{t\over 2};1;|q|^{2u_1})\cr
&\qquad+|q|^{2(N+1)(u_1+u_3)}e_N(|q|^{2u_1})e_n(|q|^{2u_3})\cr}
$$
The contributions of the last three terms in $A(s,t;S,T_0)$ are meromorphic in
the half-plane $Re\,s,\ Re\,t<2(N+1)$. As for the first term, the symmetry
$u_1\leftrightarrow u_3$ allows us to restore the region of integration
back to the full region $0\leq u_1,u_3\leq 1$, up to a globally
meromorphic function. We recognize then the contribution
of each term to $A(s,t;S,T_0)$ to be the expression $A_{kl}(s,t;T_0)$ of (E.1).

\bigskip
\no {\it Analytic Continuation of $A_{kl}(s;S,T_0)$.}
\medskip
Our strategy is to isolate within the integral expression for $A_{kl}(s;S,T_0)$
the region which produces double poles on top of cuts. For this we begin by
rewriting $A_{kl}(s;S,T_0)$ as
$$
\eqalign{A_{kl}(s;S,T_0)
=&\int_1^{\infty}{d\tau_2^2\over\tau_2^2}
\int_0^{\infty}du_1du_2\theta(1-u_1-u_2)
|q|^{(-Su_1+2l)(1-u_1-u_2)+2ku_1}F({s\over 2},{s\over 2};1;|q|^{2u_2})\cr
&\qquad\times\int_0^1{d\lambda_4\over (ln|q|)^2}\theta(\lambda_4-|q|
^{2(1-u_1-u_2)})\lambda_4^{-1-{1\over 2}T_0u_2+{1\over 2}Su_1}F({s\over 2},
{s\over 2};1;\lambda_4)\cr}
$$
Next we break up
the region of integration as follows:
$$
\eqalign{({\rm I})&:\ |q|^{2(1-u_1-u_2)}\leq \lambda_4\leq 1/2\cr
({\rm II})&:\ |q|^{2(1-u_1-u_2)}\leq1/2\leq \lambda_4\cr
({\rm III})&:\ 1/2\leq |q|^{2(1-u_1-u_2)}\cr}
$$
In the region (III), the exponential factor
 $|q|^{(-Su_1+2l)(1-u_1-u_2)+2ku_1}$
remains bounded. Thus this region contributes a meromorphic
function, and we need only consider
the regions (I) and (II).
\medskip
We divide further (II) into
$$
\eqalign{({\rm II}.1)&:\ 2|q|^{2(1-u_1)}\leq |q|^{2u_2}\leq 1/2\cr
({\rm II}.2)&:\ 2|q|^{2(1-u_1)}\leq 1/2\leq |q|^{2u_2}\cr
({\rm II}.3)&:\ 1/2\leq |q|^{2(1-u_1)}\cr}
$$
Again the third region (II.3) contributes a meromorphic function. To treat
 (II.2), we introduce the new
variable $\lambda_2=|q|^{2u_2}$.
If $T_0$ were $0$, we would be
able to carry out the $du_2$ integral explicitly and write (II.2) in terms of
$f_-(s, -\half Su_1+l)f_-(s,-\half Su_1+l) |q|^{-su_1(1-u_1)}$.
The integral in $|q|$ can then
also be carried out explicitly, yielding cuts in terms of ${\rm
 ln}(-su_1(1-u_1))$
and hence ${\rm ln}(-s)$.  When $T_0$ is not $0$, the integral $du_2$
can no longer be carried out explicitly, but the $|q|$ integral needs
to be done in order to exhibit the cut in $s$.  To handle this technical
difficulty, we have to
\medskip
\item{i.}  Integrate {\it first} in $|q|$ to get the analytic cut, and
{\it last} in $\lambda_4$;
\item{ii.}  Show that the resulting dependence on $\lambda_2$ and $u_1$
is $C^\infty$;
\item{iii.}  From this $C^\infty$ dependence in $\lambda_2$ and the
$d\lambda_2$ integral, produce finally the double poles in $s$.
\medskip
Thus we introduce $x \equiv 2\pi(1-u_1)\tau_2
= - {1\over 2} (1-u_1)ln |q|^2$ and rewrite the
contribution from (II.2) as
$$
\eqalign{
(II.2)& = {\pi\over 2}\int_0^1 du_1 (1-u_1)^3
\int_{1/2}^1 d\lambda_2 ~\lambda_2^{-1+{1\over 2} Su_1}
F_s(\lambda_2) \int_{1/2}^1 d\lambda_4~\lambda_4^{-1+({1\over 2} Su_1-l)}
F_s(\lambda_4)\cr
&\qquad \times \int_{{\rm ln}\sqrt 2}^\infty
{dx\over x^4}
exp\big({-x(-Su_1+2l+2k{u_1\over 1-u_1})-
{1\over 2x}(1-u_1)T_0({\rm ln} \lambda_2)
({\rm ln} \lambda_4)}\big)\cr}
$$

Now the integral in $x$ converges initially for $Re s\leq 0$ only. However,
as we shall see in Appendix F,
we can analytically continue it into
$$
\Psi_4(-Su_1+2l+2k{u_1\over 1-u_1},T_0(1-u_1){\rm ln}\lambda_2\,{\rm
 ln}\lambda_4)$$
for $s\in{\bf C}\setminus{\bf R}_+$, $T_0\in{\bf C}$, with $\Psi_4$ the
function
defined in Lemma F.1. Furthermore, for
$T_0\in{\bf R}_-$, the resulting expression is uniformly
bounded in $u_1$ and smooth in ${\rm ln}\lambda_2$, ${\rm ln}\lambda_4$,. Since
 we are in the range $1/2\leq \lambda_2,\lambda_4\leq1$,
this means that it is smooth in $\lambda_2$ and $\lambda_4$.
It follows from Lemma 1 that the
integrals $d\lambda_2d\lambda_4$
in (II.2) can be carried out to give, on top of the cut in $S$ from
$\Psi_4$, a meromorphic function in $s$ with double poles at
$s = 2, 3, 4, \cdots$.
\medskip
It is now clear that double poles on top of cuts in $s$ can only arise from
 regions
where both $\lambda_2\equiv |q|^{2u_2}$ and $\lambda_4\equiv |q|^{2u_4}$
can approach 1 and where the exponent in
$|q|^{-(Su_1u_3+T_0u_2u_4+2ku_1+2lu_3)}$
can be negative. This implies in particular that (II.1) is actually the {\it
 only}
region which gives rise to such singularities. For example, in (II.1),
$|q|^{2u_2}$ stays away from the radius of convergence of
$F({s\over 2},{s\over 2};1;|q|^{2u_2})$.
Thus $F({s\over 2},{s\over 2};1;|q|^{2u_2})$ is entire in $s$, and this region
 can contribute
at most $simple$ poles on top of cuts. As for region (I), we can divide it
into three subregions (I.1), (I.2), and (I.3), in complete analogy
with the division (D.) for (II). In (I.3), the exponent in
$|q|^{-(Su_1u_3+T_0u_2u_4+2ku_1+2lu_3)}$
is bounded, and hence there are no cuts. In both (I.1) and (I.2), the
term $F({s\over 2},{s\over 2};1;|q|^{2u_4})$ is entire in $s$, and hence
these regions produce again at most simple poles on top of cuts.
We provide now some details of the argument.
\medskip
Consider first (II.1). We can expand $F({s\over 2},{s\over 2};1;|q|^{2u_2})$ in
 a uniformly
convergent series, integrate first with respect to $|q|$, and then with
respect to $u_2$ and $\lambda_4$. In terms of the variables $x$ and $\kappa$
defined by
$$
\eqalign{x&=2(1-u_1)\tau_2\cr
|q|^{2u_2}&=2^{-\kappa}[2|q|^{2(1-u_1)}]^{1-\kappa}\cr}
$$
we obtain
$$
\eqalign{
({\rm II}.1) & = \sum_{k_2 = 0}^\infty C_s (k_2)2^{-k_2}
\int_0^1 du_1 (1-u_1)^{3}
\int_0^1 d\kappa~2^{({1\over 2}-\kappa)(-Su_1+2l)}\cr
&\qquad \times \int_{1/2}^1 d\lambda_4~\lambda_4^
{-1+{1\over 2} Su_1-l-{1\over 2}
T_0(1-u_1)(1-\kappa)} F({s\over 2},{s\over 2};1;\lambda_4)\cr
&\qquad \times \int_{{\rm ln} 2}^\infty {dx\over x^4} ({\rm ln} 2
- x)
exp\big({-x((- Su_1+2l)\kappa + 2(1-\kappa)k_2)+2k{u_1\over 1-u_1})}\big)\cr
&\qquad\qquad\qquad\times 2^{2(1-\kappa)k_2}
exp\big({-{1\over 8x} T_0(1-u_1)(1-2\kappa){\rm ln} (
\lambda_4 ){\rm  ln}\, 2}\big)\cr}
\eqno(E.4)
$$
Again we may rewrite the $dx$ integral in terms of its analytic continuations
$\Psi_3$ and $\Psi_4$ (c.f. Appendix F), which are holomorphic for $s\in{\bf
 C}\setminus{\bf R}_+$.
and smooth in all the other parameters $u_1,\kappa$, and $\lambda_4$.
Furthermore, in the range of integration $x\geq {\rm ln}\,2$, the factor
$$
e^{-(1-\kappa)k_2(x-ln\,2)}
$$
guarantees uniform bounds in $k_2$. Thus the series in $k_2$ can be summed, and
only simple poles due to the $\lambda_4$ integration can arise.
\medskip
We turn next to (I.2). Here $F({s\over 2},{s\over 2};1;|q|^{2u_4})$ is entire
in
 $s$ and can be expanded
in a uniformly convergent series in $|q|^{2u_4}$. Introducing the variables
$$
|q|^{2u_4}=2^{-\kappa}[2|q|^{2(1-u_1-u_2))}]^{1-\kappa}\eqno(E.5)
$$
we find
$$
\eqalign{
(I.2)=&\sum_{k_4=0}^{\infty}C_s(k_4)2^{-k_4}\int_0^{\kappa}
\int_{1/2}^1d\lambda_2\lambda_2^{-1-T_0(1-u_1)(1-\kappa)}
F({s\over 2},{s\over 2};1;\lambda_2)\cr
&\times\int_0^1du_1\theta({\rm ln}\,2-2\pi(1- u_1))\lambda_2^{{1\over
 2}((Su_1-2l)\kappa
-T_0(1-u_1)(1-\kappa))}2^{-{\kappa\over 2}(Su_1-2l)}\cr
&\times\int_1^{\infty}{dx\over x^4}(2(1-u_1)x-ln{\lambda_2\over 2})
[2e^{-4\pi x(1-u_1)}\lambda_2^{-1}]^{k_4(1-\kappa)}\cr
&\qquad\times exp\big({-2\pi x(-Su_1+2l)(1-u_1)\kappa+
{1\over 4\pi x}T_0((1-\kappa)({\rm ln}\,\lambda_2)^2
+\kappa \,{\rm ln}\,\lambda_2\,{\rm ln}\,2)}\big)\cr}
$$
As in the previous case, the analytic continuation in $S$ can be obtained
now via the holomorphic functions $\Psi_3$ and $\Psi_4$ of Appendix F.
The series in $k_4$ is easily verified to be uniformly convergent,
and there are at most simple poles on top of cuts, arising from the integration
 in $\lambda_2$.
\medskip
Finally we discuss (I.1). The bounds defining the region of integration
$$
|q|^{2(1-u_1)}\leq {1\over 4},\
|q|\leq e^{-2\pi}
$$
can be expressed using Heaviside functions as
$$
\theta(ln\,2-2\pi(1-u_1))\theta(1-4|q|^{2(1-u_1)})
+\theta(2\pi(1-u_1)-ln\,2)\theta(1-e^{2\pi}|q|)
$$
On the support of the first term we keep the same variables $u_1,|q|$,
while on the support of the second we change variables to
 $u_1,p\equiv|q|^{2(1-u_1)}$.
The net effect is to split the $du_1d\tau_2$ integral as
$$
\int du_1\int {d\tau_2\over2\pi\tau_2}
=\int_0^{1-{ln2\over 2\pi}}du_1\int_0^{e^{-2\pi}}{d|q|\over |q|}
+\int_{1-{ln2\over 2\pi}}^1du_1[2(1-u_1)]^{-1}\int_0^{1/4}{dp\over p}
$$
Both terms can be treated in the same way. The additional factor
$(1-u_1)^{-1}$ in the second term is harmless, because of the presence of
$(1-u_1)^3$ in the integrand. As an example, we discuss the first term.

We expand both $F({s\over 2},{s\over 2};1;q^{2u_2})$ and
$F({s\over 2},{s\over 2};1;q^{2u_4})$ in uniformly convergent series.
Convenient
 variables are now
$\kappa$ defined as before by (E.5), and $\mu$ defined by
$$
|q|^{2u_2}\equiv 2^{-\mu} (2|q|^{2(1-u_1)} )^{1-\mu}
= (4|q|^{2(1-u_1)})^{1-\mu} 2^{-1}
\eqno(E.6)
$$
in terms of which we have

$$
\eqalignno{
({\rm I}.1)& =  \sum_{k_2,k_4= 0}^{\infty} C_s(k_2) C_s(k_4) 2^{-(k_2+k_4)}
\int_0^1 d\kappa \int_0^1 d\mu\,\mu
\int_0^1 du_1 [2(1-u_1)]^3\cr
&\qquad\times \int_{{\rm ln} 2}^{\infty} {dx\over x^4}
({\rm ln} (4 e^{-2\lambda}))^2 2^{-\mu (Su_1-2l)\kappa}
2^{{T_0\over 2}([(1-2\mu)(1-\kappa)+\kappa](1-\mu)(1-u_1)
-(1-u_1)(1-2\mu)(1-\kappa)\mu)}\cr
&\qquad\times exp\big({-x([(-Su_1+2l)\kappa -T_0(1-\mu)
(1-u_1)(1-\kappa)]\mu+2k)}\cr
&\qquad\times (4e^{-2x})^{(1-\mu)k_2}
(4e^{-2x})^{\mu(1-\kappa)k_4}\big) &(E.7)\cr}
$$
As before, the $d\lambda$ integral can be analytically continued as a
holomorphic function in
$s~\in~{\bf C}\setminus {\bf R}_+$, with uniform bounds, and the series in
$k_2, k_4$ converge. This implies that (I.1) contributes no poles on top of
cuts.
\bigskip
In summary, the double poles on tops of cuts come solely from the term (II.2),
which we now work out explicitly.
\bigskip

\no{\it Isolating the coefficients of the double poles.}
\medskip
The integral representing the leading singularities in (II.2) is
complicated because it mixes the $\lambda_2$, $\lambda_4$ variables. However,
for the coefficients of the double poles, these variables disentangle. Indeed
we may write
$$
\eqalign{exp\big({{1\over 2x}(1-u_1)({\rm ln}\lambda_2)(
{\rm ln}\lambda_4)T_0}\big)
=&\sum_{p=0}^N{1\over p!}[{(1-u_1)({\rm ln}\lambda_2)({\rm
ln}\lambda_4)T_0\over
 2x}]^p\cr
&+{1\over N!}[{(1-u_1)({\rm ln}\lambda_2)
({\rm ln}\lambda_4)T_0\over 2x}]^{N+1}\cr
&\qquad\times\int_0^1dt(1-t)^Ne^{t{1\over
 2x}(1-u_1)(ln\lambda_2)(ln\lambda_4)T_0}\cr}
$$
If we substitute this expansion in (II.2), we note that the contribution
of the Taylor remainder on the right hand side has no poles in the half-space
$Re \,S\leq N+2$. Indeed each insertion of a factor ${\rm ln}\lambda$ in the
 integral
$$
\int_{1/2}^1d\lambda\lambda^{-1-\alpha}(ln\lambda)^pF_s(\lambda)
$$
pushes out the location of the first pole in $s$ by 1, since ${\rm ln}\lambda$
 vanishes
of first order at $\lambda=1$. This means that for $S$ in any fixed half-space,
the leading singularities of $A_{kl}(s;S,T_0)$ are of the form
$$
\eqalign{A(s;S,T_0)
&={\pi\over 2}\sum_{p=0}^N{T_0^p\over p!}\int_0^1du(1-u)^{3+p}
\big(\int_{1/2}^1d\lambda\lambda^{-1+{1\over 2}Su-k}
({\rm ln}\,\lambda)^pF_s(\lambda)\big)^2\cr
&\qquad\int_{{\rm ln}\sqrt 2}^{\infty}{dx\over 2^px^{p+4}}
exp\big({-x(-Su+2l+2k{u\over 1-u})}\big)
\cr}
$$
Since we are presently concerned only with the poles on top of cuts
 (equivalently
the imaginary part of the coefficient or the decay rate), we can replace
the $dx$ integral by the singular part of the function $\Psi(s,0)$ of Appendix
A, i.e., by $c_ns^{n-1}ln\,(-s)$. This means that the leading coefficients
of the poles on top of cuts can be written entirely in terms of the
coefficients
$d_{pm}$ defined by
$$
\int_{1/2}^1d\lambda\lambda^{-1-\alpha}({\rm ln}\lambda)^pF_s(\lambda)
=\sum_{m=2+p}^N{d_{pm}\over s-m}+\cdots
$$

\bigskip
\bigskip
\bigskip

\centerline{\bf Appendix F: LOGARITHMIC CUTS}

\bigskip

In this appendix we treat analytic continuations producing logarithmic cuts. We
 also formulate some
bounds which are needed for convergence issues.
The integrals which concern us here are of the form
$$L_n(s,t)=\int_{\alpha}^{\infty}{dx\over x^n}e^{-2\pi xs+{1\over 8\pi x}t}
$$

\no{\bf Lemma F.1.} {\it a)
The function $L_n(s,t)$ can be analytically continued as a holomorphic function
 of
 both
$s$ and $t$ for
$$s\in{\bf C}\setminus{\bf R}_+,\ t\in {\bf C}$$
b) There exist constants $C$ and, for each fixed compact set $K$, constants
 $C_K$
so that
$$
\eqalign{
|L_n(s,t)|&\leq C({|s|\over|Im\, s|})^{n-1}e^{|st|/2\pi\alpha|\Im s|},\ n>1\cr
|L_n(s,t)|&\leq C_K(1+{|Re\, s|\over |Im\, s|}+{\rm ln}\,|s|),\ s,t\in K\cr}
$$
c) Furthermore}
$$
\eqalign{
|L_n(s+\mu,t)|&\leq Ce^{-2\pi\mu\alpha},\ \mu\geq 0,n>1\cr
& \leq Ce^{-2\pi\mu\alpha}{1\over\mu}{\rm ln}(1+{\mu\over\alpha}),\mu\geq
 0,n=1\cr}
$$
\bigskip
\no {\it Proof}. To obtain the analytic continuation in (a), we note that
for $s$ a positive number, the integral can be rewritten as
$$
e^{-2\pi\alpha s}s^{n-1}\int_0^{\infty}{d\lambda\over(\lambda+\alpha s)^n}
e^{-2\pi\lambda+{st\over 8\pi(\lambda+\alpha s)}}\eqno(F.1)
$$
It is evident that for $s,t$ in the region described in (a), this expression
 converges,
giving the desired analytic continuation. The remaining statements (b) and (c)
 are also easy to verify
starting from (F.1).
\medskip
For $t=0$ we can write down explicit formulas for $L_n$ in terms of logarithms.
Thus
$$
\int_1^{\infty}{dx\over x}e^{-xs}=-{\rm ln}\,s+\int_0^1{dx\over x}(1-e^{-xs})
+\int_0^1{dx\over x}(e^{-x}-1)+\int_1^{\infty}{dx\over x}e^{-x}
$$
as can be seen by differentiating in $s$ and integrating back. Similarly
$$
\int_1^{\infty}{dx\over x^2}e^{-xs}=-s\,{\rm ln}\,s+(1+c_2)s-1+\int_0^1{dx\over
 x^2}(e^{-xs}-1
+xs)\eqno(F.2)
$$
For later reference, we note that
$$
\int_0^{\infty}dx\big({x^2\over (x+s)^2}-1+{2s\over x+1}\big)=s+2s{\rm ln}\,s
$$
so that (A.2) can be rewritten as
$$
\int_1^{\infty}{dx\over x^2}e^{-xs}
=-\int_0^1{dx\over x^2}(e^{-xs}-1+xs)+{1\over 2}\int_0^{\infty}dx
\big({x^2\over (x+s)^2}-1+{2s\over x+1}\big)-(c_2+{3\over 2})s+1
\eqno(F.3)
$$
More generally, we have
$$
\int_1^{\infty}{dx\over x^n}e^{-xs}={(-1)^{n-1}s^{n-1}{\rm ln}\,s\over
\Gamma(n)}+ ~{\rm entire ~function}
$$
The function $L_2(s,0)$ is the one appearing in (7.15).

\vfill\break

\centerline{{\bf ACKNOWLEDGEMENTS}}

\bigskip

E. D'H. gratefully acknowledges the hospitality extended to him
by the Institute for Theoretical Physics at Santa Barbara,
and by the Yukawa Institute for Theoretical Physics at Kyoto
University.  In particular, thanks are due to
Takeo Inami, Satoshi Matsuda and Hirosi Ooguri.
D.H. P. would like to thank the Institute for Theoretical and Experimental
Physics in Moscow,
the organizers of the Alushta conference, and especially Andrei Marshakov
and Alexei Morozov
for the very warm hospitality extended to him during his
stay in Russia and Crimea.

\bigskip
\bigskip
\bigskip

\centerline{{\bf FIGURE CAPTIONS}}

\bigskip

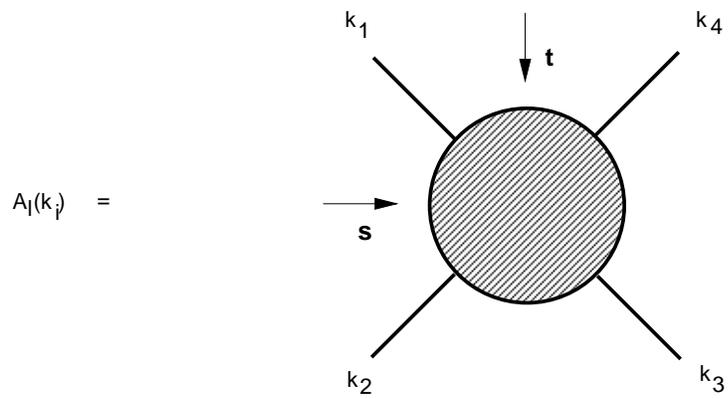
\item{Fig. 1} One loop superstring amplitude as an integral of vertex
operators over the torus.
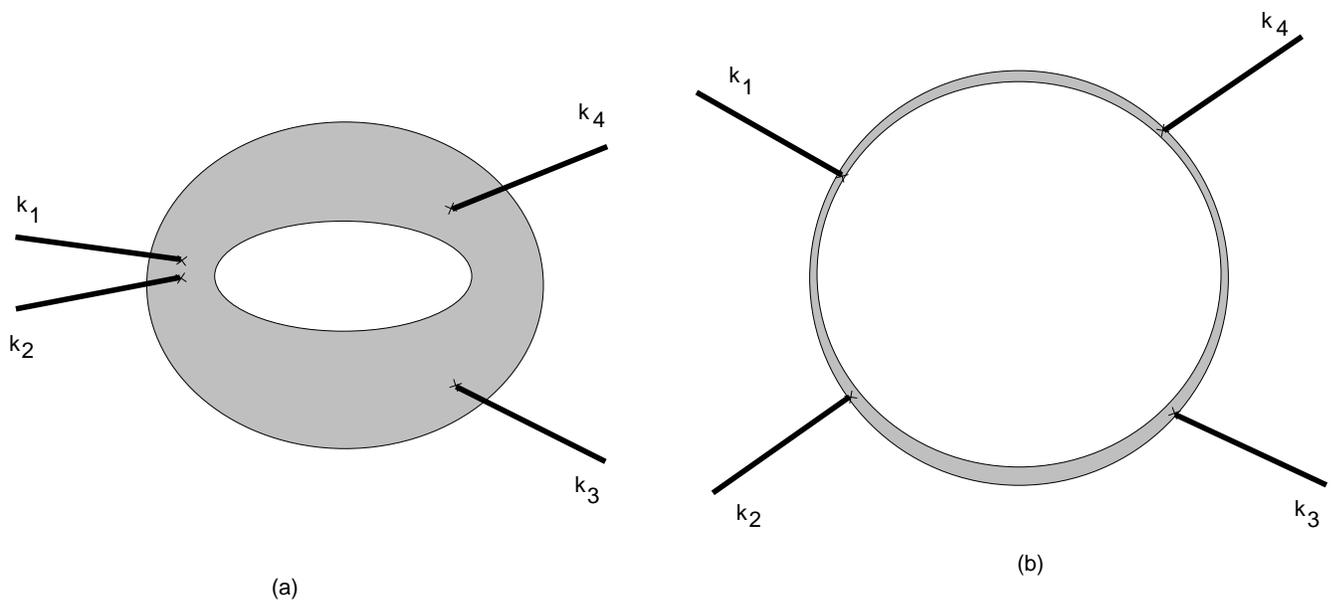
\item{Fig. 2} Singularities in the one loop integral representation
\itemitem{(a)} when two vertex operators collide and $z_i \sim z_j$
\itemitem{(b)} when $\tau _2 \mapsto \infty$ and the torus degenrates to a thin
wire.
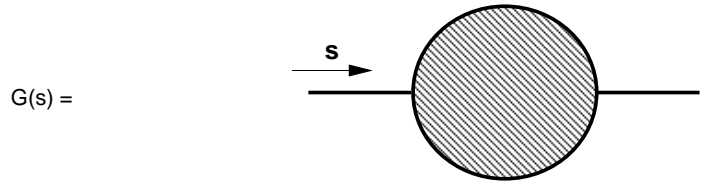
\item{Fig. 3} The off-shell two point function in quantum field theory.
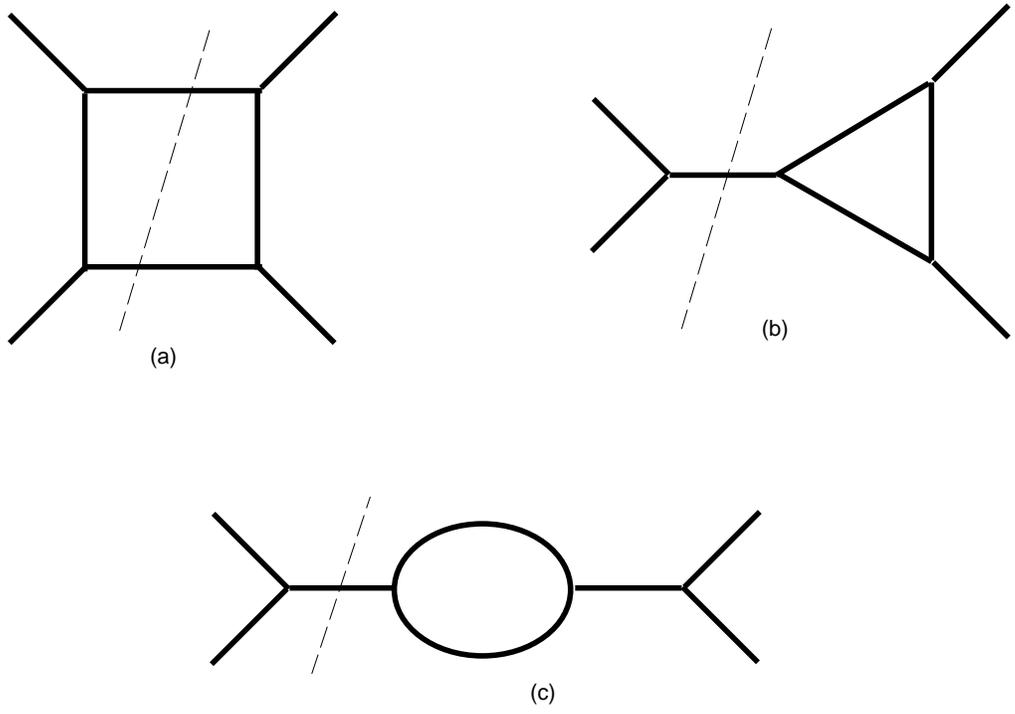
\item{Fig. 4} One loop superstring singularities expected in the analytically
continued amplitudes.
\itemitem{(a)} branch cuts
\itemitem{(b)} single poles with branch cuts on top of the pole
\itemitem{(c)} double poles with real and imaginary parts to the two point
function.
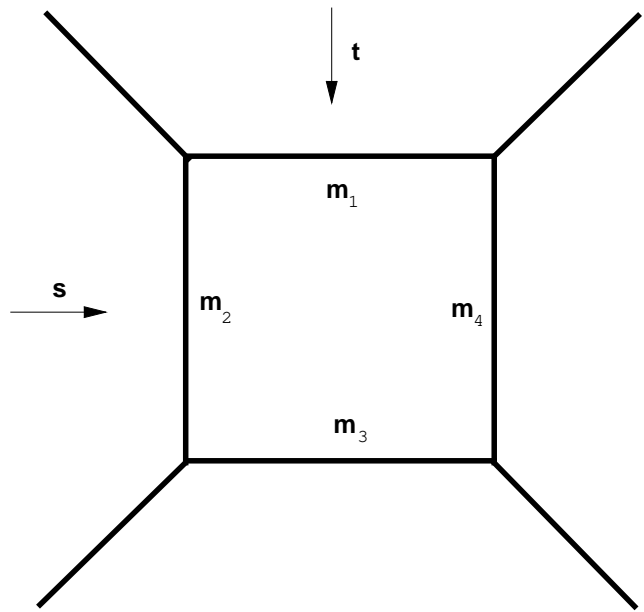
\item{Fig. 5} The box diagram in a $\phi ^3$ like theory with arbitrary mass
assignments on all propagators.
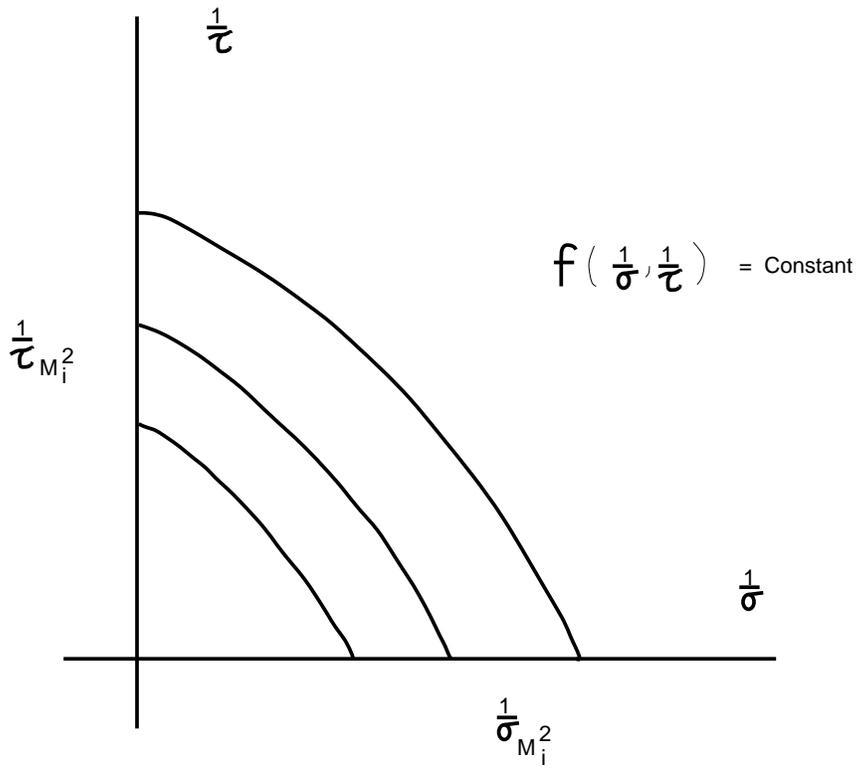
\item{Fig. 6} Constant $f$ contours and location of extrema.
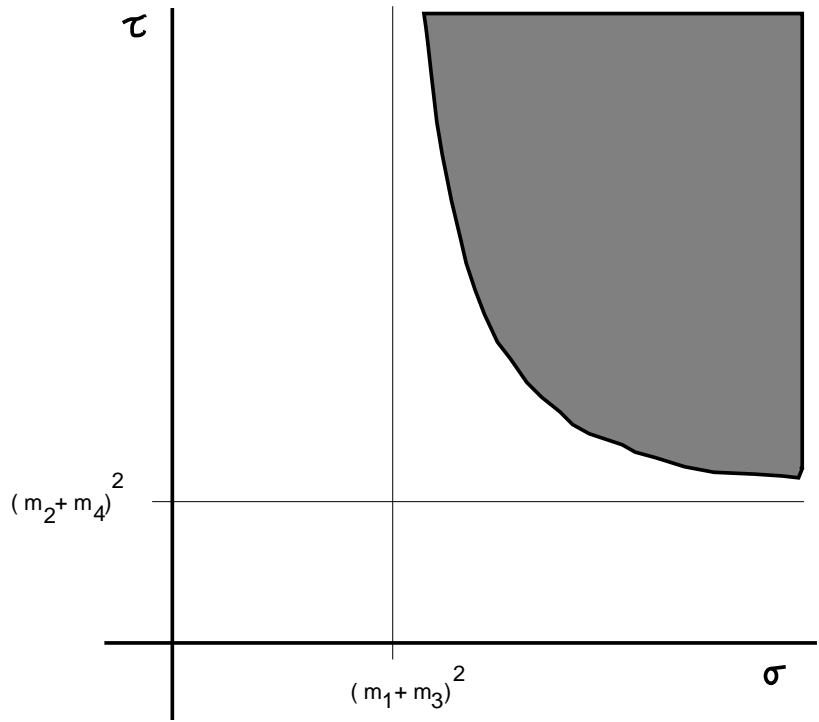
\item{Fig. 7} Support of the spectral density function $\rho _{\{ m_i ^2 \}}$.
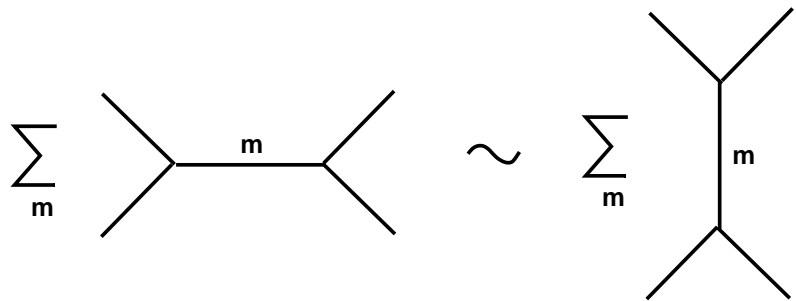
\item{Fig. 8} Duality and its connection with analytic continuation to tree
level.
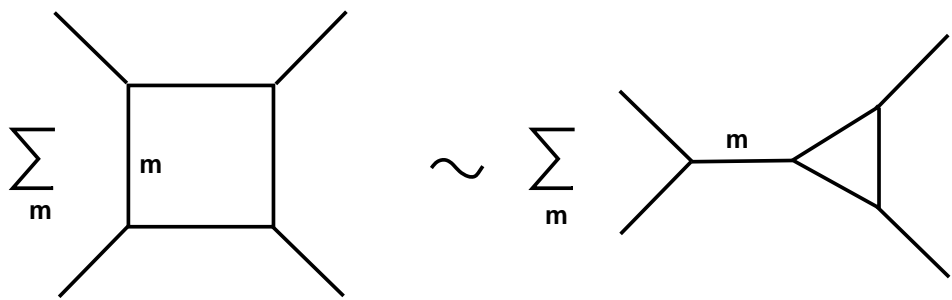
\item{Fig. 9} Appearance of poles due to analytic continuation of an infinite
sum of box diagrams.

\bigskip
\bigskip
\bigskip

\centerline{\bf REFERENCES}

\bigskip

\item{[1]} M.B. Green, J. Schwarz, and E. Witten, {\it Superstring Theory},
	Cambridge University Press, 1987;
	A.M. Polyakov, {\it Gauge Fields and Strings},
	Harcourt Brace Jovanovich, New York, 1987
\item{[2]} E. D'Hoker and D.H. Phong, Rev. Mod. Physics 60 (1988) 917.
\item{[3]}  K. Aoki, E. D'Hoker, and D.H. Phong, Nucl. Phys. B 342 (1990) 149.
\item{[4]} S. Mandelstam in {\it Unified String Theories, Proceedings
	of the 1985 Santa Barbara Workshop} eds. M.B. Green and D.J. Gross,
	World Scientific Publishers, 1986;
	S. Mandelstam, in {\it Second Nobel Symposium on Elementary
	Particle Physics}, Mastrand, 1986.
\item{[5]} N. Berkovits, Phys. Lett. 300B (1993) 53;
 Nucl. Phys. B408 (1993) 43
\item{[6]} G. Chew, {\it The Analytic S-matrix}, W. A. Benjamin (1966);
	R.J. Eden, P.V. Landshoff, D.I. Olive and J.C. Polkinghorne, {\it The
	Analytic S-Matrix}, Cambridge University Press (1966);
 I.T. Todorov, {\it Analytic Properties of Feynman Diagrams in Quantum Field
Theory}, Pergamon Press, 1971
\item{[7]} E. D'Hoker and D.H. Phong, Phys. Rev. Lett. 70 (1993) 3692;
 E. D'Hoker and D.H. Phong, ``Dispersion Relations in String
	Theory'', to be published in Theor. Math. Phys. and
 Columbia-YITP-UCLA/93/TEP/45 preprint (1993).
\item{[8]} M.B. Green and J. Schwarz, Phys. Lett. 109 B (1982) 444
\item{[9]}	J. Polchinski, Comm. Math. Phys. 104 (1986) 37;
	E. D'Hoker and D.H. Phong, Nucl. Phys. B278 (1986) 225;
	G. Moore, P. Nelson and J. Polchinski, Phys. Lett.  B169 (1986) 47
\item{[10]} D. Gross, in {\it Proceedings of the Twenty-Fourth
	International Conference on High Energy Physics, Munich 1988},
	ed. R. Kotthaus and J.H. Kuhn (Springer Verlag, 1989)
\item{[11]} J.L. Montag and W.I. Weisberger, Nucl. Phys. B 363 (1991) 527;
	J.L. Montag, Nucl.Phys. B393 (1993) 337
\item{[12]} K. Amano, Nucl. Phys. B328 (1989) 510.
\item{[13]} B. Zwiebach, Nucl. Phys. B390 (1993) 33;
  H. Hata and B. Zwiebach, Ann. Phys. 229 (1994) 177;
  B. Zwiebach, in {\it Les Houches Summer School lecture notes, 1992}
  hep-th 9305026 preprint (1993)
\item{[14]} M.B. Green, J.H. Schwarz and L. Brink, Nucl. Phys. B198 (1982) 474.
\item{[15]} A. Berera, Nucl. Phys. B411 (1994) 157;
            A. Berera, Phys. Rev. D49 (1994) 6674
\item{[16]} S. Mandelstam, Phys. Rev. 112 (1958) 1344; 115 (1959) 1741.
\item{[17]} D.J. Gross, J.A. Harvey, E. Martinec and R. Rohm, Nucl. Phys.
	B256 (1985) 253; Nucl. Phys. B267 (1986) 75.
\item{[18]} V. A. Miranski, V. P. Shelest, B. V. Struminsky,
	and G. M. Zinovjev,
	Phys. Lett. B43 (1973) 73; C. B. Chiu and S. Matsuda,
	Nucl. Phys. B134 (1978) 463.
\item{[19]} K. Amano and A. Tsuchiya, Phys. Rev. D39 (1989) 565;
	B. Sundborg, Nucl. Phys. B319 (1989) 415;
	D. Mitchell, N. Turok, R. Wilkinson, and P. Jetzer,
	Nucl. Phys. B315 (1989) 1;
	D. Mitchell, B. Sundborg, N. Turok, Nucl.Phys. B335 (1990) 1990;
	N. Marcus, Phys. Lett. B219 (1989) 265
\item{[20]} G. Chalmers, E. D'Hoker and D.H. Phong, to be published
\item{[21]} C. Itzykson and J.-B. Zuber, {\it Quantum Field Theory},
 Mc Graw Hill publ. 1980; J.D. Bjorken and S.D. Drell, {\it Relativistic
Quantum
Fields}, Mc Graw Hill (1965)
\item{[22]} E. D'Hoker and D.H. Phong, Comm. Math. Phys. 125 (1989) 469
\item{[23]} E. Martinec, Phys. Lett. 171B, (1986) 189

\end